\documentclass[iop, onecolumn]{emulateapj}
\usepackage{amsmath}
\usepackage{rotating}
\usepackage{natbib}
\shorttitle{Atmospheric Circulation and Clouds of HD~80606b}
\shortauthors{Lewis et al.}

\begin{document}

\title{Atmospheric Circulation and Cloud Evolution on the Highly Eccentric Extrasolar Planet HD~80606b}

\author{
Nikole K. Lewis\altaffilmark{1,2}, Vivien Parmentier\altaffilmark{3,8}, 
Tiffany Kataria\altaffilmark{4}, 
Julien de~Wit\altaffilmark{5}, Adam P. Showman\altaffilmark{3}, 
Jonathan J. Fortney\altaffilmark{6}, Mark S. Marley\altaffilmark{7}
}

\altaffilmark{1}{Space Telescope Science Institute, Baltimore, MD 21218, USA;
\email{nlewis@stsci.edu}}
\altaffiltext{2}{Department of Earth and Planetary Sciences, Johns Hopkins University, Baltimore, MD 21218, USA}
\altaffiltext{3}{Department of Planetary Sciences and Lunar and Planetary Laboratory, The University of Arizona, Tucson, AZ 85721, USA}
\altaffiltext{4}{Jet Propulsion Laboratory, California Institute of Technology, 4800 Oak Grove Drive,
Pasadena, CA 91109, USA}
\altaffiltext{5}{Department of Earth, Atmospheric and Planetary Sciences, Massachusetts Institute of Technology, Cambridge, MA 02139, USA}
\altaffiltext{6}{Department of Astronomy \& Astrophysics, University of California, Santa Cruz, CA 95064, USA}
\altaffiltext{7}{NASA Ames Research Center 245-3, Moffett Field, CA 94035, USA}
\altaffiltext{8}{Sagan Postdoctoral Fellow}

\begin{abstract}
Observations of the highly-eccentric (e$\sim$0.9) hot-Jupiter HD~80606b with {\it Spitzer} have 
provided some of best probes of the physics at work in exoplanet atmospheres.  By observing 
HD~80606b during its periapse passage, atmospheric radiative, advective, and chemical timescales 
can be directly measured and used to constrain fundamental planetary properties such as 
rotation period, tidal dissipation rate, and atmospheric composition (including aerosols).  
Here we present three-dimensional general circulation models for HD~80606b that aim to 
further explore the atmospheric physics shaping HD~80606b's observed Spitzer phase curves.  
We find that our models that assume a planetary rotation period twice that of the pseudo-synchronous 
rotation period best reproduce the phase variations observed for HD~80606b near periapse passage with {\it Spitzer}.   
Additionally, we find that the rapid formation/dissipation and vertical transport of clouds 
in HD~80606b's atmosphere near periapse passage likely shapes its observed phase variations.   
We predict that observations near periapse passage at visible wavelengths could constrain 
the composition and formation/advection timescales of the dominant cloud species in 
HD~80606b's atmosphere.   The time-variable forcing experienced by exoplanets on 
eccentric orbits provides a unique and important window on radiative, dynamical, 
and chemical processes in planetary atmospheres and an important link between exoplanet 
observations and theory.

\end{abstract}


\section{Introduction}

HD~80606b is a `hot Jupiter' ($M_p$=3.94 M$_J$, $R_p$=0.98 R$_J$)
on an extremely eccentric orbit ($e$=0.9332) \citep{pon09b}.   
First discovered by \citet{nae01} using radial velocity observations, HD~80606b 
was later determined to be eclipsed by \citep{lau09} and transit \citep{mou09} 
its host star.  The extreme eccentricity of the orbit of HD~80606b is the result of 
``Kozai Migration" in the presence of a binary star system \citep{wu03}. 
During its $\sim$111 day orbit, HD~80606b experiences extreme ($\sim800\times$) 
shifts in the amount of incident flux it receives from its host star from 
apoapse to the periapse of its orbit. These variations in stellar insolation 
are likely to cause not only dramatic changes in the thermochemical structure of 
the planet, but also in global scale wind and cloud patterns.  

Observations with the {\it Spitzer Space Telescope} \citep{wer04} presented in 
\citet{dewit2016} and \citet{lau09} have shown that HD~80606b exhibits large 
variations in its flux as a function of orbital phase near periapse passage.  
Such flux variations were predicted by atmospheric models for exoplanets on eccentric orbits, 
which exhibit significant variations in atmospheric temperature and wind speeds that directly inform atmospheric radiative and dynamical timescales \citep{lan08, lew10, cow11, kat13, lew14}.
\citet{lau09} were the first to directly measure the atmospheric radiative timescale 
for an exoplanet with their {\it Spitzer} 8~$\mu$m observations of HD~80606b.   
Through their 30~hour observation of HD~80606b near secondary eclipse, when the planet 
passes behind its host star, and periapse passage, \citet{lau09} estimated a radiative 
timescale of 4.5~hours near the 8~$\mu$m photosphere of the planet.  

Building on the original analysis of HD~80606b presented in \citet{lau09}, 
\citet{dewit2016} revisited the 8~$\mu$m {\it Spitzer} observations given the 
recent evolution in data reduction techniques \citep[e.g.][]{lew13} and performed 
a combined analysis with HD~80606b observations taken in the 4.5~$\mu$m channel  
of {\it Spitzer} in 2009. The HD~80606b observations taken at 4.5~$\mu$m 
provided a significantly longer temporal baseline around the periapse of 
HD~80606b's orbit, with 80~hours worth of data in total.  This increased 
baseline allowed \citet{dewit2016} to not 
only estimate the radiative timescales in HD~80606b's atmosphere near its 4.5 and 
8~$\mu$m photospheres ($\sim$4~hours), but also the atmospheric absorptivity 
at the pressures probed and the bulk rotation rate of the planet.  \citet{dewit2016}
estimate the bulk rotation period of HD~80606b to be 93~hours, significantly 
longer than the predicted `pseudo-synchronous' rotation period \citep{hut81}
of $\sim$40 hours.

The interpretation of the HD~80606b phase variations observed by {\it Spitzer} 
presented in \citet{lau09} and \citet{dewit2016} relied on semi-analytic 
\citep[e.g.][]{cow11, dewit2016} and two-dimensional hydrodynamical models \citep{lan08}.  
Such models allow for rapid exploration of how specific changes in model parameters 
might provide a better match to the observed atmospheric flux variations.   However, 
such models do not fully capture the interplay of radiative, dynamical, and chemical 
processes in a planet's atmosphere. Here we present new atmospheric models for 
HD~80606b that more fully explore the physical processes at work in HD~80606b's 
atmosphere.  Our models are three-dimensional (latitude, longitude, and pressure) 
and include full radiative transfer calculations that consider equilibrium chemistry 
processes in HD~80606b's atmosphere throughout its highly eccentric orbit.  We 
specifically explore how assumed rotation period plays a role in shaping global 
circulation patterns and how the formation and evolution of clouds might 
play a role in shaping the observed flux variation of HD~80606b.  The following 
sections outline the specifics of our modelling approach, explore the 
complex atmospheric physics at work in HD~80606b, and provide observational 
predictions that give further insights into the current {\it Spitzer} 
observations and guide future atmospheric characterization 
observations of the HD~80606 system.

\section{Models}

The results presented here for our atmospheric modeling effort rely on 
a combination of simulation tools that link the physical processes potentially at work in 
HD~80606b's atmosphere and what can be observed.  In the following 
sections we present an overview of the global circulation, cloud, and 
observational phase curve models employed in this study.  

\subsection{Global circulation model}
In the simulations presented here we employ the Substellar and Planetary Atmospheric 
Radiation and Circulation (SPARC) model first presented in \citet{sho09} as applied 
to HD~189733b and HD~209458b and subsequently used in a number of other exoplanet 
studies \citep[e.g.][]{lew10, kat13, lew14, kat2014, kat2015, sho2015, kat2016, par16}.  
At the core of the SPARC model is the MITgcm \citep{adc04}, which solves the primitive 
equations with pressure as the vertical coordinate.  We determine the amount of 
heating/cooling at each grid point by the divergence of radiative fluxes in each model layer.
Radiative fluxes are calculated using the two-stream non-gray radiative transfer model 
of \citet{mar99}. We use the correlated-k method with 8 k-coefficients inside each of 
our 11 wavelength bins.  These simulations of HD~80606b utilize the cubed-sphere 
grid \citep{adc04} with a horizontal resolution of C16 ($\sim$32$\times$64 in 
latitude and longitude) and a vertical pressure range from 200~bar to 0.20~mbar broken 
down into 39 layers with even log-pressure spacing with a top layer that extends to 
zero pressure.  

In the simulations presented here we assume an atmospheric composition of 1$\times$ solar values 
and have excluded TiO and VO from the opacity tables since it is likely that these species are 
`cold trapped' deep in the atmosphere \citep{for08}. Clouds are neglected in the global
circulation model. The effect of clouds is, however, considered in the post-processing 
of the simulation.  We define our planetary and stellar parameters using the 
values from \citet{pon09b}.  The nominal rotation period of HD~80606b is determined to be 
40.4761~hours assuming the pseudo-synchronous rotation relationship presented in \cite{hut81}.  
Because it is possible that HD~80606b's rotation period might significantly deviate 
from this nominal value, we construct additional atmospheric models where the 
rotation period is assumed to be half (20.2380~hours) and twice (80.9522~hours) the nominal
pseudo-synchronous rotation period. We initialize the model with wind speeds set to zero 
and each column of the grid assigned a pressure-temperature profile derived from 
one-dimensional radiative equilibrium calculations assuming that the planet is at periapse.  
The planet loses `memory' of this initial temperature distribution fairly rapidly and 
we have found no significant difference between simulations started near periapse and 
those initialized near apoapse with the corresponding pressure-temperature profile.

We find that in order to maintain numeric stability near periapse that the timestep used to solve 
the relevant momentum and energy equations must be very short ($\sim$5~s).  There is a high 
computational overhead associated with solving for our radiative fluxes along each grid column. 
We elect to update our radiative fluxes using a radiative timestep of 5~s near periapse, 20~s 
in the region 3.5 to 17.7 hours away from periapse, and 50~s for the remainder of the orbit 
to increase our computational efficiency.  This radiative timestep scheme allows us to update 
the radiative fluxes in our model more frequently near periapse where the incident flux on the
planet is changing rapidly and less frequently near apoapse where the relative change in the
incident flux on the planet is smaller. We determine the time-varying incident flux on 
HD~80606b from its host star by calculating the planet-star distance at each radiative 
timestep using Kepler's equation \citep{mur99}.    We integrate our models for up to 1000
simulated days, but it is clear that the velocity profile for each orbit has reached a 
stable configuration after just a few orbits of the planet ($\sim$300 simulated days).  
Integrating our simulations beyond this point results in only 
small changes in planetary wind and thermal patterns that are confined 
to pressures well below photospheric pressures ($\sim$300~mbar).

\begin{figure*}[ht]
\centering
\includegraphics[width=0.49\textwidth]{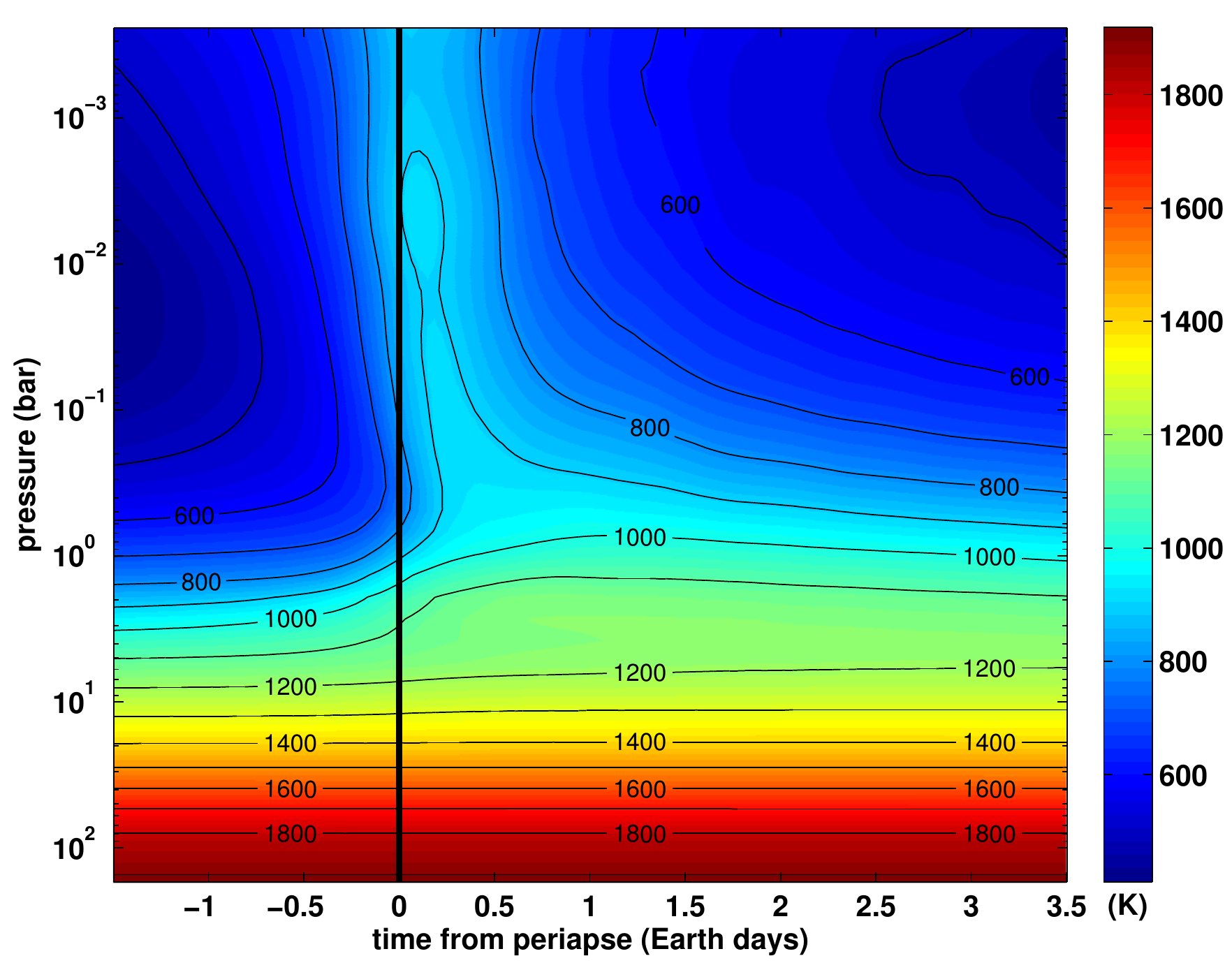}
\includegraphics[width=0.49\textwidth]{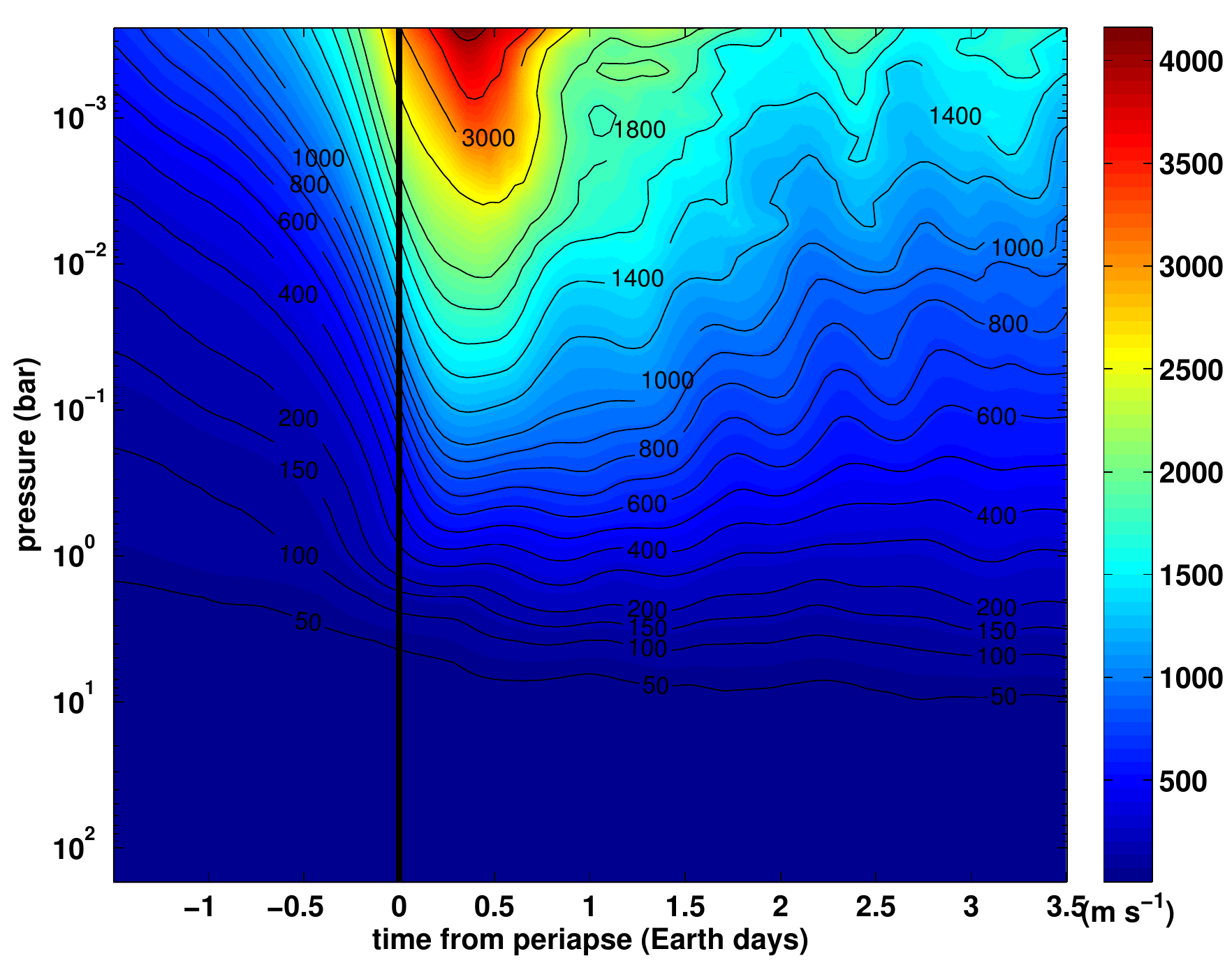}\\
\includegraphics[width=0.49\textwidth]{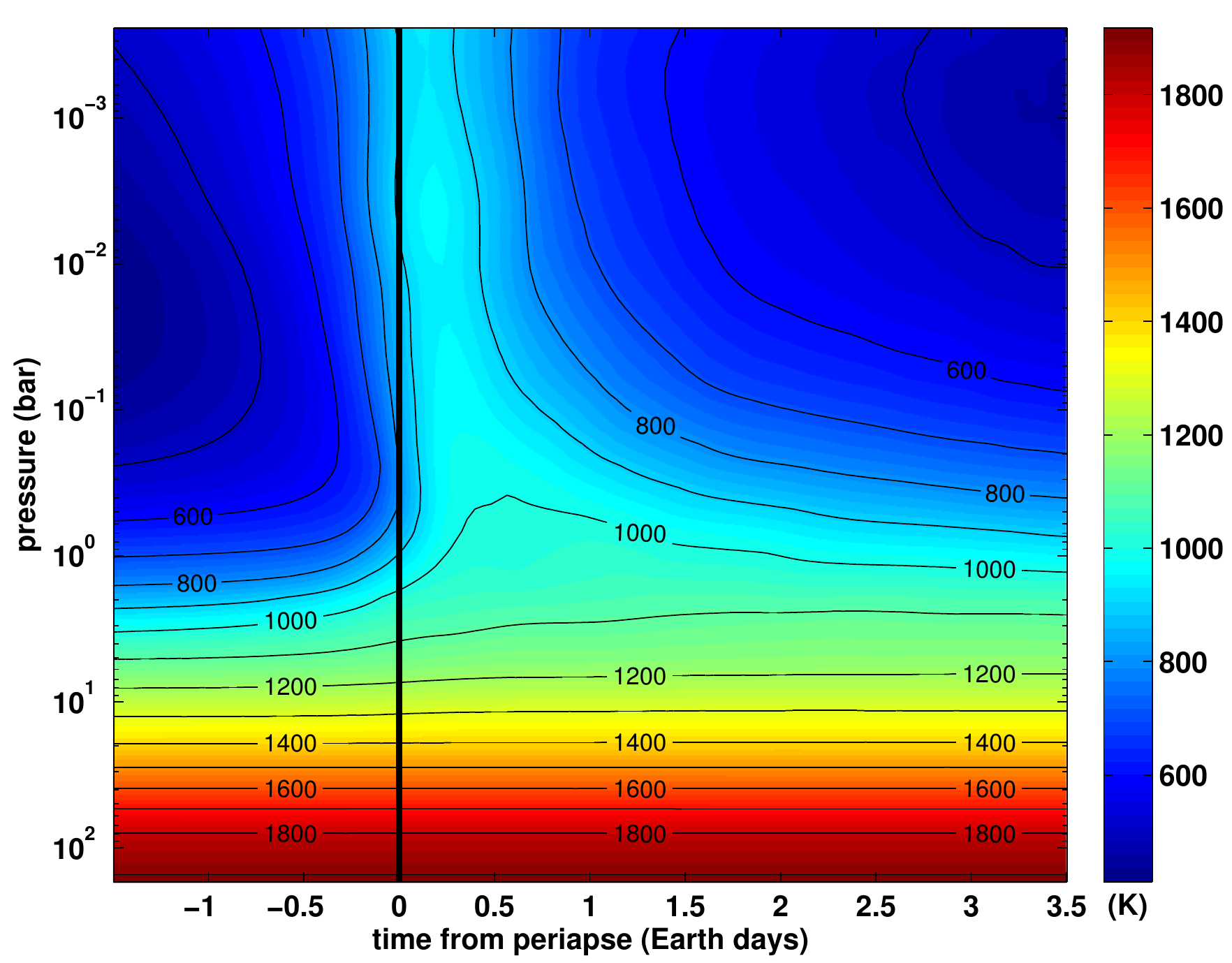}
\includegraphics[width=0.49\textwidth]{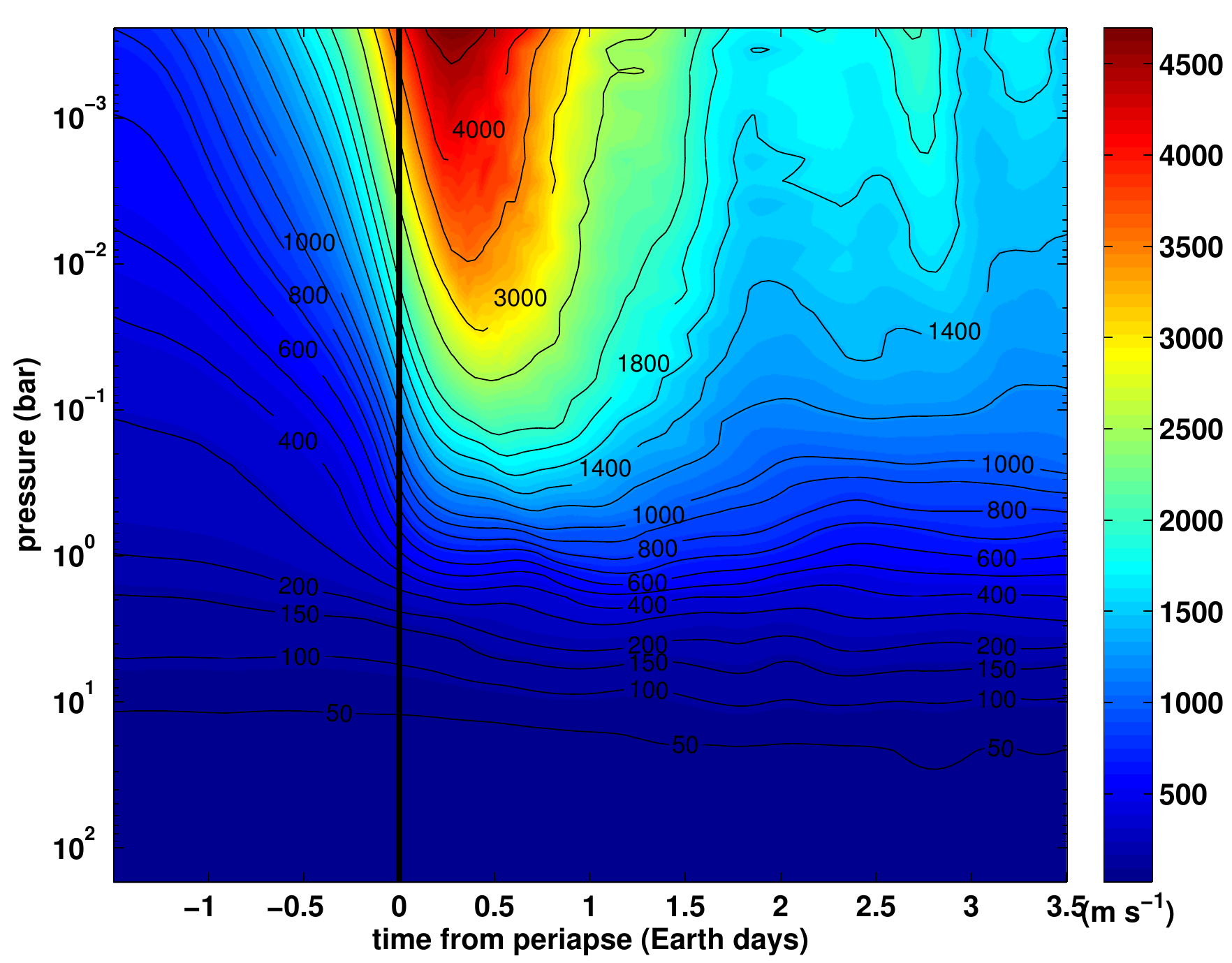}\\
\includegraphics[width=0.49\textwidth]{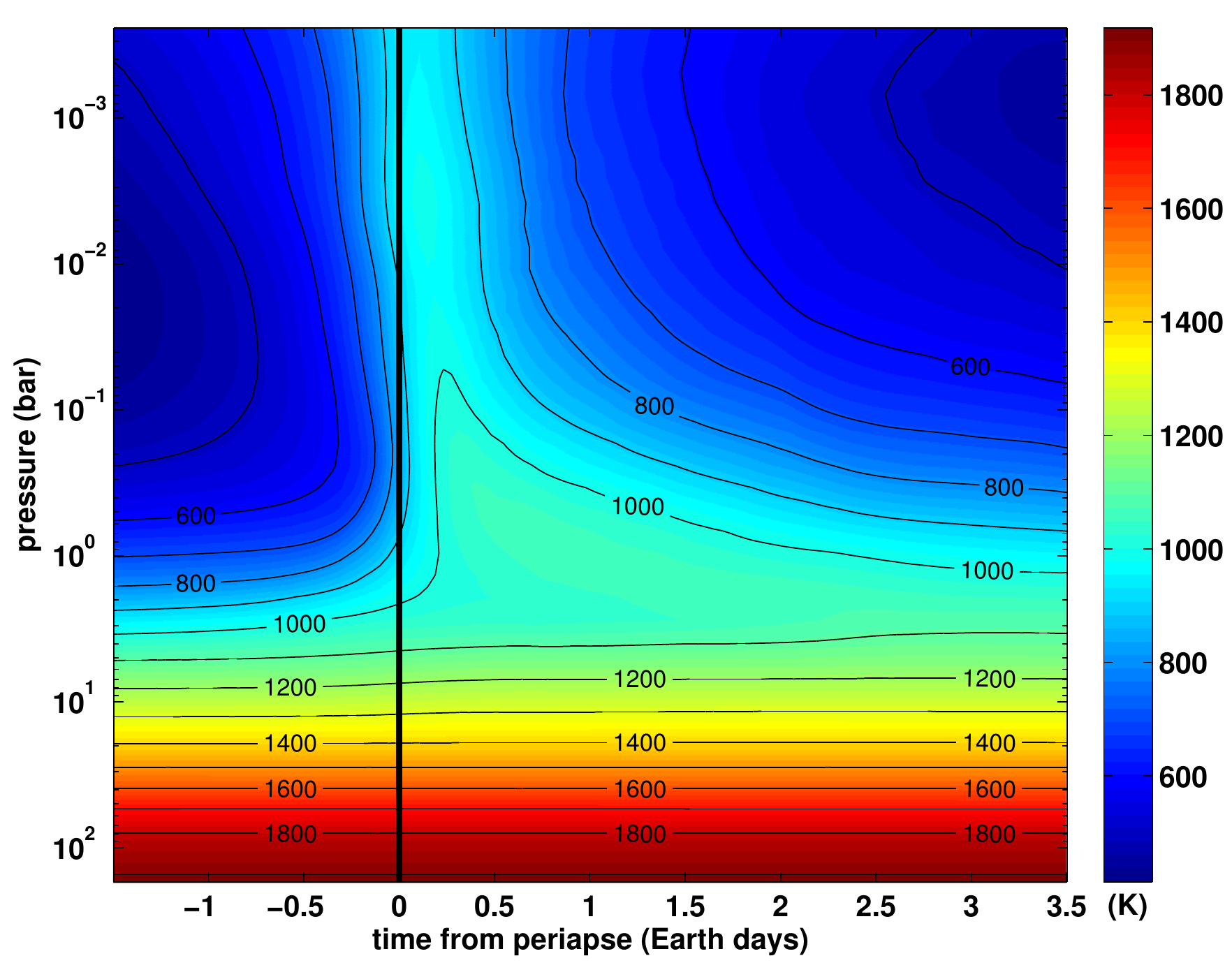}
\includegraphics[width=0.49\textwidth]{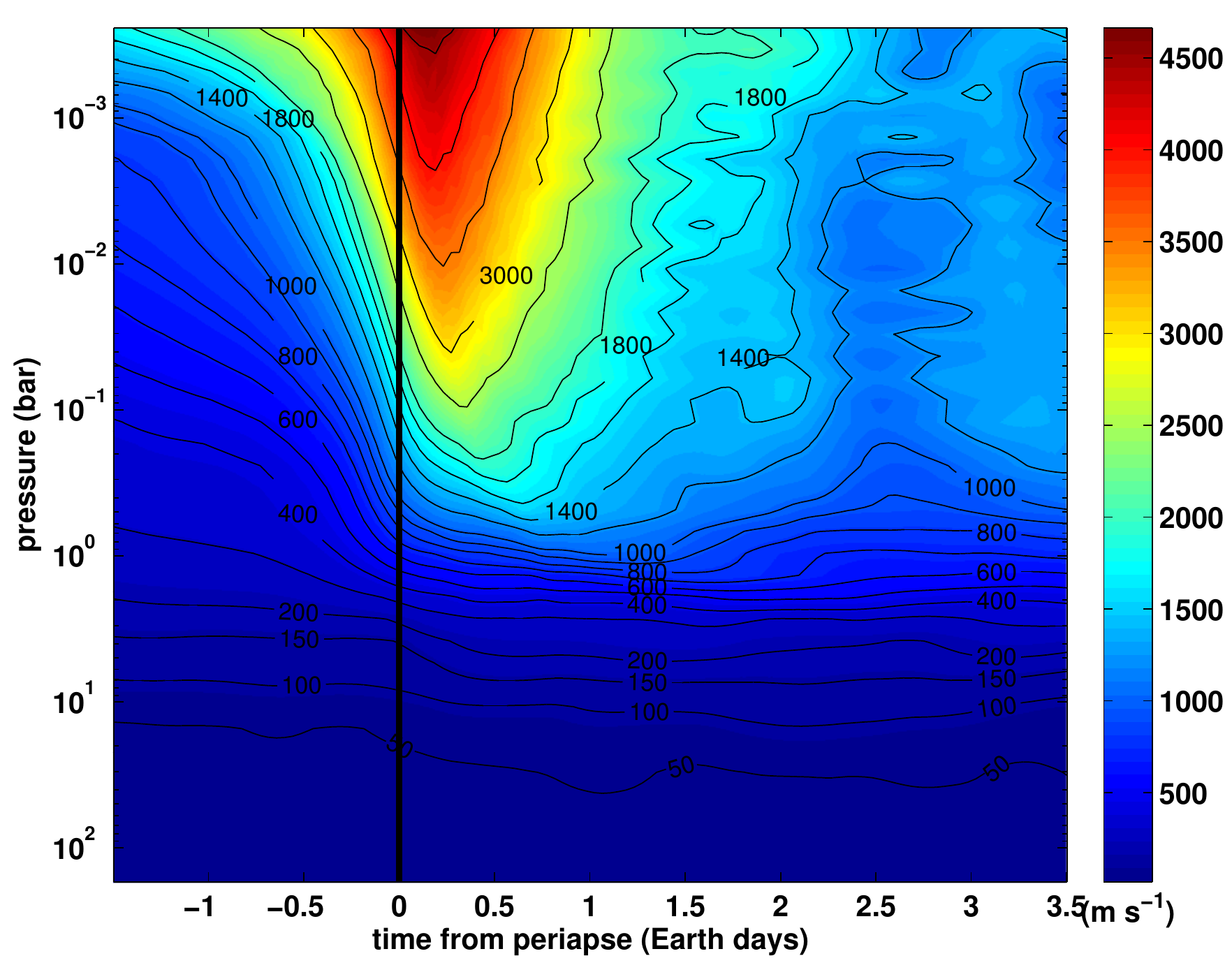}\\
\caption{Average temperature (left) and RMS horizontal velocity (right) as a function of time 
relative to periapse passage for our half-nominal (top), nominal (middle), and 
twice-nominal (bottom) rotation period models.  The temperatures and RMS velocities 
represent averages over latitude and longitude as a function of pressure.}\label{806_tave_vrms}
\end{figure*}

\subsection{Cloud modeling}\label{cldmod}

As shown in Section~\ref{results}, our global circulation model predicts that the dayside temperature 
of the planet can vary by more than 700~K between periapse and apoapse in the 
observable portion of the atmosphere. As such, the photospheric layers (1~bar to 1~mbar) 
can cross the condensation curves of several potential cloud species 
\citep[e.g.][]{mor2012, mar2013}. To simulate the formation, 
dissipation and general evolution of clouds in HD~80606b's atmosphere, we use 
the cloud model presented in \citet{par16}. In a given atmospheric column, 
a material is condensed if the temperature is cooler than its condensation temperature 
at a given pressure level in the atmospheric general circulation model.
Clouds are assumed to condensed with a uniform particle size of $a$, which is a free 
parameter ranging from 0.1$\mu$m to 10$\mu$m. In this study of HD~80606b, we consider 
the following possible cloud species: MgSiO$_3$ (enstatite), MnS (manganese sulfide), 
and Na$_2$S (sodium sulfide). Cloud species ZnS and KCl are not considered in this study
as they are fully evaporated from the planet's dayside during periastron passage and 
are expected to form significantly less massive, and less opaque, clouds compared with 
MgSiO$_3$, MnS, and Na$_2$S \citep[e.g.][]{for2005}.

\subsection{Theoretical Phase Curve Calculation}

To calculate theoretical phase curves, planetary flux as a function of time, 
we solve the two-stream radiative transfer equations 
along the line of sight for each atmospheric column and for each planetary 
orbital position. In our radiative transfer calculations, we consider 
absorption, emission and scattering. This method, 
similar to the calculations of \citet{for06}, naturally takes into account 
geometrical effects such as limb darkening and the variation of the stellar 
and planetary orientations with respect to each other and a earth-based observer. 
The stellar flux is assumed to be a collimated flux propagating in each atmospheric 
column with an angle equal to the angle between the local vertical and the direction 
of the star. We use 196 frequency bins ranging from 0.26 to 300~$\mu$m and integrate 
the resulting outgoing flux over a variety of observational bandpasses spanning the 
visible to infrared wavelengths (0.3-28~$\mu$m). The gaseous opacities used to 
calculate the values of the k-coefficients are the same as the ones used for the 
global circulation model calculations. Rayleigh scattering by the gas and 
Mie scattering by the clouds are combined with the gaseous opacities to 
produce our theoretical spectra as a function of time for HD~80606b.

In this study, cloud opacities are included in post-processing and not self-consistently 
included within the framework of the global circulation model.   This is done to maintain 
the computational feasibility of our atmospheric models.  A single SPARC model can 
require weeks to months of computational time in a cluster computing environment 
to reach a stable solution, largely limited by computational requirements of the 
fully non-gray radiative transfer calculations.
Incorporation of cloud formation, evolution, and radiative feedback is not impossible within 
the framework of the SPARC model, but would currently require prohibitively long 
computation times.  As shown by \citep{par16}, the cloud distribution in hot 
Juptiers is primarily determined by the thermal structure of the planet, with 
radiative feedback and the dynamical mixing of the clouds being secondary effects. 
Therefore, the cloud distributions we derive here from our models should 
be a good approximation for the spatial and wavelength dependent opacities we could expect 
to shape HD~80606b's phase curve.
We are currently working on further optimization to the SPARC and cloud formation models 
to more robustly treat clouds without requiring highly tuned or parameterized schemes.

\begin{figure*}
\centering
\includegraphics[width=0.49\textwidth]{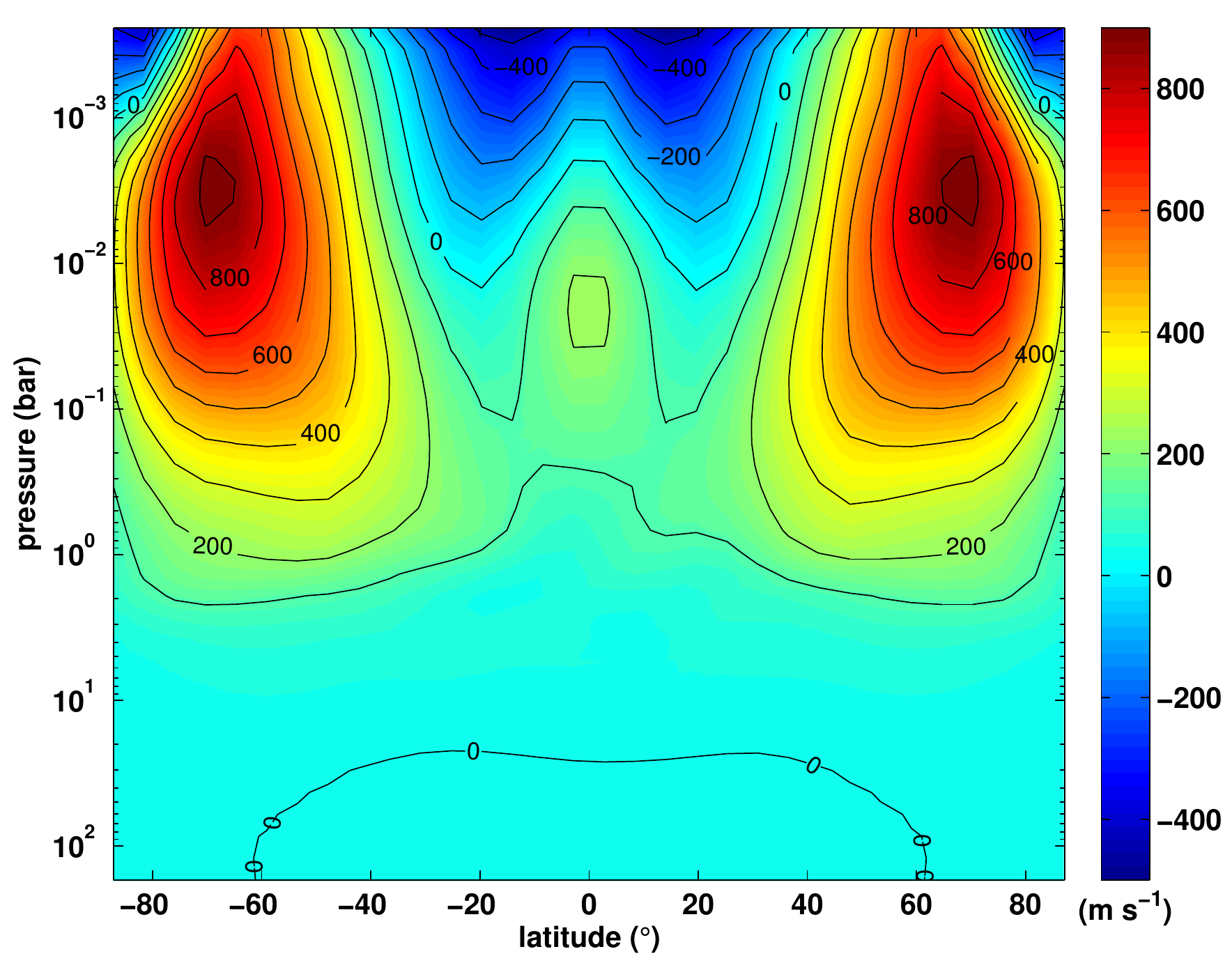}
\includegraphics[width=0.49\textwidth]{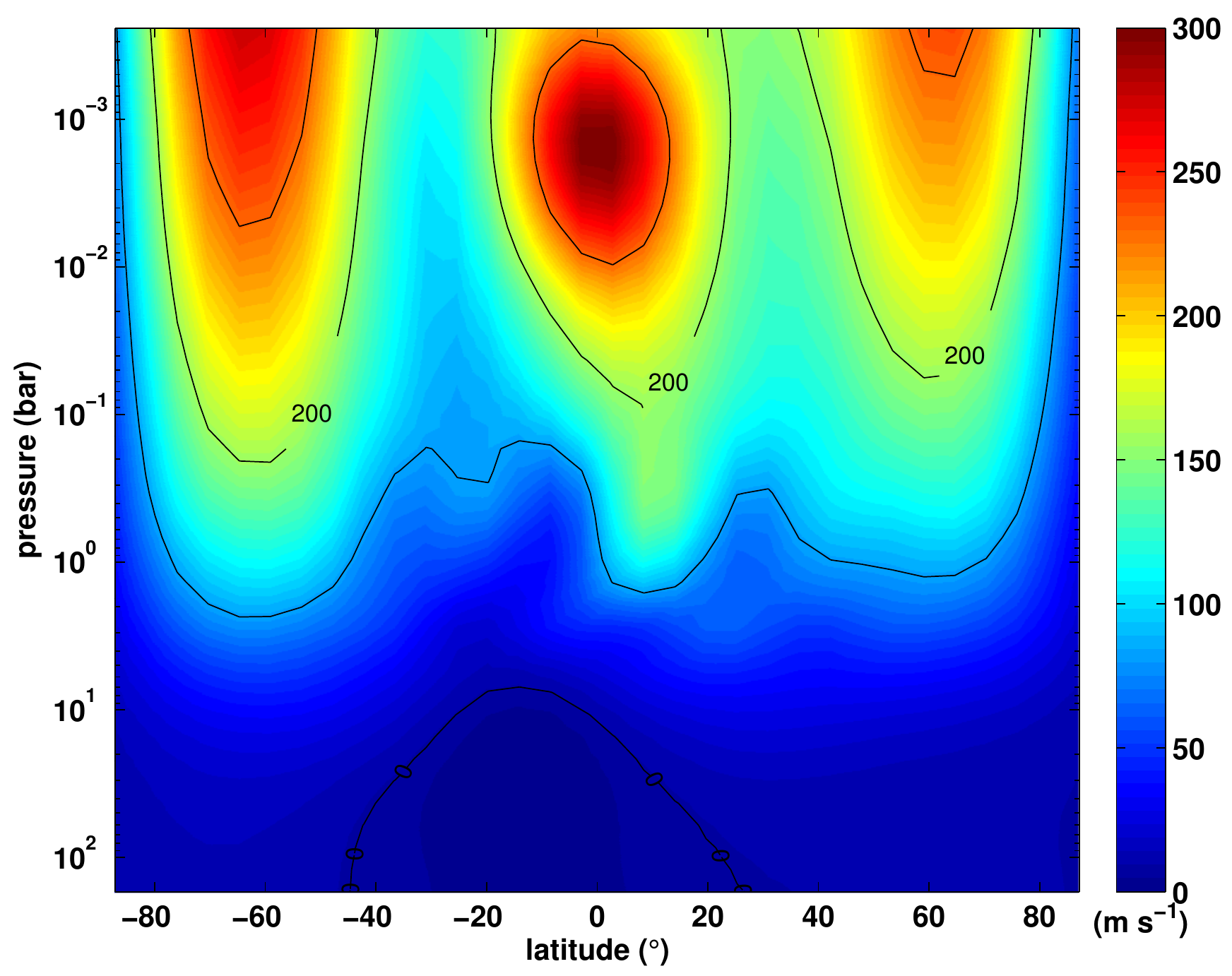}\\
\includegraphics[width=0.49\textwidth]{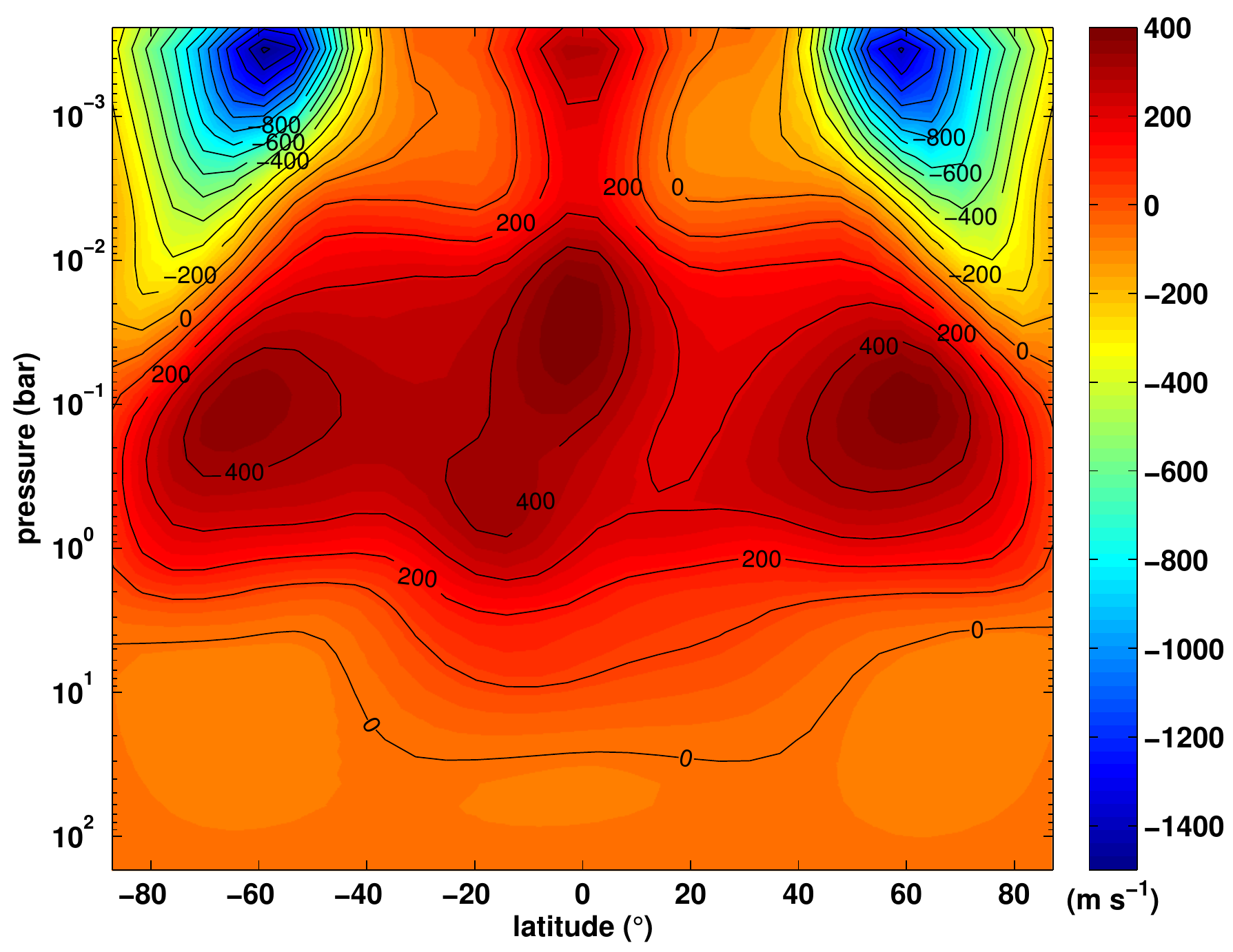}
\includegraphics[width=0.49\textwidth]{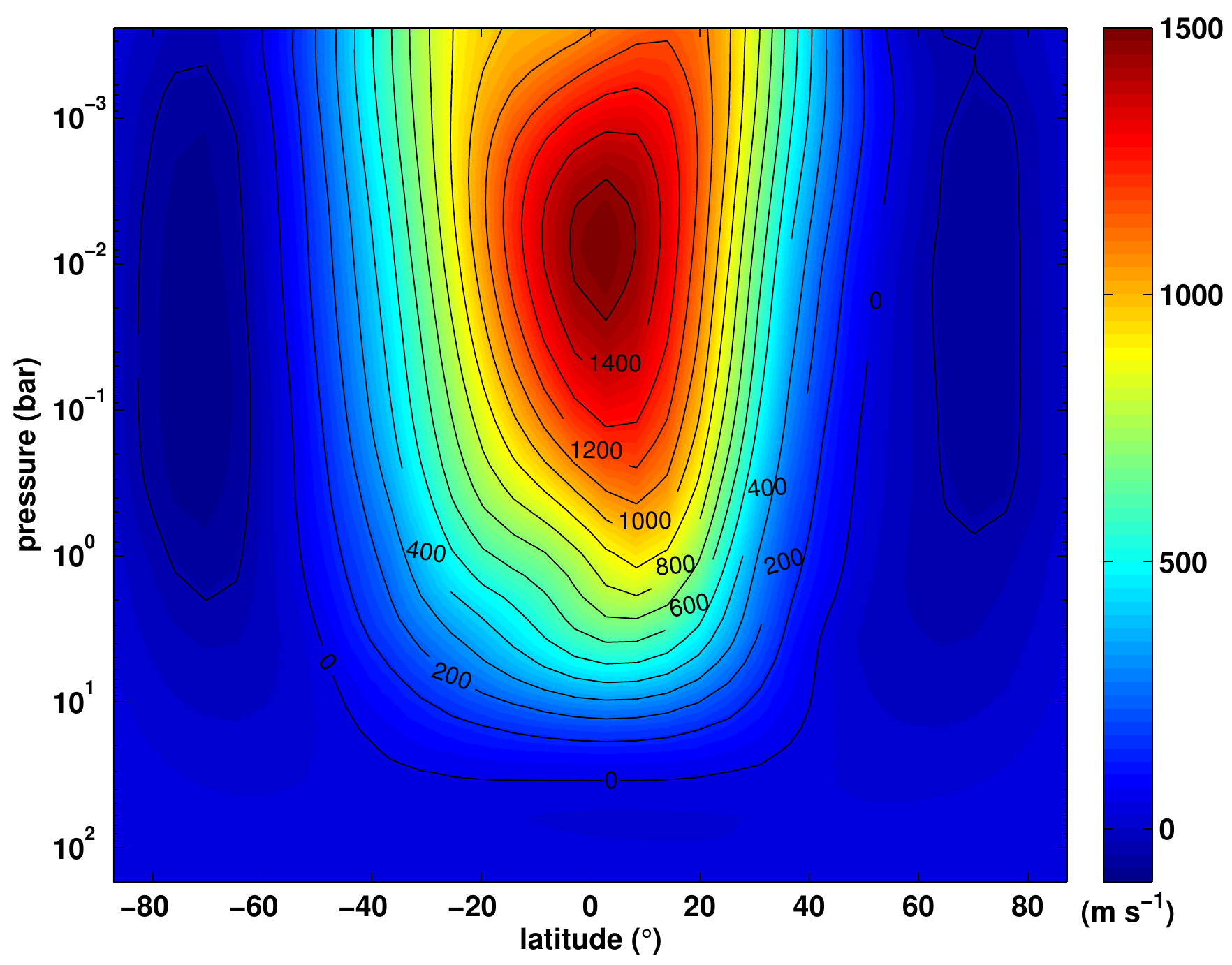}\\
\includegraphics[width=0.49\textwidth]{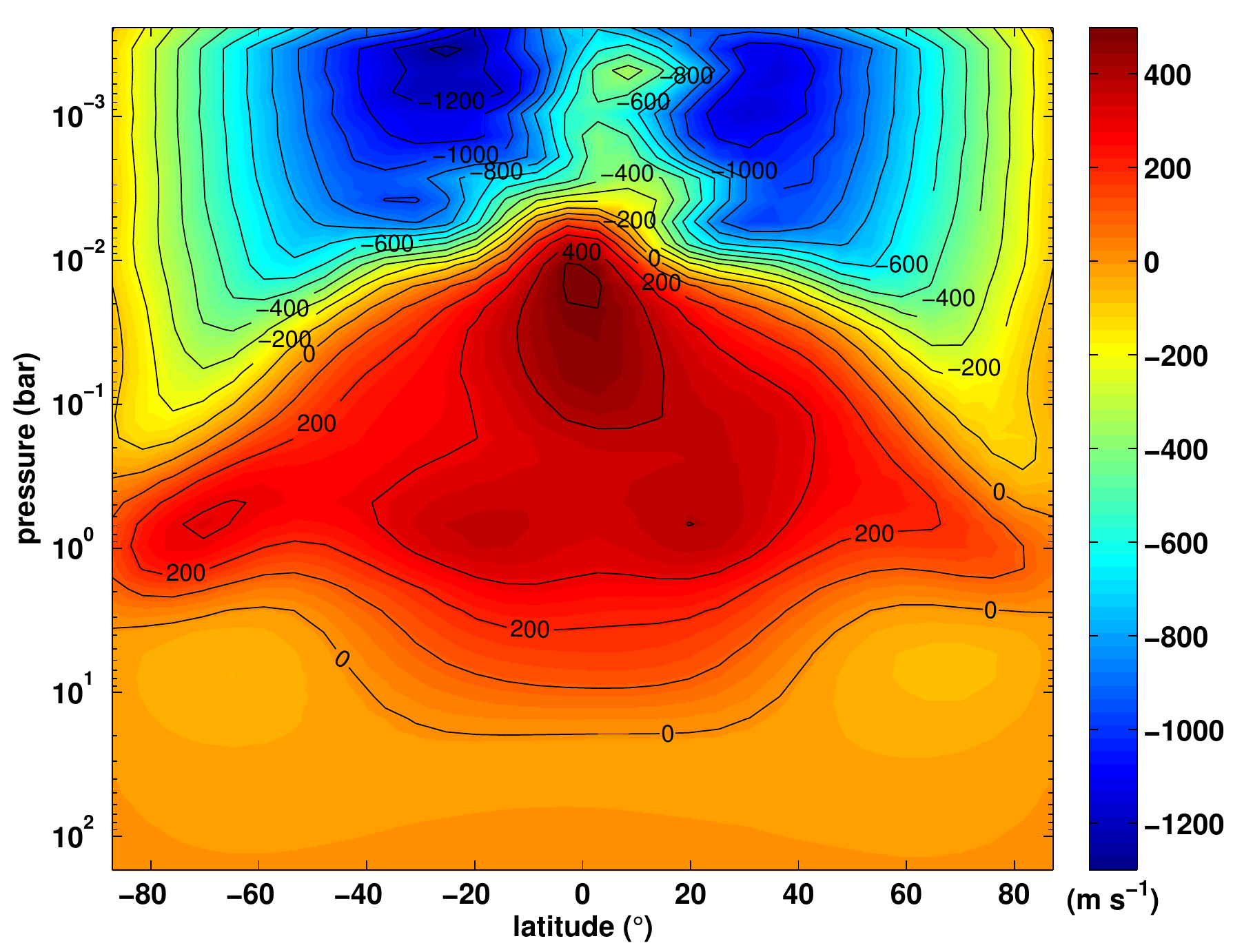}
\includegraphics[width=0.49\textwidth]{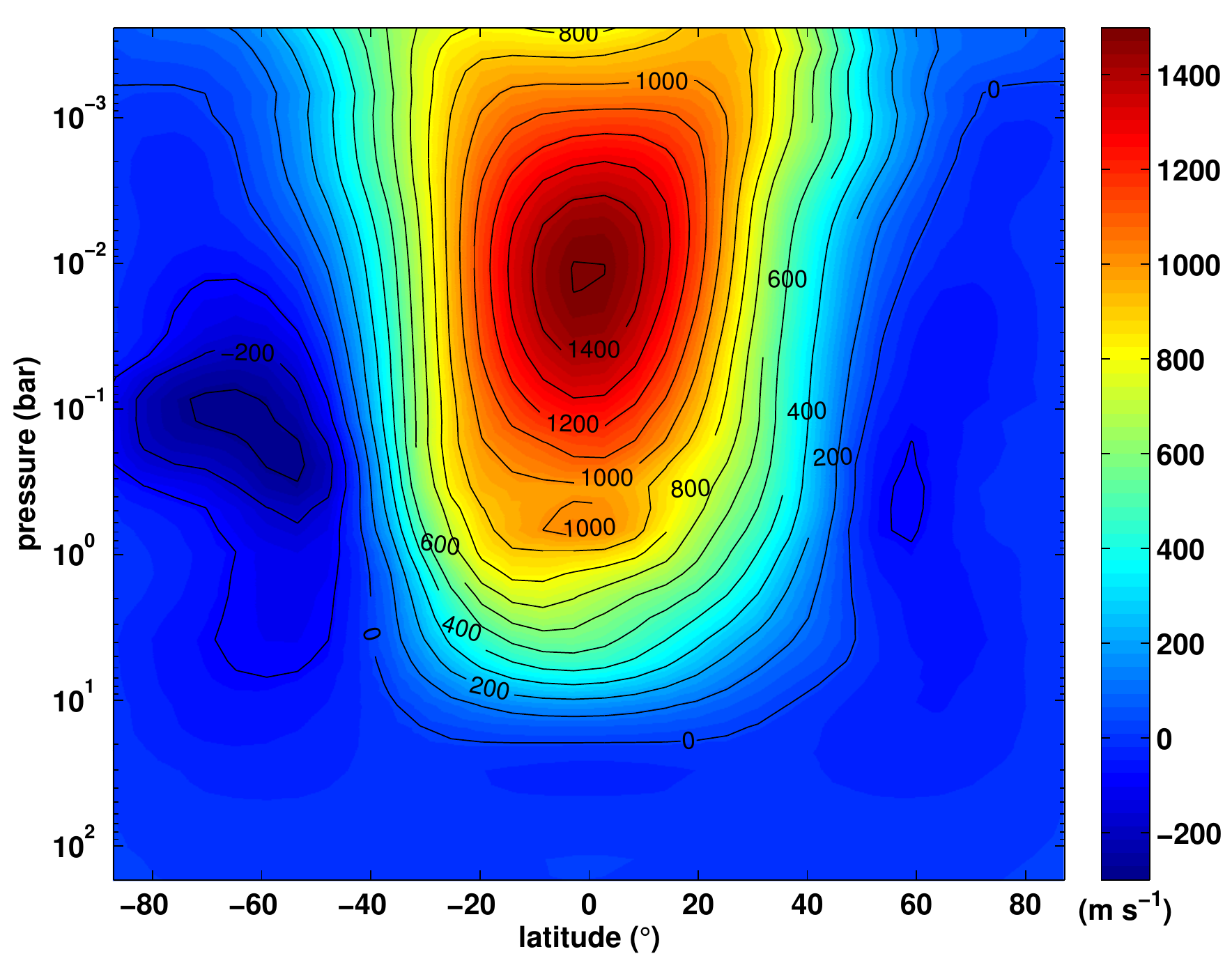}\\
\caption{Zonal-mean zonal winds for HD~80606b near periapse (left) and apoapse (right) 
for our models that assume the half-nominal (top), nominal (middle), and twice-nominal (bottom) 
rotation period for the planet.   Apoapse and periapse occur at true anomalies ($f$) of
180$^{\circ}$ and 0$^{\circ}$ respectively.  The colorbar shows the strength of the zonally
averaged winds in m s$^{-1}$.  Contours are spaced by 100 m s$^{-1}$. Positive wind speeds 
are eastward, while negative wind speeds are westward.  Note the significant change in 
the jet structure as a function of orbital position.}\label{806_uzonal}
\end{figure*}

\begin{figure*}
\centering
\includegraphics[width=0.49\textwidth]{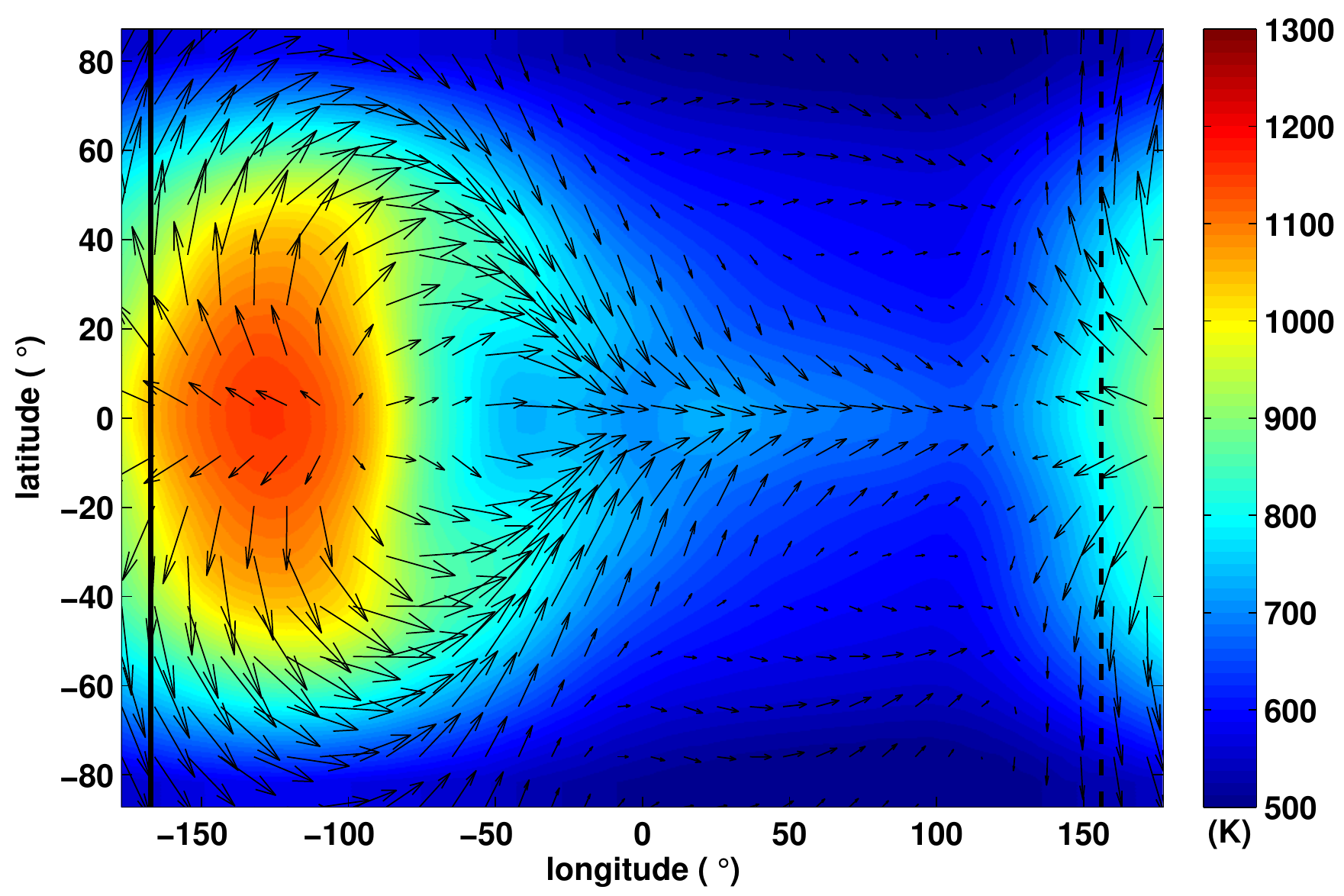}
\includegraphics[width=0.49\textwidth]{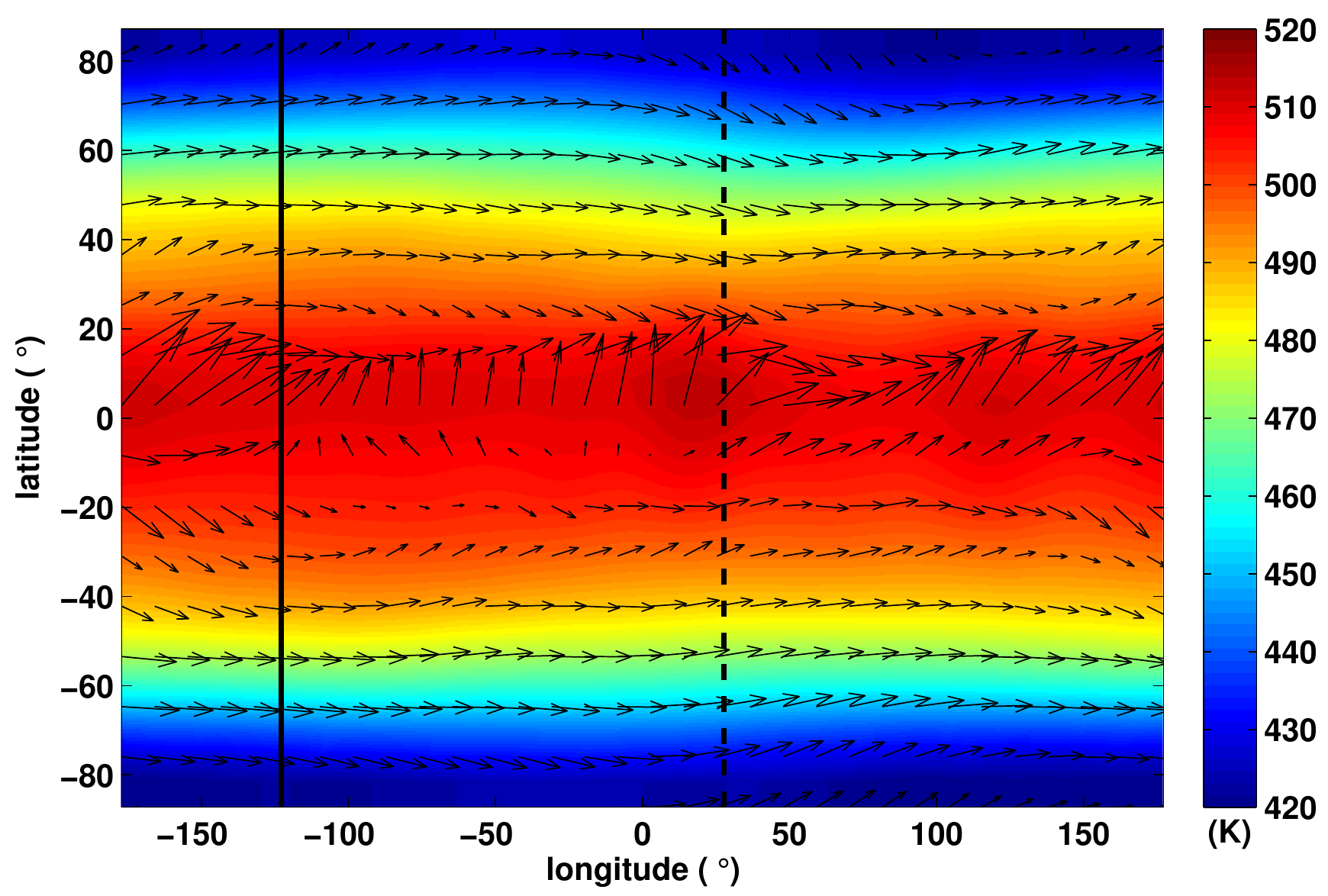}\\
\includegraphics[width=0.49\textwidth]{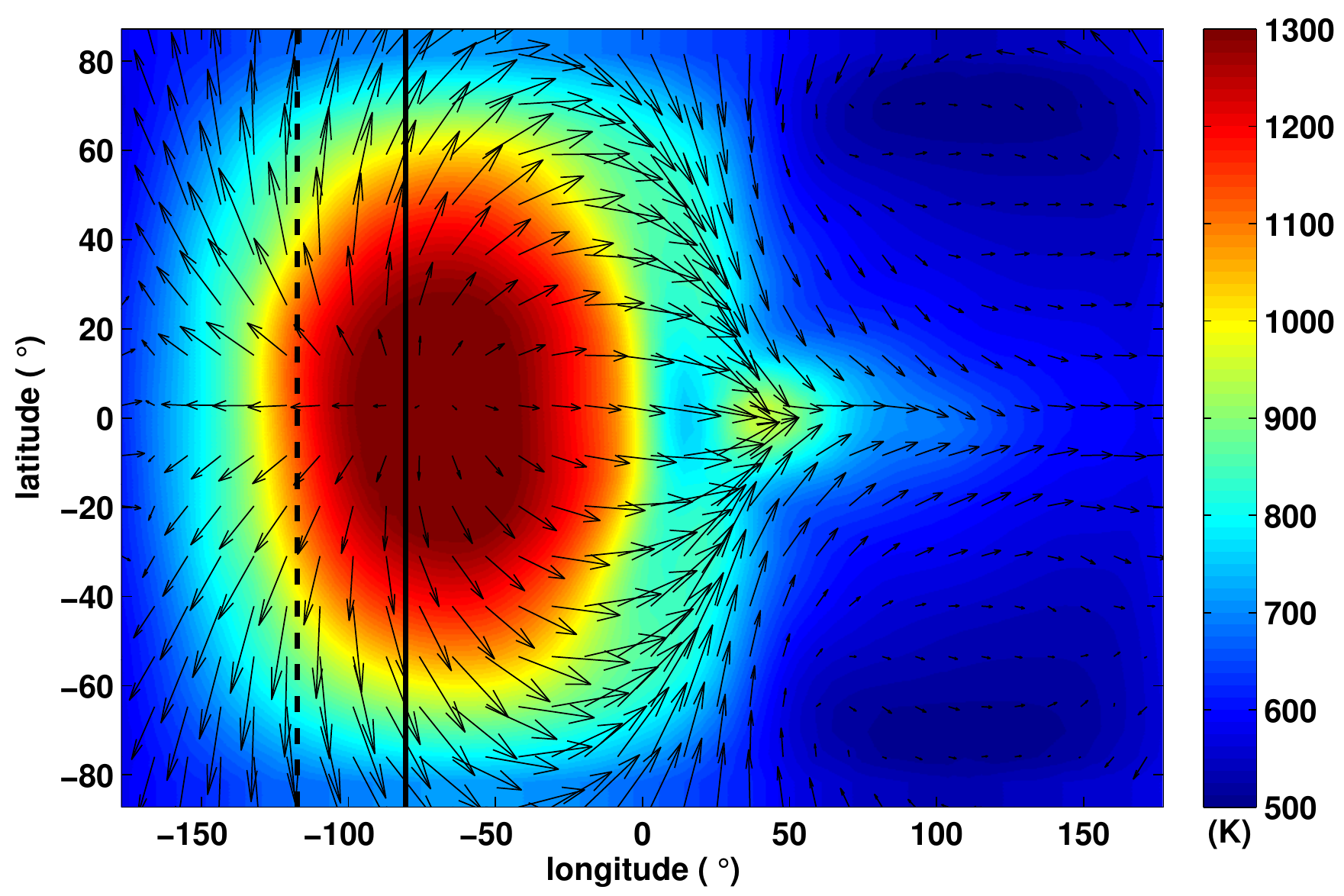}
\includegraphics[width=0.49\textwidth]{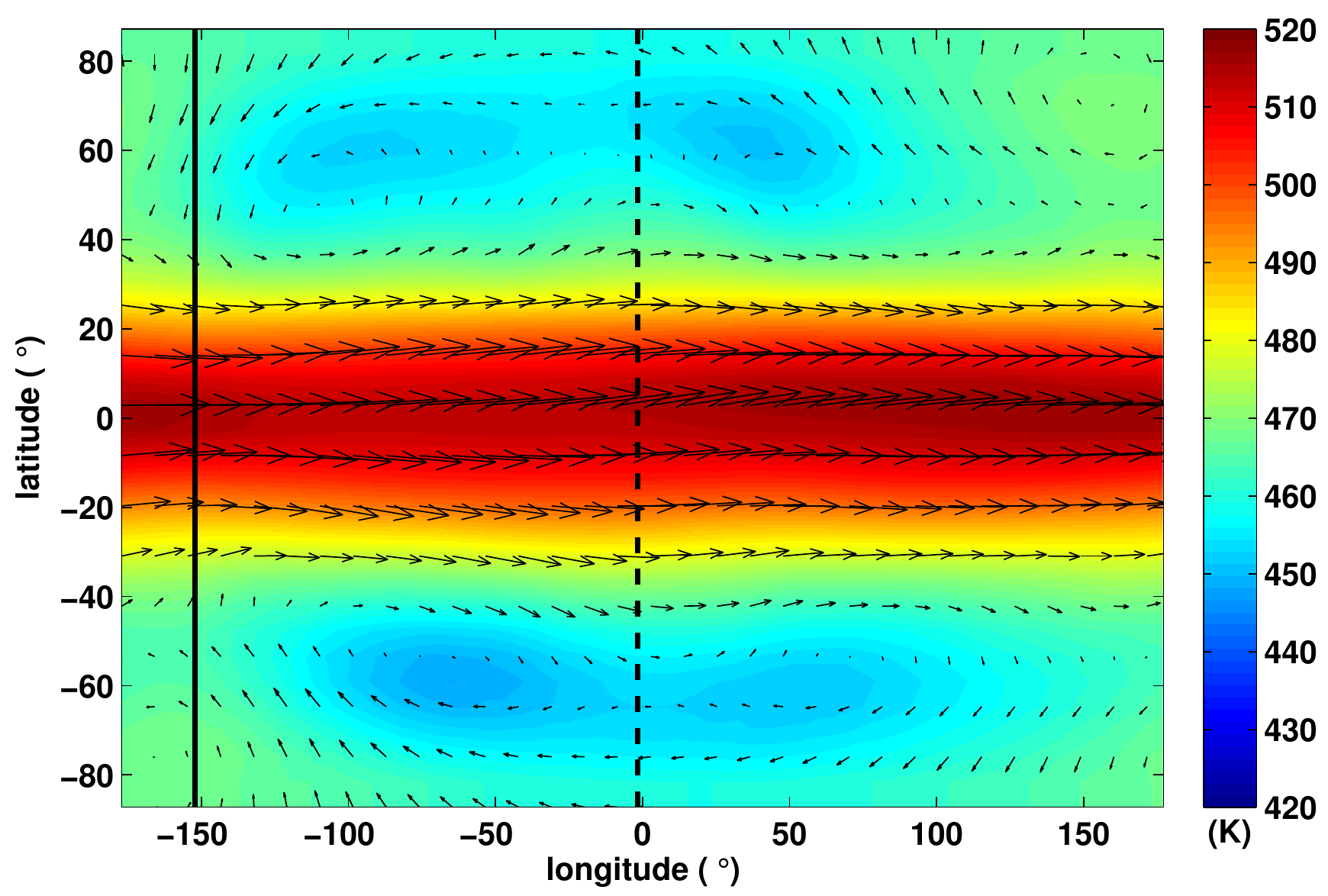}\\
\includegraphics[width=0.49\textwidth]{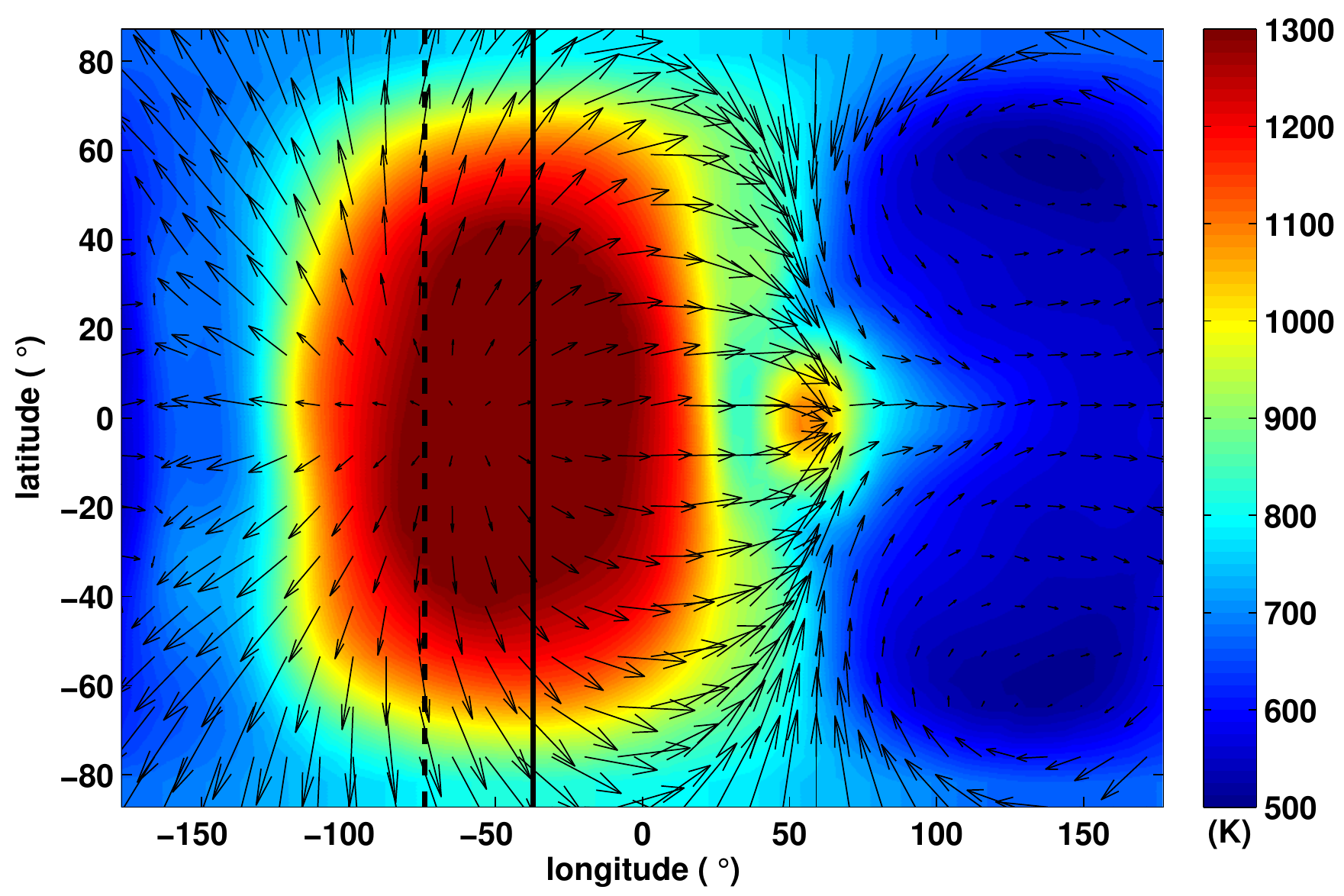}
\includegraphics[width=0.49\textwidth]{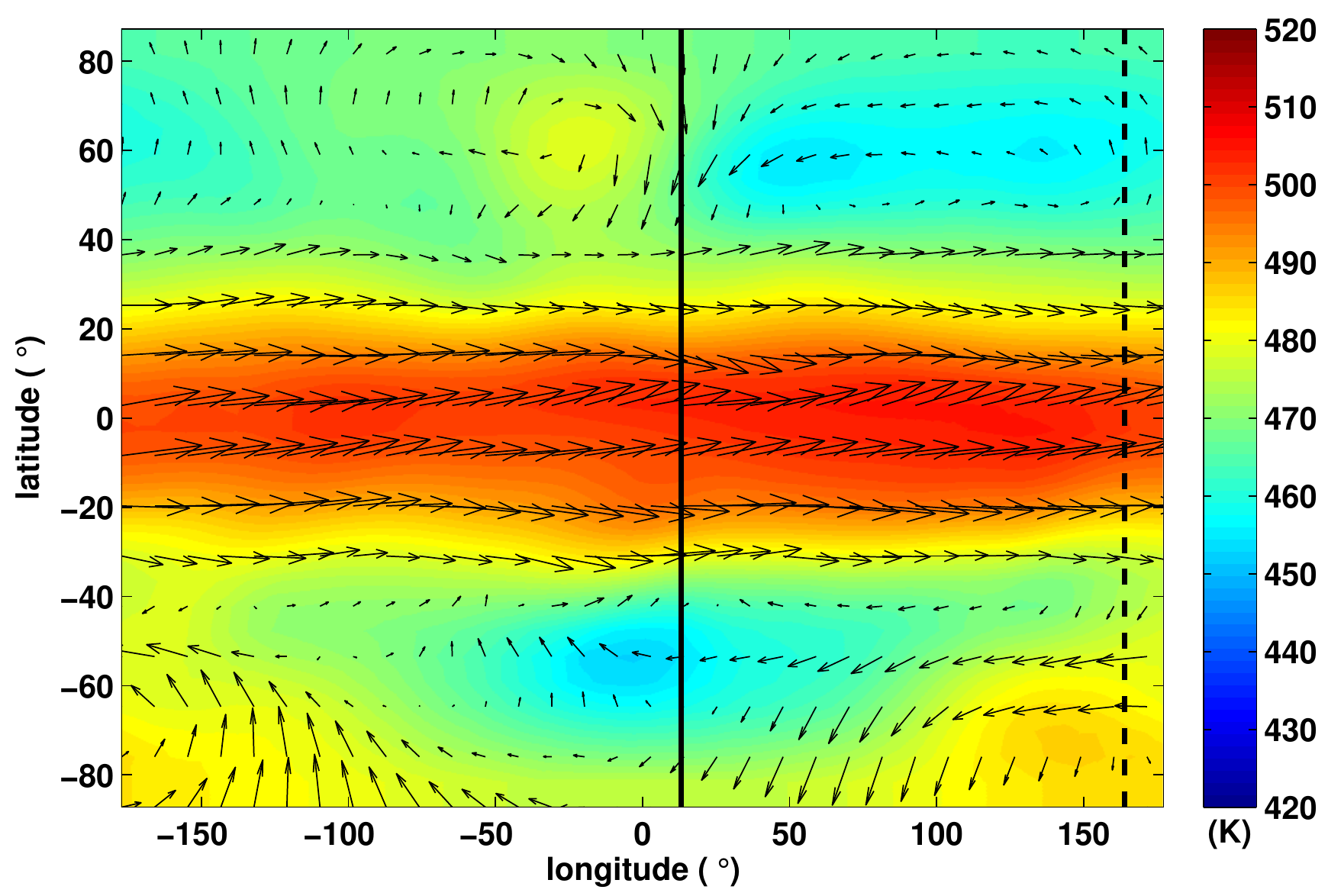}\\
\caption{Temperature (colorscale) and horizontal winds (arrows) at the 
340~mbar level of our HD~80606b model near periapse (left) and apoapse (right) for 
our half-nominal (top), nominal (middle), and twice-nominal (bottom)
rotation period cases.  The length of the arrows represent the strength of 
horizontal winds.  The longitude of the substellar point is indicated by the solid 
vertical line.  The Earth facing longitude is indicated by the dashed vertical
line.}\label{806_uvt}
\end{figure*}

\section{Results}\label{results}

The following sections overview the key results from our atmospheric modelling effort for HD~80606b.  
We first present changes in the global scale winds and temperatures of HD~80606b as a 
function of both pressure and time for our half-nominal, nominal, and twice-nominal 
rotation period models.  We then compare the thermal patterns and winds predicted 
for HD~80606b near the periapse and apoapse of its orbit from our half-nominal, nominal, 
and twice-nominal rotation period atmospheric models.  We further focus on the 
thermal structure and vertical wind profiles as a function of longitude that develop 
in our models near periapse. Using the post-processing method described in 
Section~\ref{cldmod}, we predict the evolution of cloud coverage in HD~80606b.  
Finally, we present theoretical phase curves derived from our simulations 
for each rotational period case assuming a range of cloud properties 
and compare them with the {\it Spitzer} observations at 4.5 and 8~$\mu$m 
presented in \citet{dewit2016}.

\subsection{Global Scale Winds and Temperatures}

Since our simulations are performed in three-dimensions, we can 
investigate changes in the overall temperatures and wind speeds of HD~80606b 
as a function of depth in the atmosphere as well as time.
Figure \ref{806_tave_vrms} presents globally averaged temperatures 
and root mean square (RMS) horizontal velocities as a function of pressure 
and time relative to the periapse of HD~80606b's orbit for each of our rotation 
period cases.  The peak global average atmospheric temperatures near 
the 340~mbar level in our simulations occurs on average 11~hours after 
periapse passage.  Near the 10~mbar level of our simulation where radiative 
timescales are shorter, peak temperatures are 
reached on average 4.5~hours after periapse passage.  In all cases peak wind speeds are 
achieved $\sim$2 hours after the peak in the planetary temperature.
It is also interesting to note that both temperatures and wind speeds remain elevated 
with respect to their pre-periapse values for several days to weeks depending 
on the pressure level for all of our models regardless of the assumed rotation 
period (Figure~\ref{806_tave_vrms}).   HD~80606b transits its host star as seen from 
earth $\sim$5.6 days after periapse passage \citep{mou09}, so these elevated 
temperatures and wind speeds could be relevant to transit observations of this planet.
 
Because HD~80606b is subject to large variations in the amount of flux it receives from 
its host star it is likely to show distinct changes in wind and thermal patterns as a function 
of orbital phase.  Figure \ref{806_uzonal} shows the zonal-mean zonal winds for HD~80606b 
near the apoapse and periapse of its orbit for each of our rotation period cases.  The term 
zonal wind refers to the component of the wind vector that is in the east-west direction 
while the zonal mean is an average over all longitude for a given latitude.  For most its orbit,
HD~80606b maintains a fairly steady jet structure similar to the apoapse jet structures 
presented in the right half of Figure~\ref{806_uzonal}.  The width and number of jets that 
develop in our simulations of HD~80606b near the apoapse of its orbit is dependent to a 
large degree on the assumed rotation period of the planet \citep{sho02, sho08, sho10, sho2015}.  
As HD~80606b approaches the periapse of its orbit, its jet structure dramatically changes.  
At low pressures in all of our rotation period cases strong westward flows develop as air 
rushes from the day to the night side of the planet.  In our nominal and twice-nominal 
rotation period cases, predominately eastward flow is maintained at pressures higher 
than $\sim$10~mbar. In the half-nominal rotation period case, the high latitude eastward 
jets are maintained through out the orbit, strengthening  and weakening with the amount 
of incident flux on the planet.

The rapid heating that HD~80606b experiences as it passes through periapse results not only in 
changes in the planet's jet structure, but also produces significant changes in its horizontal 
temperature distribution.  Figure \ref{806_uvt} shows the temperature distribution and 
wind vectors at the 340~mbar level of our simulations near the periapse and apoapse of 
HD~80606b's orbit.  Near apoapse, HD~80606b's atmospheric temperature distribution at the 
340~mbar level of our simulations is fairly uniform with longitude but shows an equator to pole 
temperature difference that increases with decreasing rotation period.    The efficiency 
of equator to pole heat transport is strongly influenced by the assumed rotation period 
of the planet.  The larger latitudinal gradient in temperature in our half-nominal rotation 
period case results from the fact that equator to pole heat transport becomes less efficient 
as the rotation period is reduced \citep[e.g.,][]{sho2015}.  

Near the periapse of its orbit, the temperature distribution of HD~80606b becomes dominated 
by a strong day-night temperature contrast (Figure~\ref{806_uvt}).  The strength of this 
day-night temperature contrast is determined in part by the path travelled by the longitude 
of the substellar point as determined by a combination of the orbital geometry and the 
rotation period of the planet.  In our nominal rotation period case, the longitude of 
the substellar point ($\phi_{\star}(t)$) becomes stationary near periapse and the heating 
becomes confined to the region near that longitude.  For our half-nominal rotation period 
case, the time rate of change of $\phi_{\star}$ slows, but continues in a westward 
motion heating other longitudes and muting the day/night temperature contrast.  In our
twice-nominal rotation period case, $\phi_{\star}(t)$ changes in an eastward motion near 
periapse, revisiting previously heated regions of the atmosphere and increasing the width 
of the hot spot on the dayside.  This, in turn, affects the magnitude of the day-night 
temperature contrast and also the efficiency of day-night versus equatorial flow.  
In all of our rotation period cases, we see some convergence of the flow from the planet's dayside in the region between the eastern terminator and the nightside of the planet 
(left side of Figure~\ref{806_uvt}).


\begin{figure*}
\centering
\includegraphics[width=0.49\textwidth]{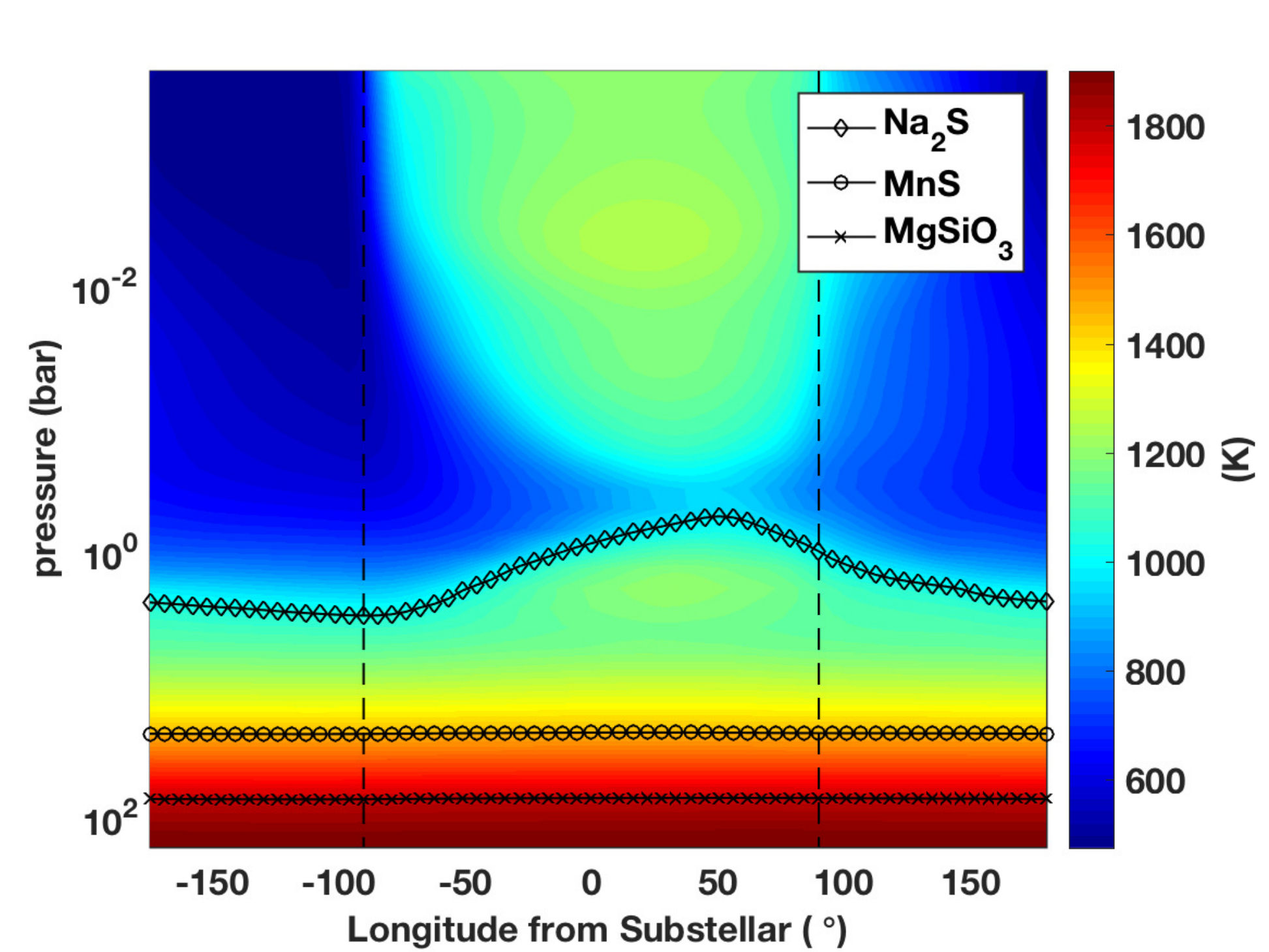}
\includegraphics[width=0.49\textwidth]{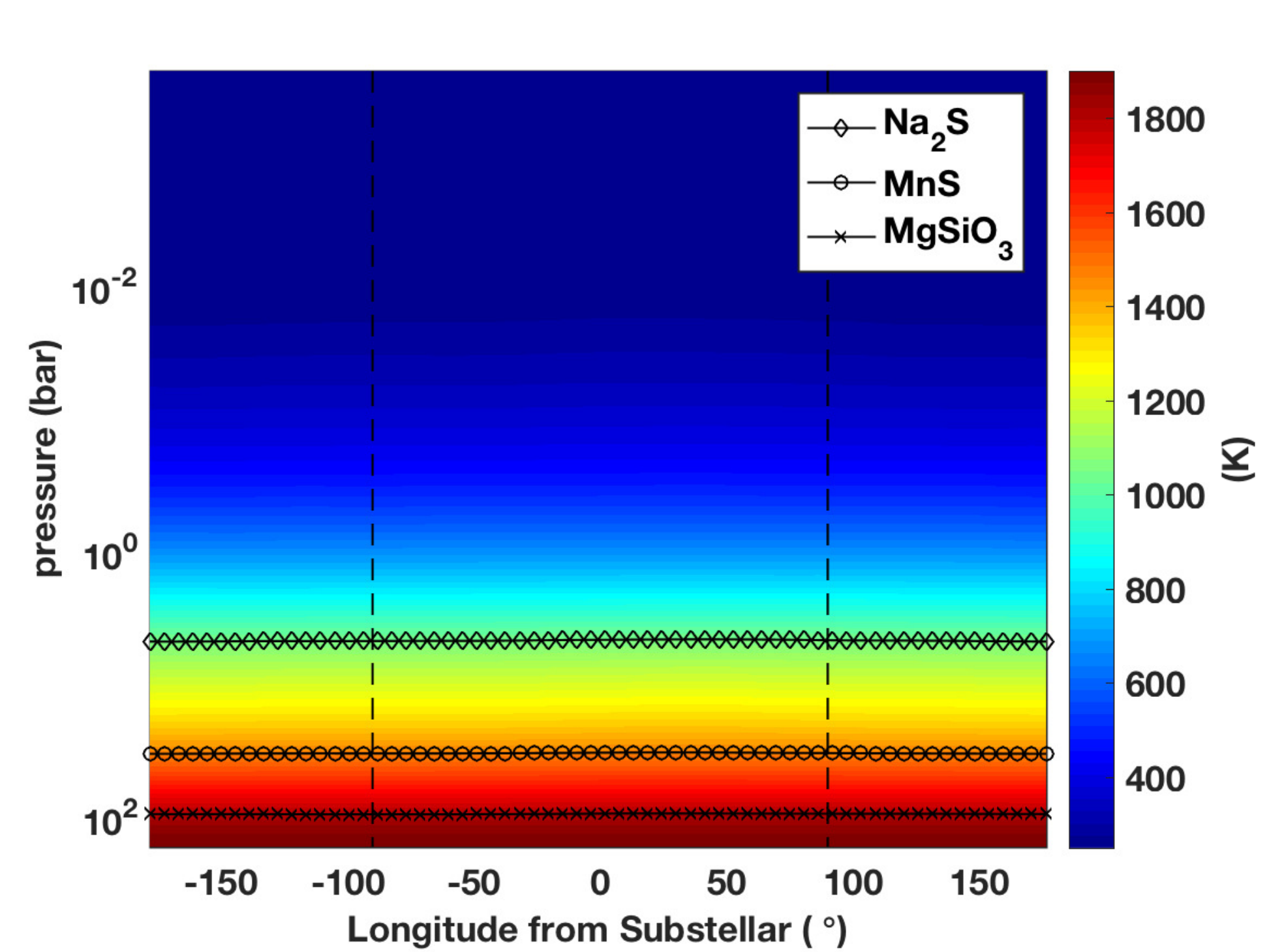}\\
\includegraphics[width=0.49\textwidth]{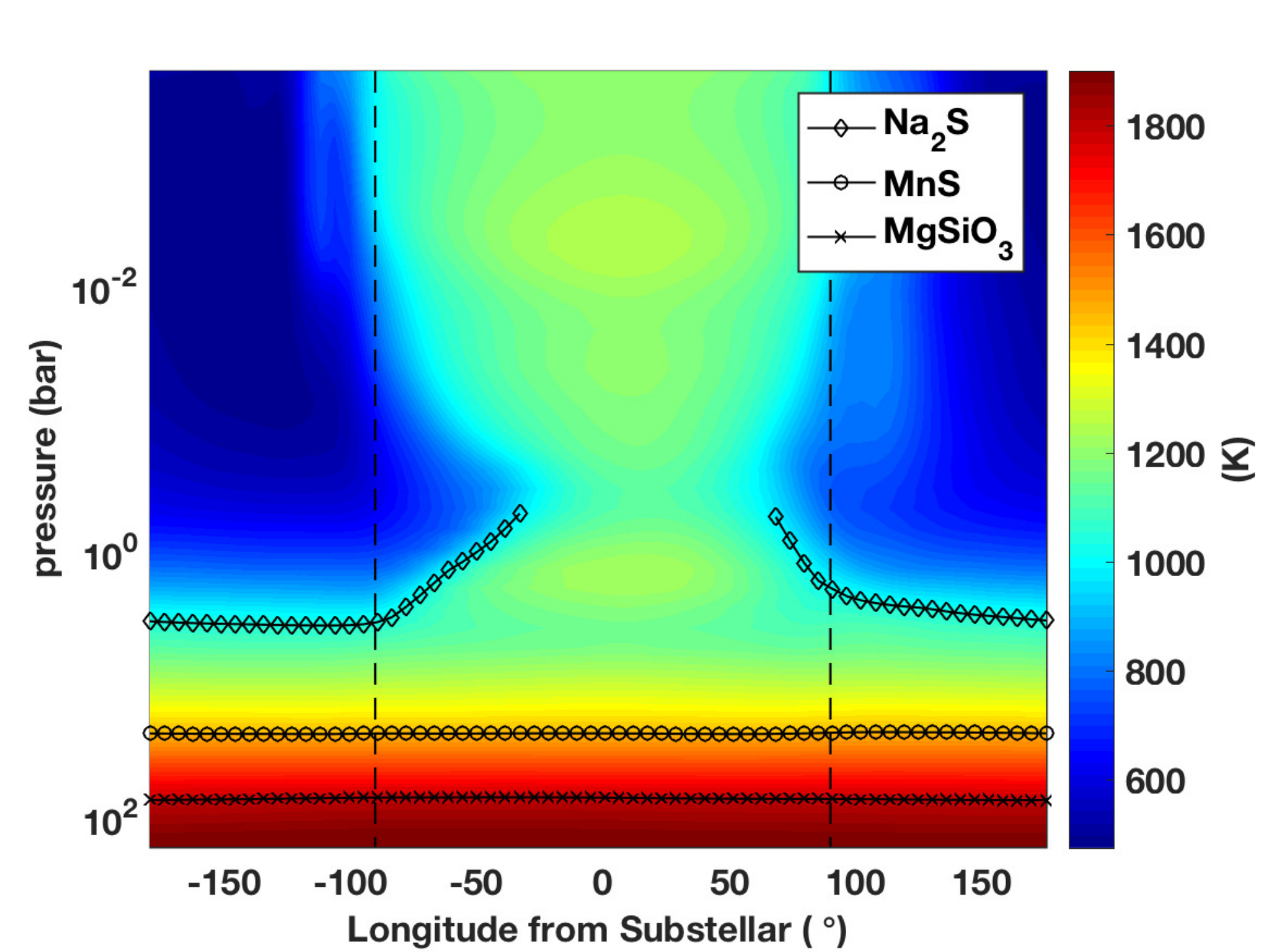}
\includegraphics[width=0.49\textwidth]{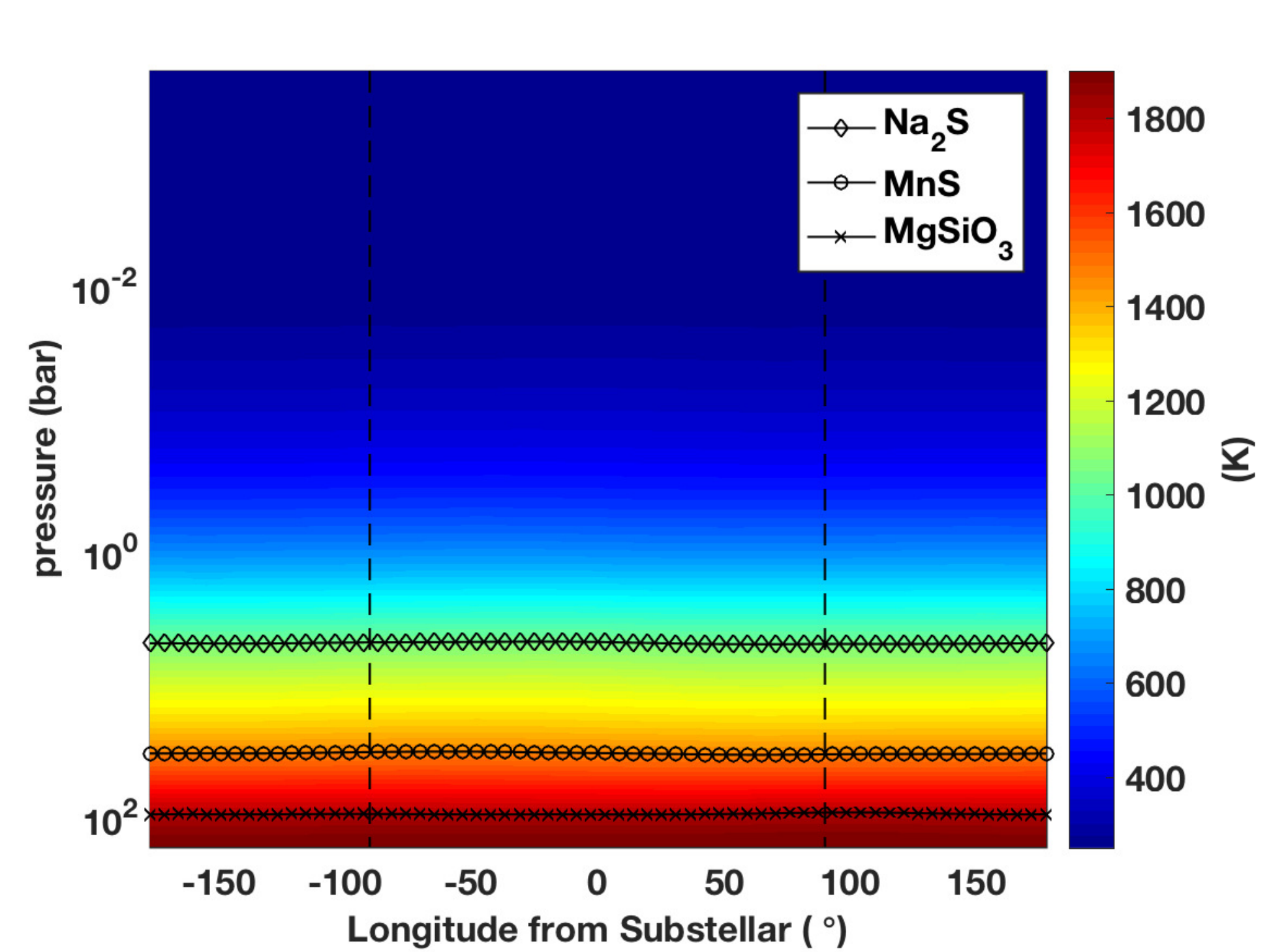}\\
\includegraphics[width=0.49\textwidth]{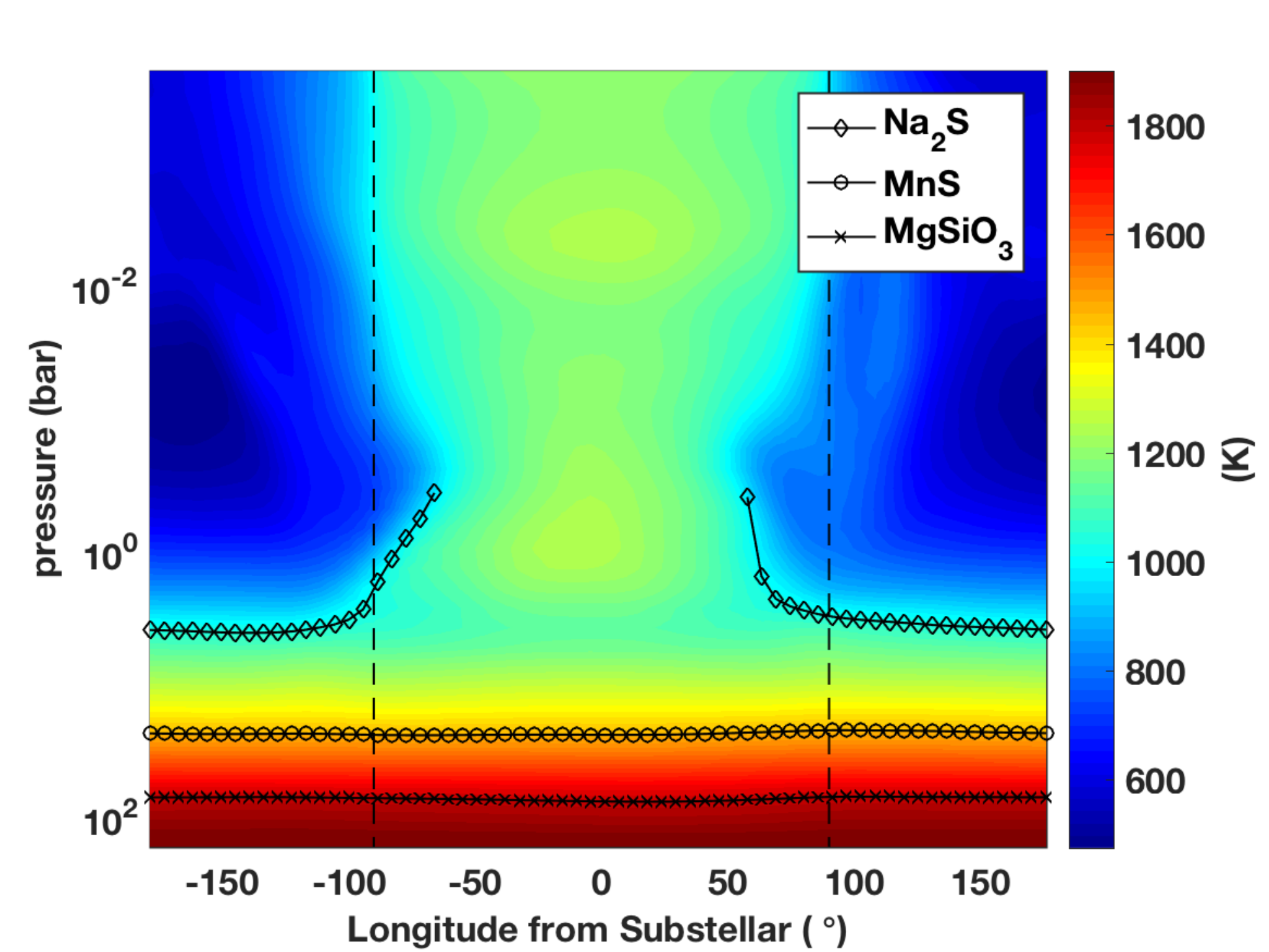}
\includegraphics[width=0.49\textwidth]{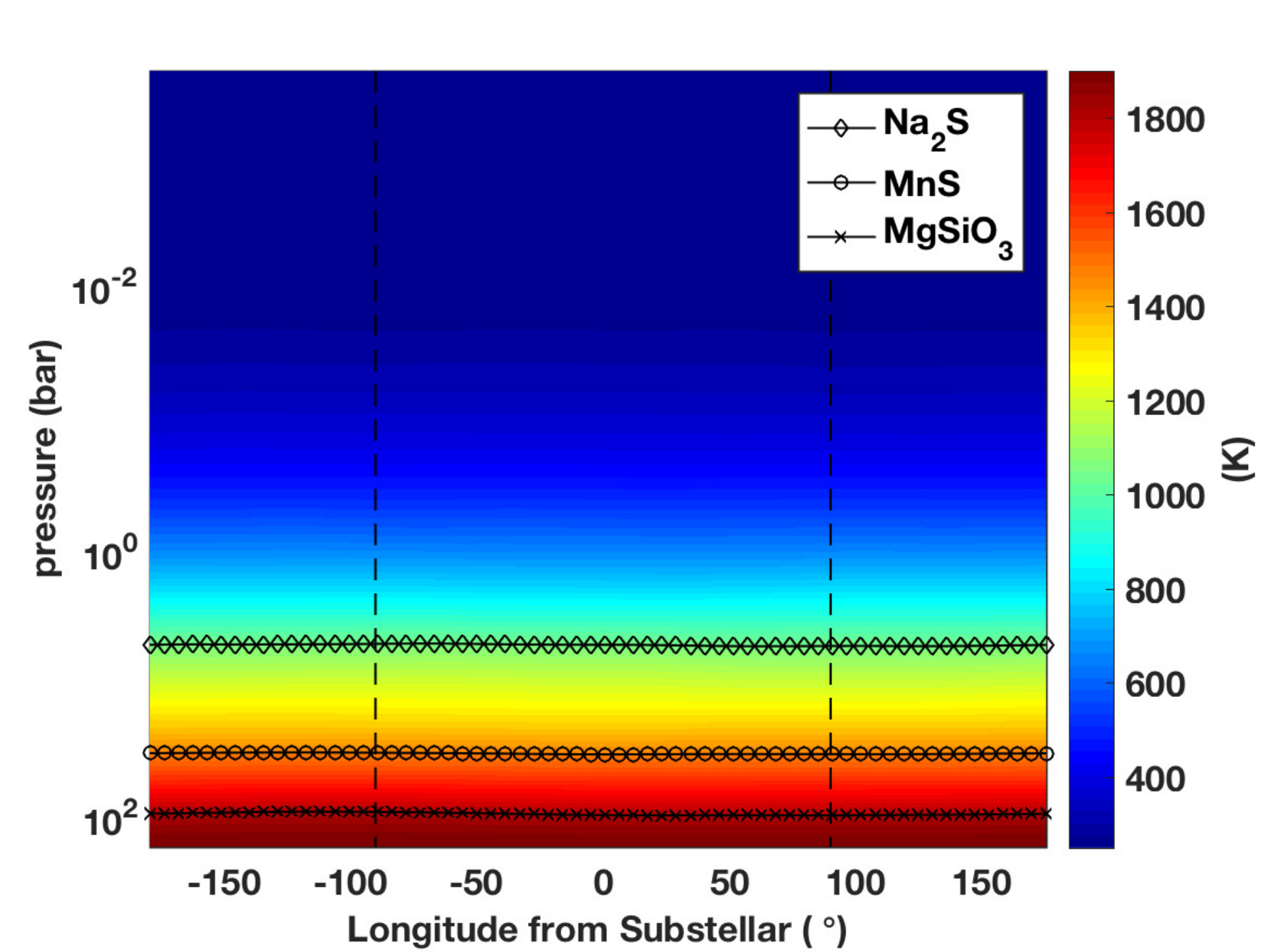}\\
\caption{Temperature (colorscale) averaged in latitude as a function of pressure and 
degrees from the substellar longitude near periapse (left) and apoapse (right)
for our half-nominal (top), nominal (middle), and twice-nominal (bottom)
rotation period cases near periapse.  Temperatures represent average values 
weighted by $\cos{\phi}$, where $\phi$ is latitude.  Contours in panels 
represent condensation points, in pressure-temperature space, 
for the cloud species considered in this study. Note the thermal 
inversion that develops at pressure less than 1~bar near periapse.}\label{806_templon}
\end{figure*}

\begin{figure*}
\centering
\includegraphics[width=0.49\textwidth]{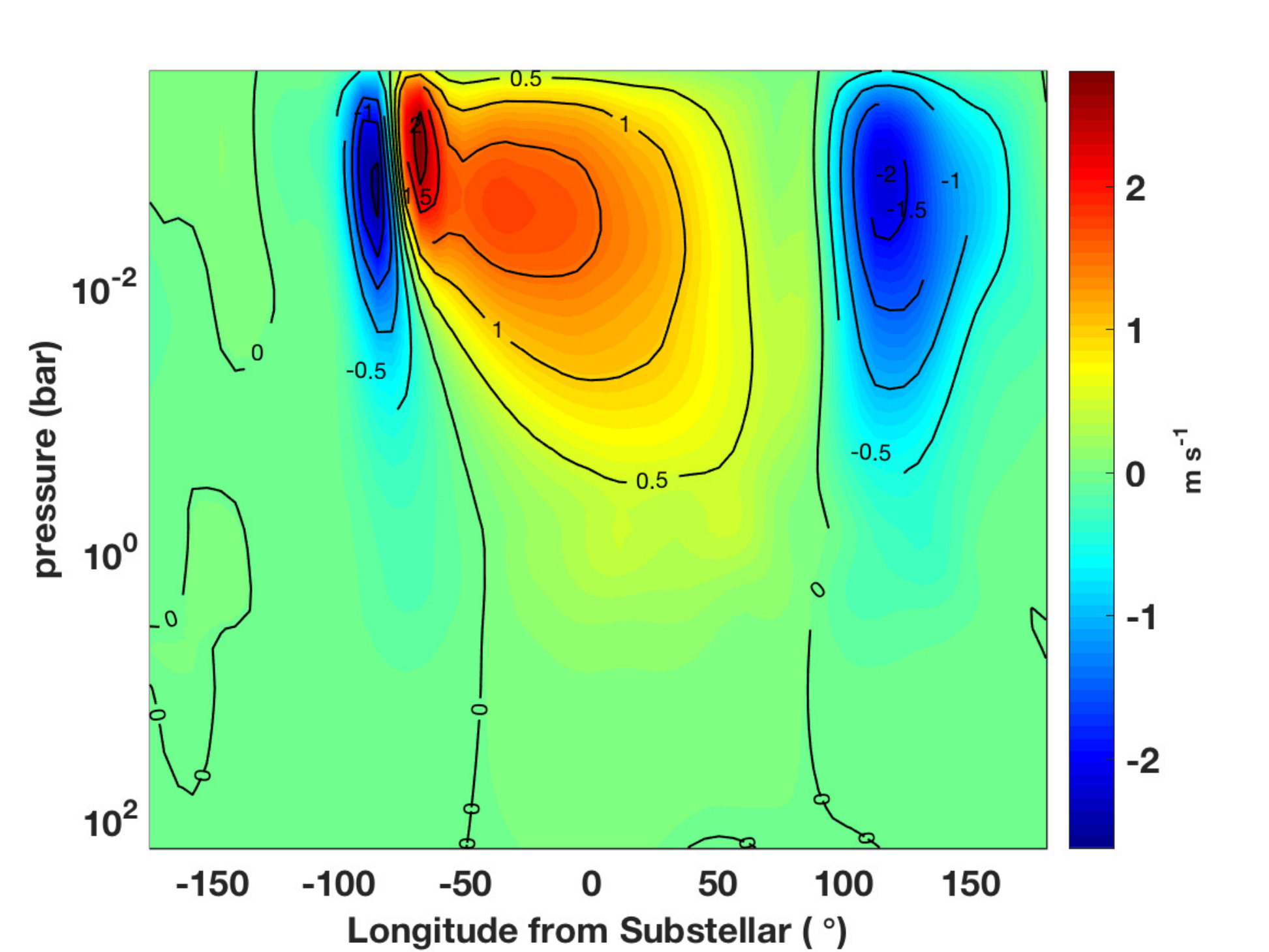}
\includegraphics[width=0.49\textwidth]{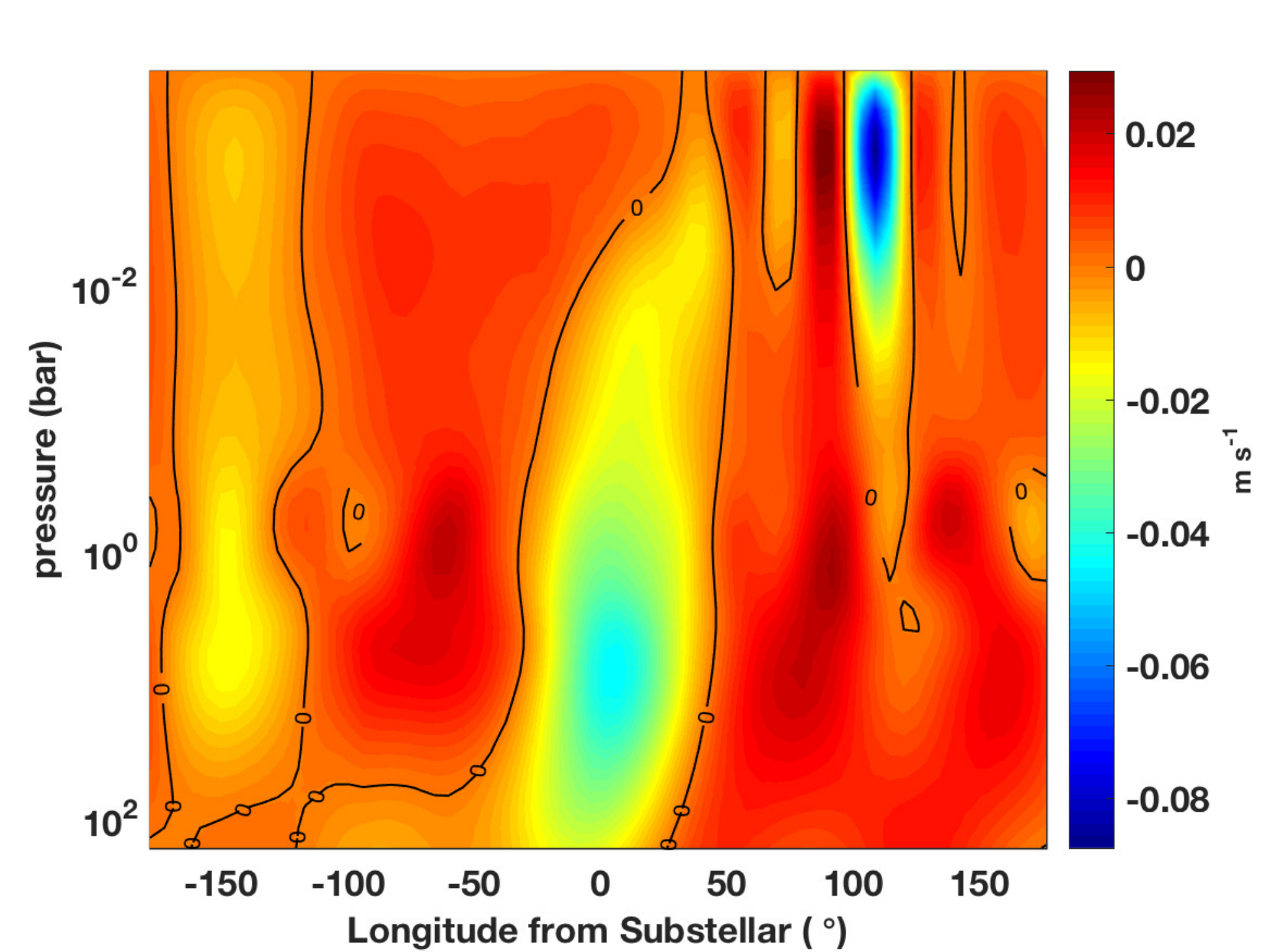}\\
\includegraphics[width=0.49\textwidth]{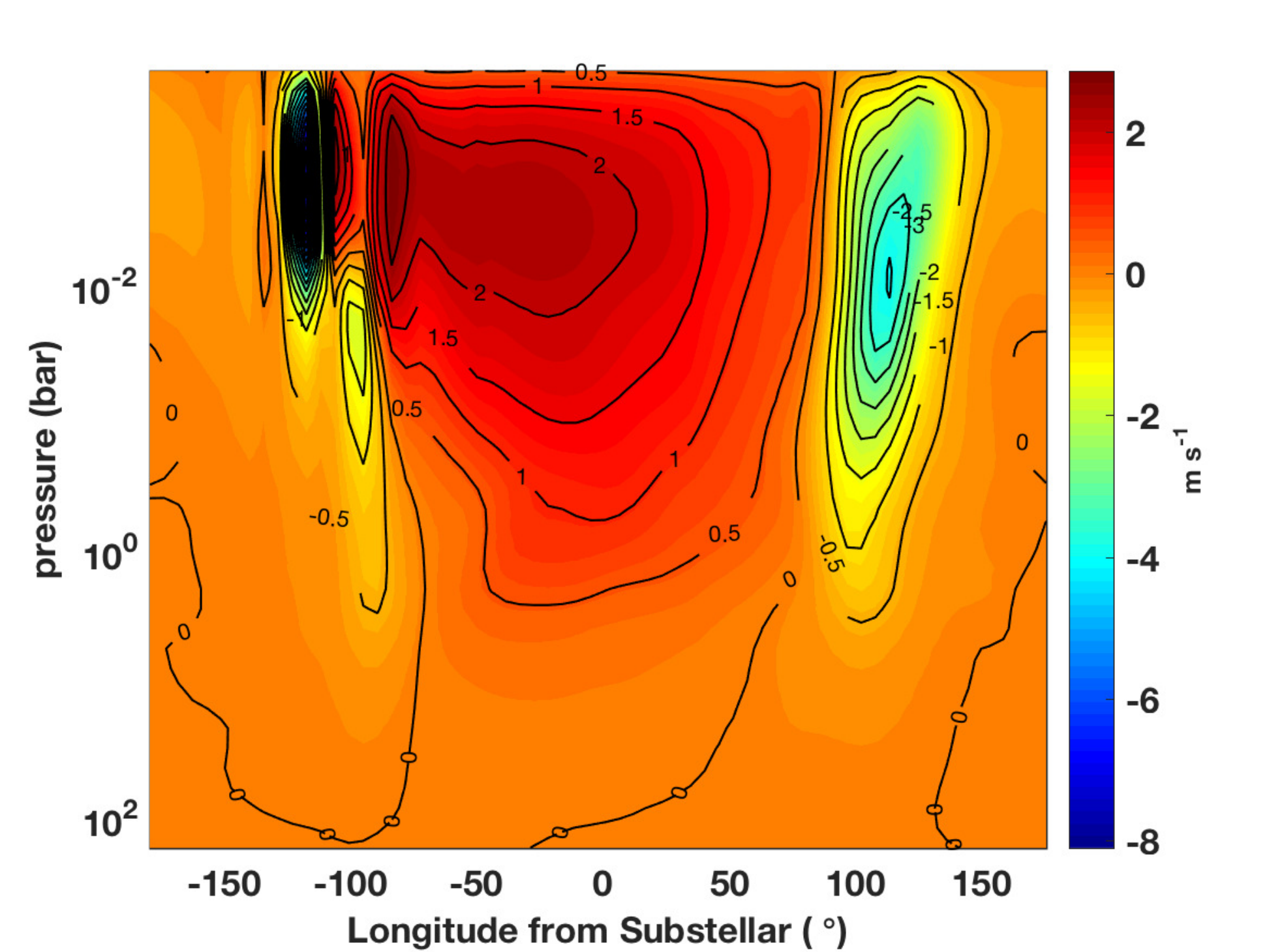}
\includegraphics[width=0.49\textwidth]{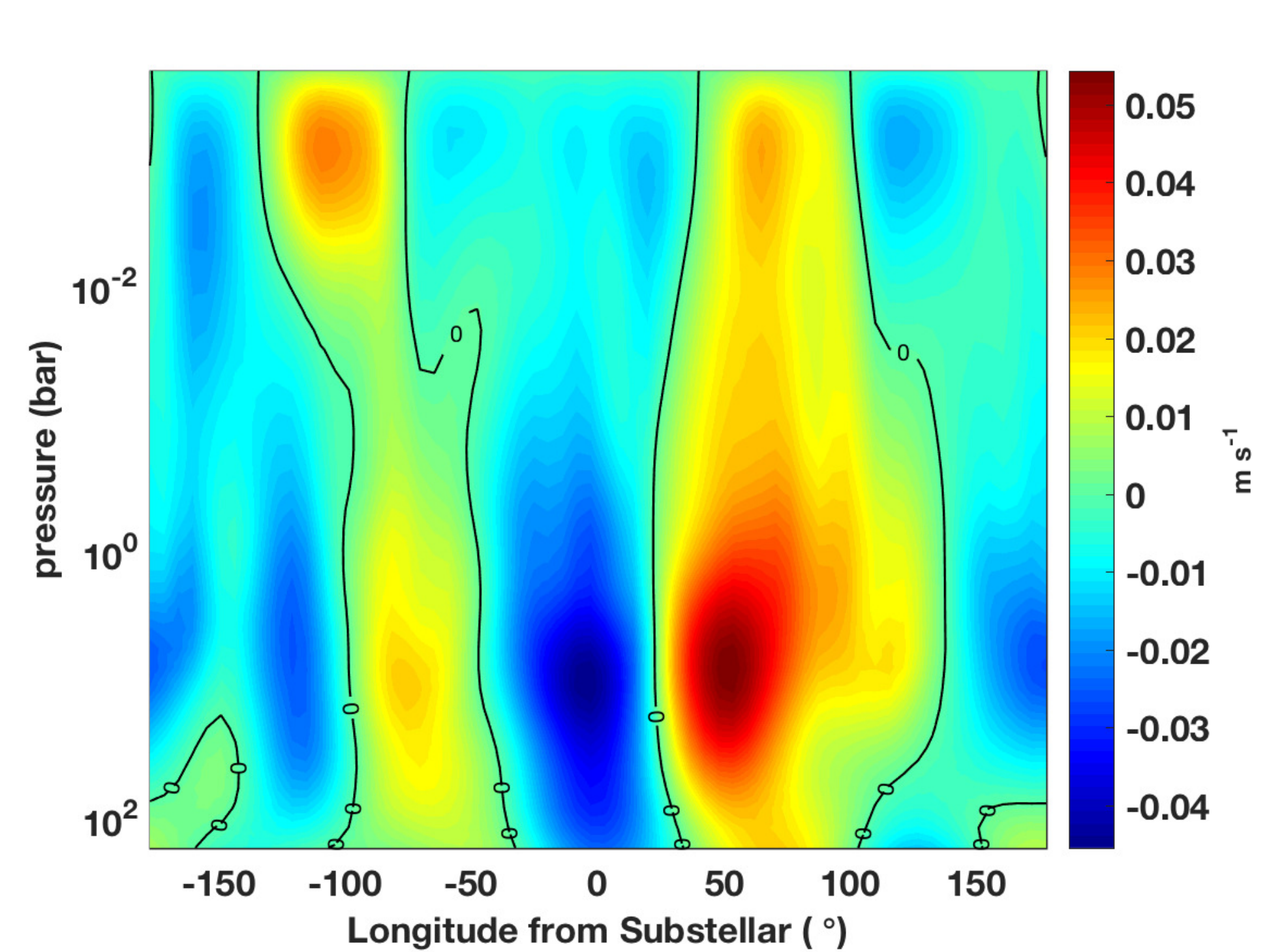}\\
\includegraphics[width=0.49\textwidth]{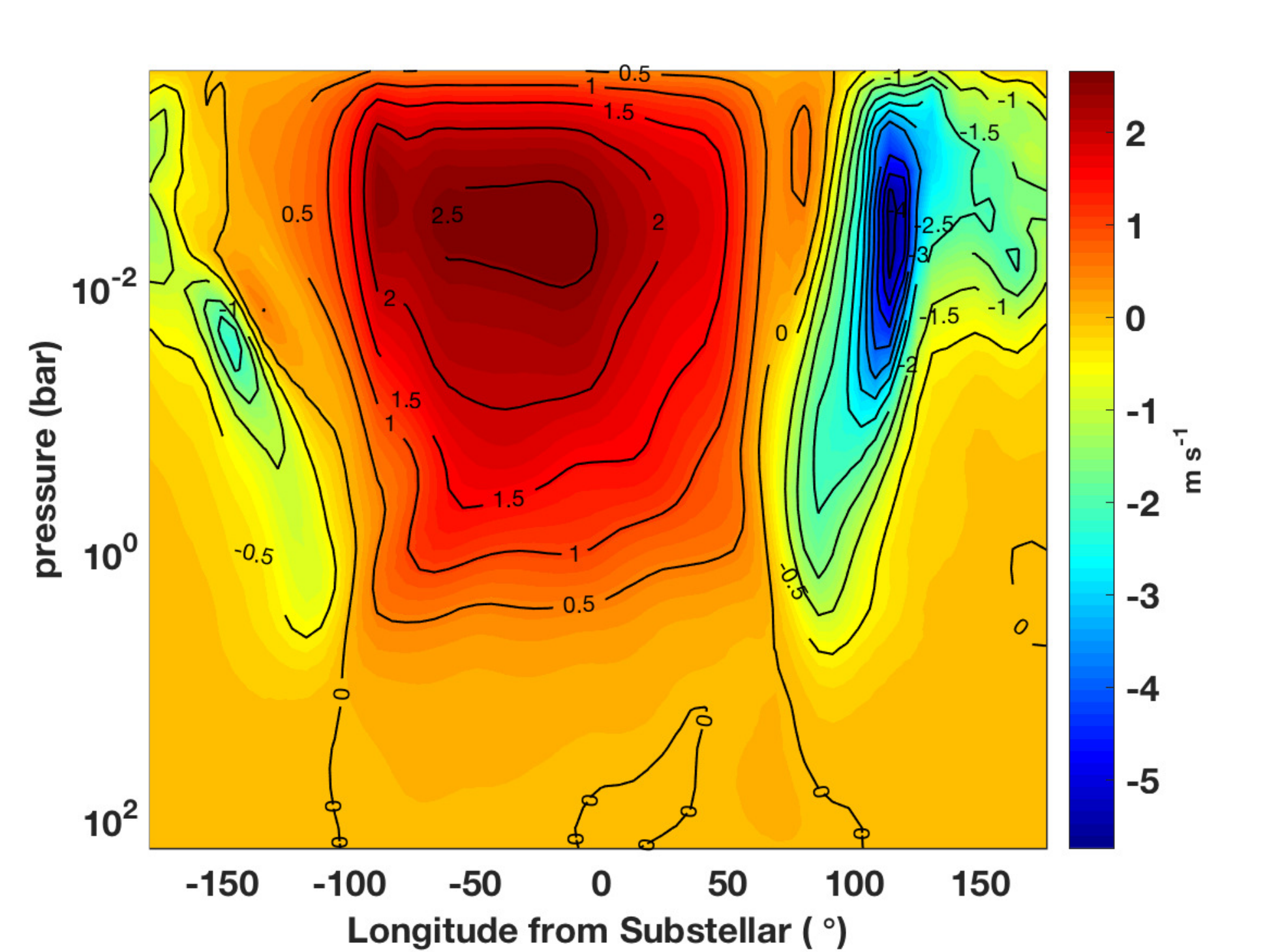}
\includegraphics[width=0.49\textwidth]{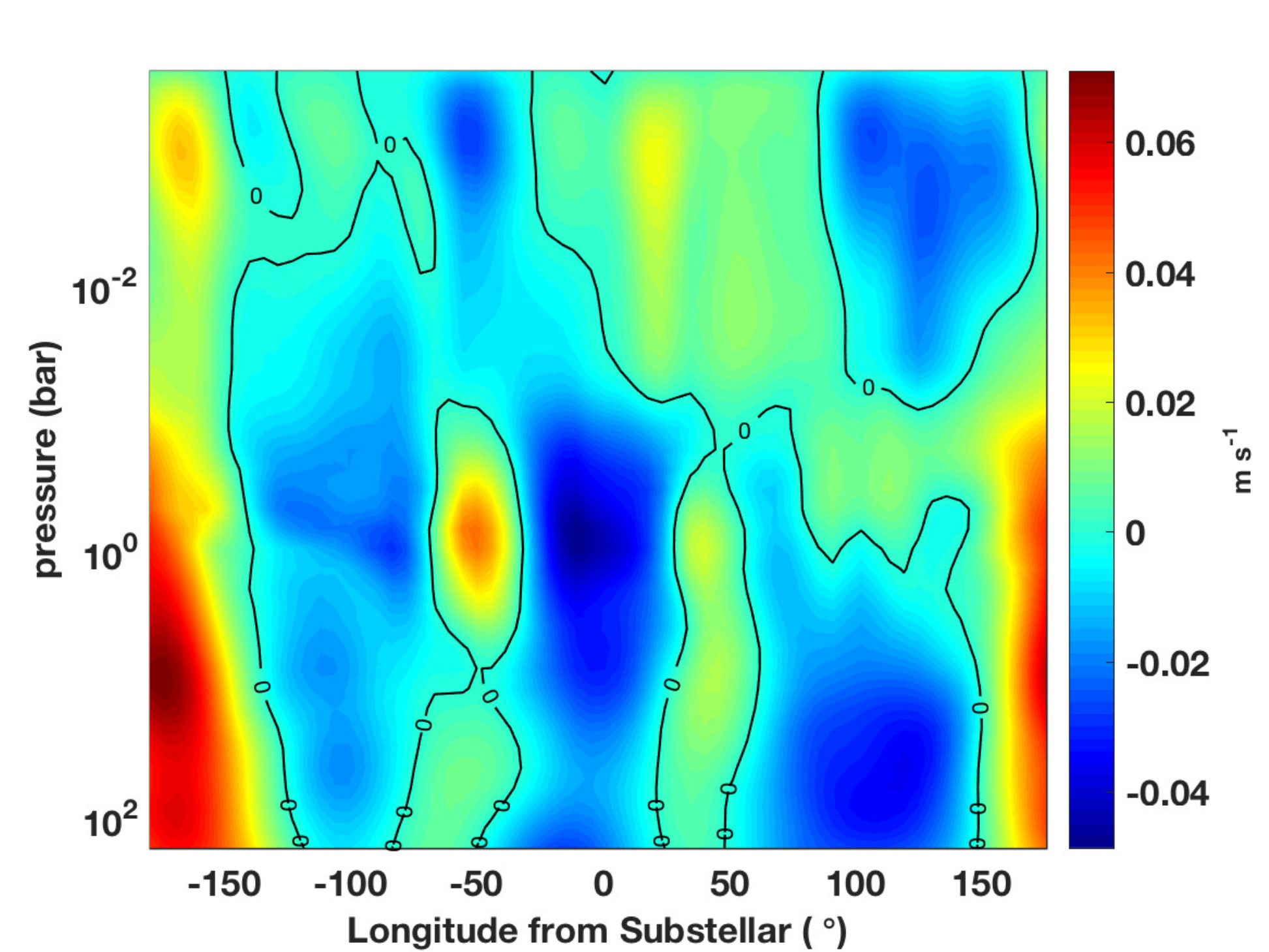}\\
\caption{Vertical velocity averaged in latitude as a function of pressure and 
degrees from the substellar longitude near periapse (left) and apoapse (right)
for our half-nominal (top), nominal (middle), and twice-nominal (bottom)
rotation period cases.  Vertical velocities represent average values 
weighted by $\cos{\phi}$, where $\phi$ is latitude.  Positive vertical 
velocity values represent updrafts while negative vertical 
velocities represent downdrafts.  Note the significant updrafts that 
develop on the dayside of HD~80606b in our models near periapse passage.
}\label{806_wlon}
\end{figure*}

The orientation of HD~80606b's orbit ($e = 0.93$ and $\omega=300.80^{\circ}$, \citet{pon09b}) 
is such that secondary eclipse, when the planet passes behind the 
host star as seen from Earth, occurs just three hours before periastron passage.  
This is an observationally advantageous orbital configuration for studies of 
planetary atmospheric response near periapse passage and has been leveraged in 
the phase-curve studies of \citet{lau09} and \citet{dewit2016}.  Phase-curve 
observations largely probe longitudinal brightness variations \citep{cow08}, 
which in the case of HD~80606b vary strongly as a function of time.  Figure~\ref{806_templon}
shows the latitude-weighted average temperature as a function of longitude from 
substellar point and pressure that evolves in our models of HD~80606b near periapse and apoapse. 
Near apoapse, our atmospheric models show virtually no variation in temperature from the 
planet's dayside to nightside, with temperatures ranging from $\sim$400-600~K above 
1~bar.  Near periapase, however, our atmospheric models of HD~80606b manifest 
large ($\sim$500 K above 10~bar) temperature variations with longitude.  The longitudinal 
location the peak temperatures and the depth to which the thermal structure of the planet is 
altered during periapse passage is a function of the assumed rotation period 
(Figure~\ref{806_templon}).  The offset of peak temperatures away from the substellar 
point increases with decreasing rotation period of the planet as the strength of the 
day-night vs equatorial flow decreases.  Similarly, the depth to which the planet's 
thermal structure is significantly altered increases with increasing rotation period, 
which reflects the duration that particular longitudes are exposed to incoming 
stellar radiation during periapse passage.

The transport of cloud material in HD~80606b's atmosphere near periapse passage will 
depend critically on both horizontal (see Figures~\ref{806_uzonal} and \ref{806_uvt}) 
and vertical transport in the planet's atmosphere.  Figure~\ref{806_wlon} shows the 
latitude-weighted average vertical wind as function of longitude from the substellar point 
and pressure in our HD~80606b models near periapse and apoapse.  Near apoapse, vertical 
velocities in our HD~80606b models are fairly weak ($\sim$0.05~m~s$^{-1}$), with patterns 
of updrafts and downdrafts that vary based on the assumed rotation period of the planet.  
Near periapse, strong updrafts develop on the dayside and strong downdrafts on the nightside 
of the planet in our models of HD~80606b's atmosphere.  These regions of updrafts and 
downdrafts near periapse correspond to regions of horizontal wind divergence and convergence 
seen in Figure~\ref{806_uvt}.  The location and depth of the strongest regions of updrafts and 
downdrafts vary with assumed rotation period, with the twice-nominal rotation period 
model manifesting the deepest updraft that is mostly closely located near the substellar 
longitude.  Under all rotation period assumptions, the presence of strong dayside updrafts
could allow for cloud material located at depth in HD~80606b's atmosphere to 
be lofted into the visible regions of the planet's atmosphere.

\begin{figure}
\centering
\includegraphics[width=\textwidth]{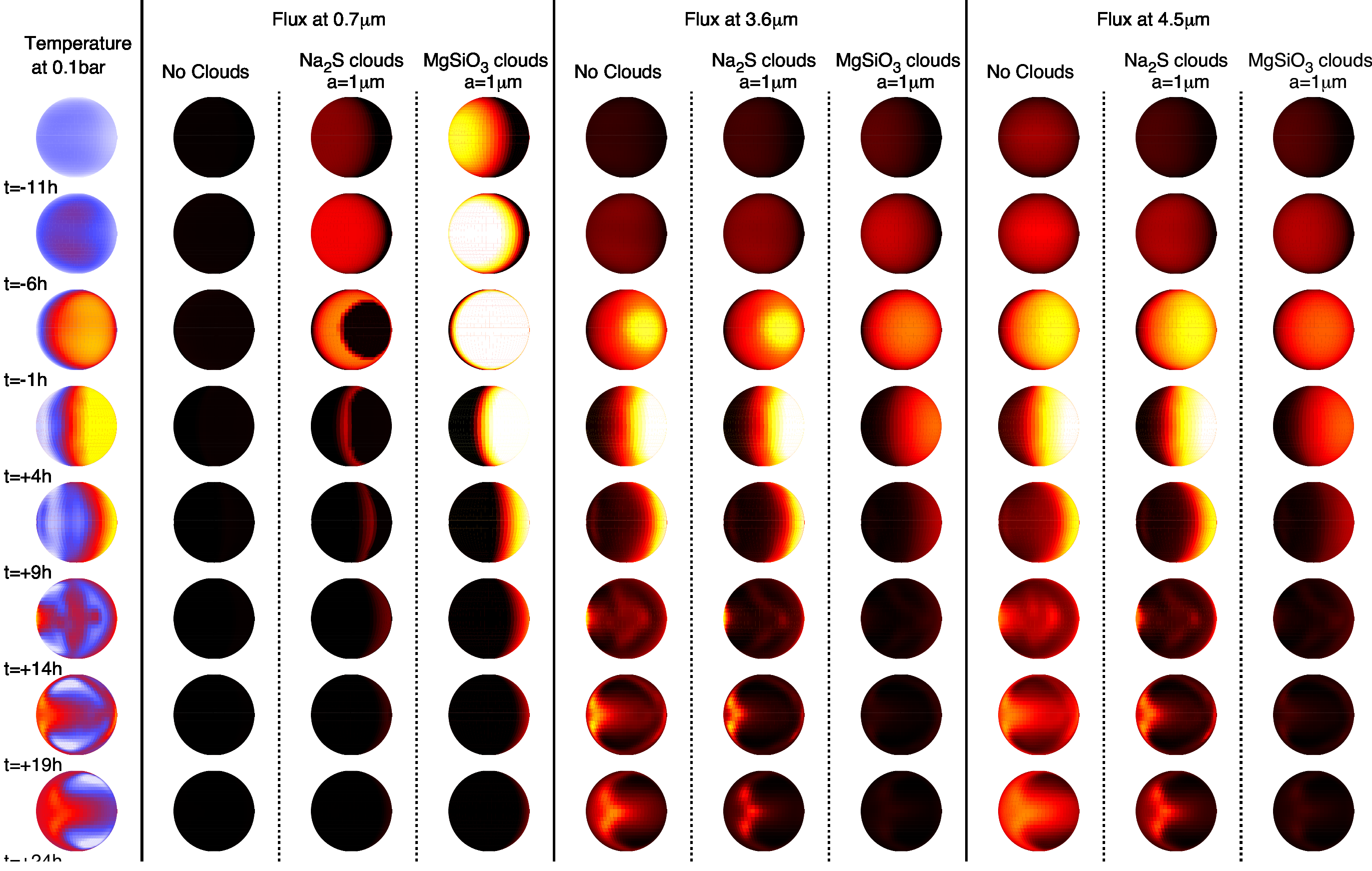}\\
\caption{Temperature at $0.1 bar$ (left column) and flux emerging from the Earth-facing planet hemisphere at three different wavelengths assuming either that the atmosphere is cloudless or that $\rm Na_2S$ or $\rm MgSiO_3$ clouds are present. The different rows are for different times, with $t=0$ being the periapse passage. The temperature scale ranges linearly from 500K to 1200K. The brightness scale is linear and normalized for each wavelengths to the brightest point simulated, so brightness maps can be compared between different times and different cloud species but not between different wavelengths. At $0.7\mu m$ the planetary flux is dominated by reflected light and cloudy parts appear bright and cloudless parts appear dark. At longer wavelengths, dominated by thermal emission, the presence of clouds raises the photosphere to higher, cooler layers and the planet appears dimmer. $\rm Na_2S$ clouds disappear partially from the dayside at $t=-1h$ when the planet becomes hotter than the condensation temperature. On the contrary, $\rm MgSiO_3$ clouds are always present.}
\label{fig::FluxMaps}
\end{figure}

\subsection{Cloud Formation and Evolution}\label{cld_evol}

Rapid formation/dissipation and transport of clouds is likely to occur in the atmosphere of 
HD~80606b during periapse passage. As shown in Figure~\ref{fig::FluxMaps} the evolution of the 
cloud coverage through periapse passage depends on the cloud species' condensation temperatures. 
When HD~80606b's atmospheric temperatures do not exceed the cloud condensation temperature,
such as for $\rm MgSiO_3$ or $\rm MnS$ clouds, the planet is fully cloudy all the time. 
When the atmospheric temperatures get hotter than the condensation temperature, 
such as for $\rm Na_2S$ clouds, then a hole in the clouds forms when the planet 
gets close to periapse (see Figure~\ref{806_templon}), qualitatively affecting the flux maps shown in Figure~\ref{fig::FluxMaps} . 
By measuring the evolution of the cloud coverage in HD~80606b, 
one could possibly determine the most probable cloud composition. For all cloud compositions 
considered in this study, the nightside should remain cloudy during periapse passage, 
leading to a dimming of the atmosphere at infrared wavelengths.

Atmospheric transport of cloud particles, not self-consistently considered here, could 
affect the picture presented in Figure~\ref{fig::FluxMaps}. At apoapse, all the clouds 
modeled here have cloud bases well below the photosphere near 300~mbar. If vertical 
mixing is small, then these species are trapped near the cloud base and should not be 
present in the observable portion of the atmosphere. During periapse, strong updrafts 
on the dayside can transport the cloud material from its original cloud base to the 
observable atmosphere (see Figure~\ref{806_wlon}).  Horizontal winds can also potentially 
play a role in transporting material lofted on the dayside to the nightside of the planet 
(see Figures~\ref{806_uzonal} and \ref{806_uvt}).

The depth of the cloud bases near apoapse strongly depends on the under-constrained 
internal temperature profile of HD~80606b. Tidal dissipation \citep[e.g.][]{bod2001} 
and other atmospheric processes (e.g. Ohmic dissipation, \citet{bat2010}) could 
significantly increase the deep temperature profile of HD~80606b and raise 
the cloud deck closer to the photosphere.  
In our models we have assumed an internal temperature ($T_{\rm int}$) of 100~K, which 
is consistent with the internal temperature of Jupiter \citep{for2011}. 
If the assumed internal temperature of HD~80606b was 
raised to 500~K or 1000~K the temperature at 200~bar level in our model atmospheres 
would be increased to roughly 3000~K and 4000~K respectively.  This increase in the 
internal temperature would cause an `upward' shift in the thermal structure of the 
models presented in Figure~\ref{806_templon}, which would result in the cloud base 
pressure for MgSiO$_3$ to move from roughly 100~bars to 10~bars and 1~bar under the 
assumption that $T_{\rm int}$ equals 500~K and 1000~K, respectively.

\begin{figure}
\centering
\includegraphics[width=0.49\textwidth]{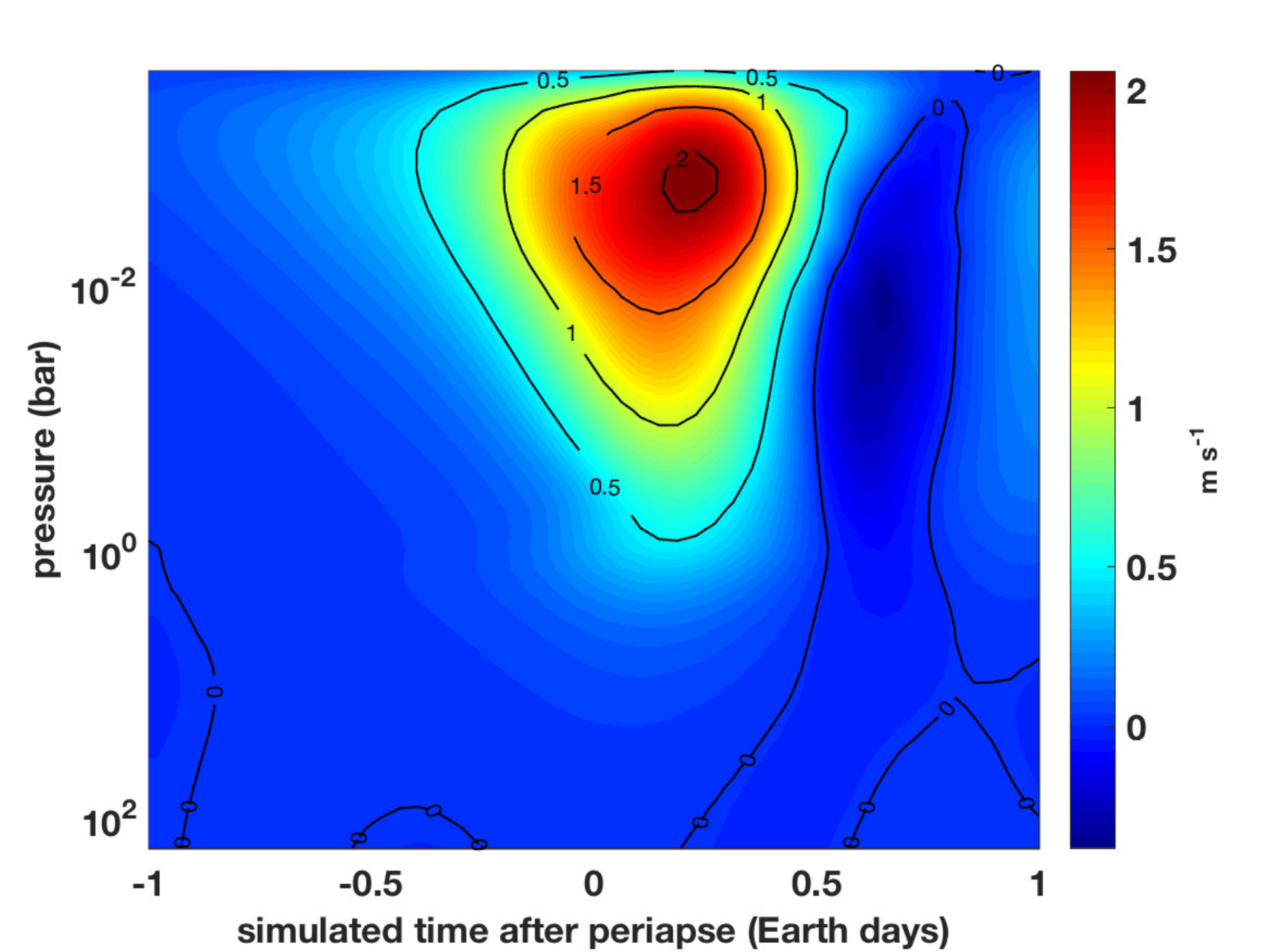}\\
\includegraphics[width=0.49\textwidth]{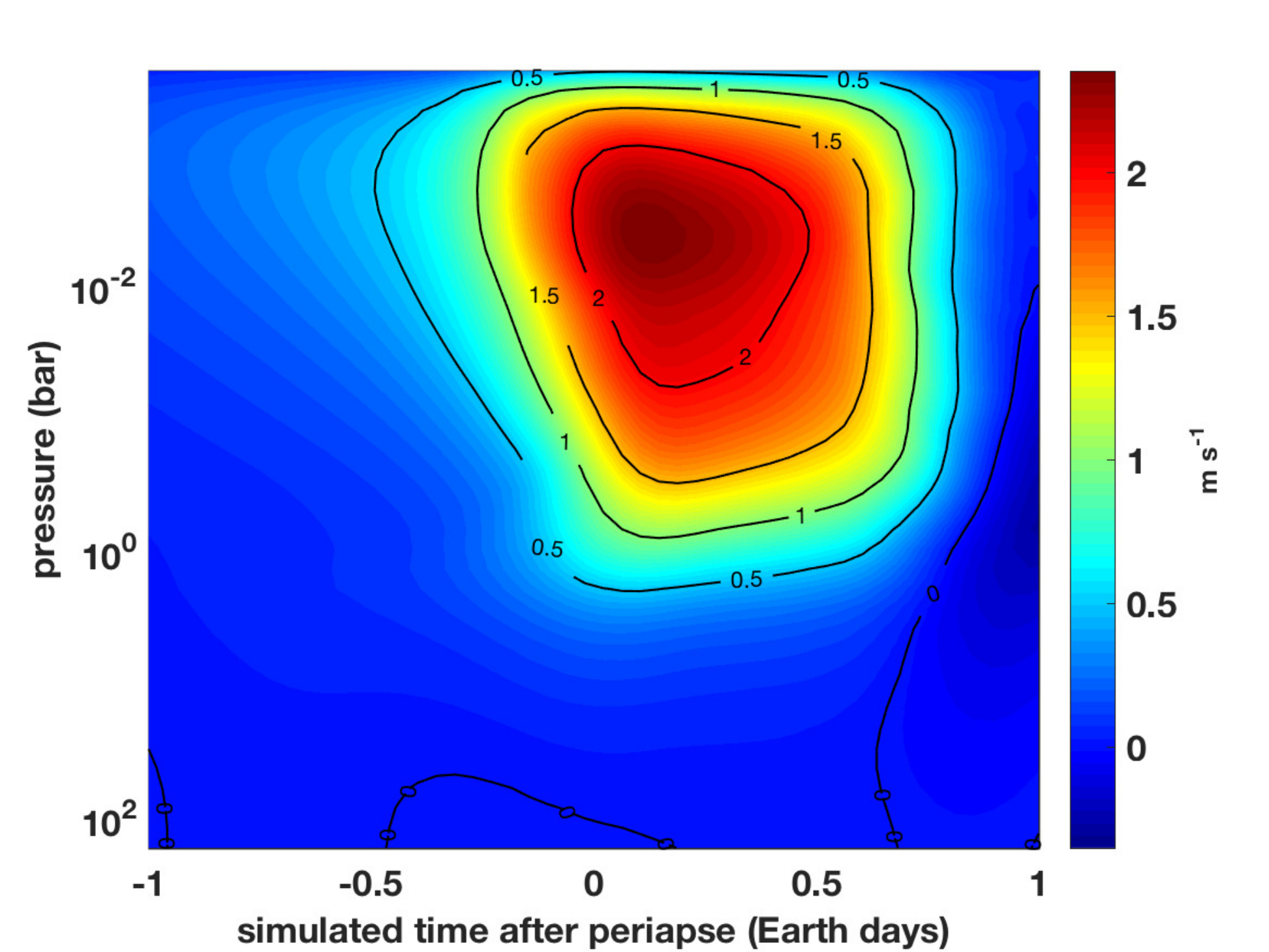}\\
\includegraphics[width=0.49\textwidth]{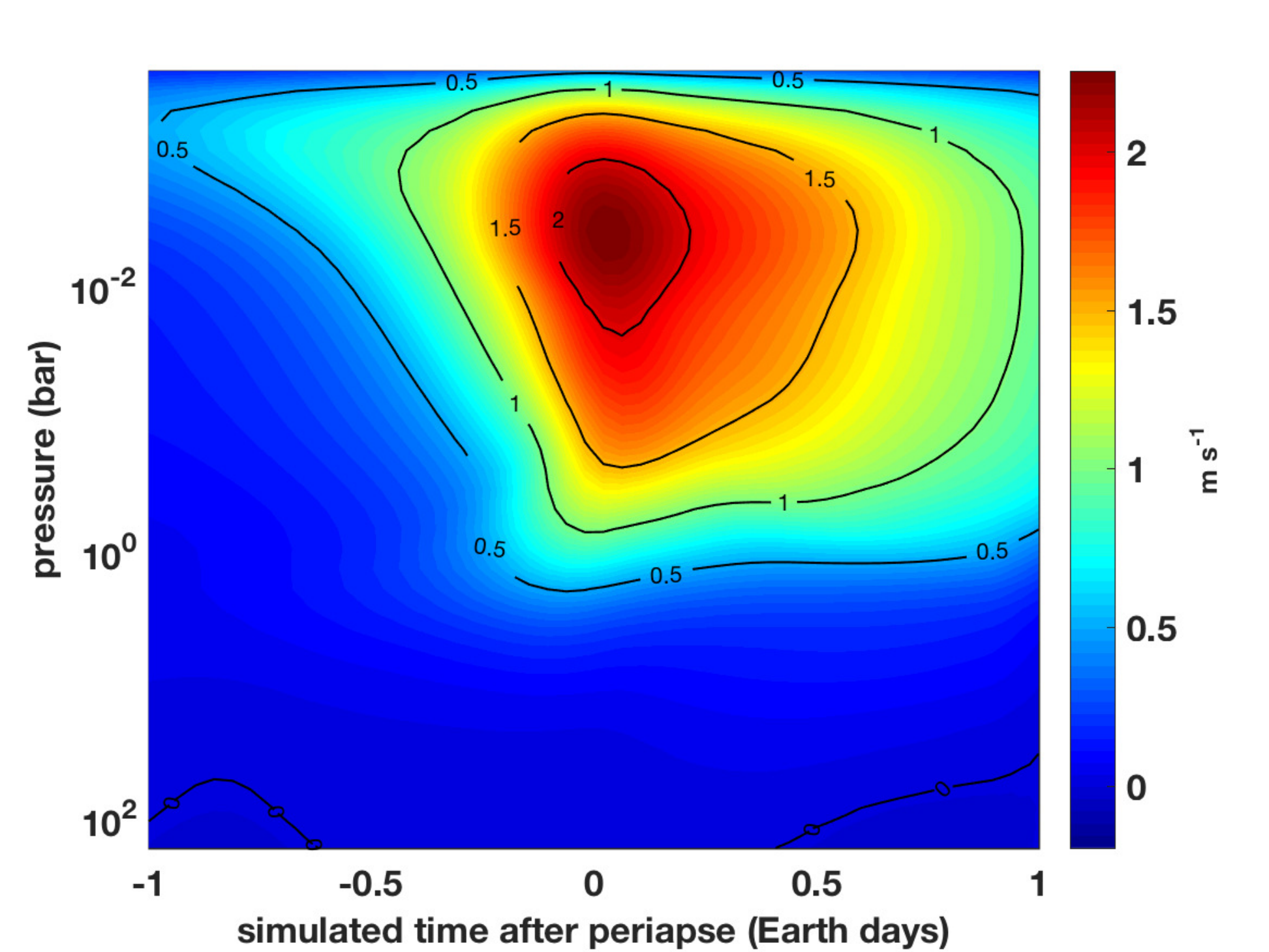}\\
\caption{Dayside average vertical velocity (m~s$^{-1}$)  as a 
function of pressure and time from periapse passage for our 
half-nominal (top), nominal (middle), and twice-nominal (bottom)
rotation period cases.  Positive vertical 
velocity values represent updrafts while negative vertical 
velocities represent downdrafts.
The strong updraft that develops in the dayside of our HD~80606b models near periapse 
persists in the atmosphere for $\sim$1~day. Note that the strength and duration of the 
dayside updraft depends on the rotation period assumed for the planet.}\label{806_vertvel_time}
\end{figure}

During periapse passage, an updraft with a vertical velocity  
of on average 1~m~s$^{-1}$ develops on the dayside 
of our HD~80606b atmospheric models that persists for $\sim$1~day 
(Figure~\ref{806_vertvel_time}). This dayside updraft can lift 
particles over $\sim$100~km in atmospheric height, corresponding to 
two scale heights and thus one order of magnitude in pressure. As a consequence, 
only particles from pressures shallower than $\sim$3~bars can be lofted above 
the photosphere during periapse to form an observable cloud. If we assume that at 
apoapse the cloud layers are approximately one scale height thick, then only clouds 
that have a cloud base pressures shallower than $\sim$10~bar can be present 
in the observable portion of HD~80606b's atmosphere. In the case of a cool deep 
temperature profile (T$_{\rm int}$=100~K), only $\rm Na_2S$ clouds can be lofted 
above the photosphere during periapse passage. If deep atmosphere is hotter 
($T_{\rm int}$=500~K), then $\rm MgSiO_3$ and MnS clouds could also be present. 
Observational determination of the composition of clouds present in HD~80606b's 
atmosphere during periapse passage could help constrain the deep pressure 
temperature profile of the planet. 

\subsection{Theoretical Light Curves}

The SPARC model is uniquely equipped to produce theoretical light curves 
and spectra for HD~80606b that account for spatial thermal variations, 
cloud coverage and dynamics within the atmosphere \citep[see][]{for06, sho09}.  
This capability allows us 
to make observational predictions based on a variety of assumptions about the 
planet's atmospheric composition (in this case clouds) and rotation period.  
In this section, we focus our model-observation comparisons on the {\it Spitzer} 
4.5 and 8~$\mu$m bandpasses used in \citet{dewit2016} to study HD~80606b.  
We consider both cloud-free and cloudy models when exploring derived 
theoretical phase curves and comparing them with the phase variations 
observed for HD~80606b near periapse.

\begin{figure*}
\centering
\includegraphics[width=0.61\textwidth]{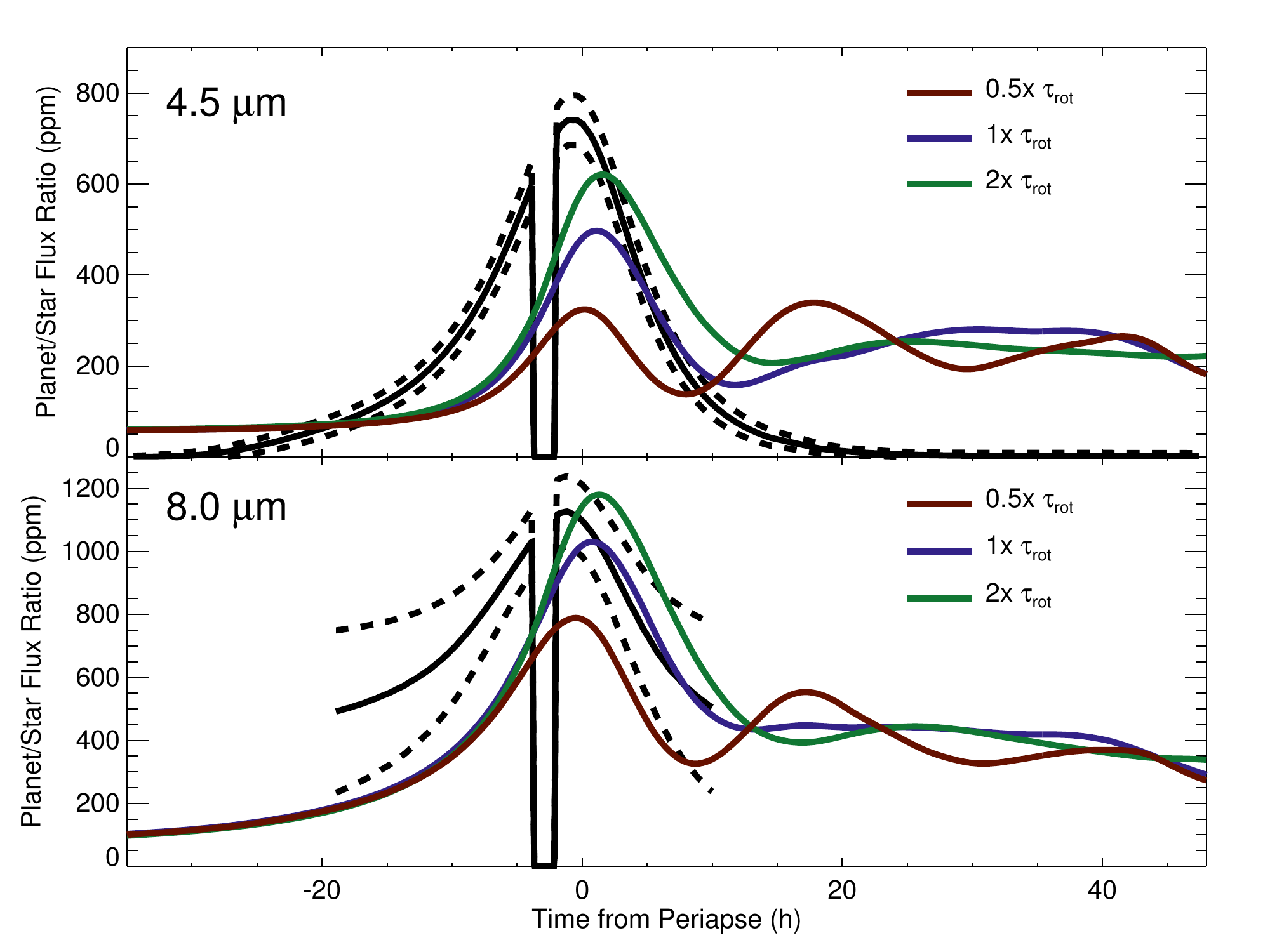}
\includegraphics[width=0.37\textwidth]{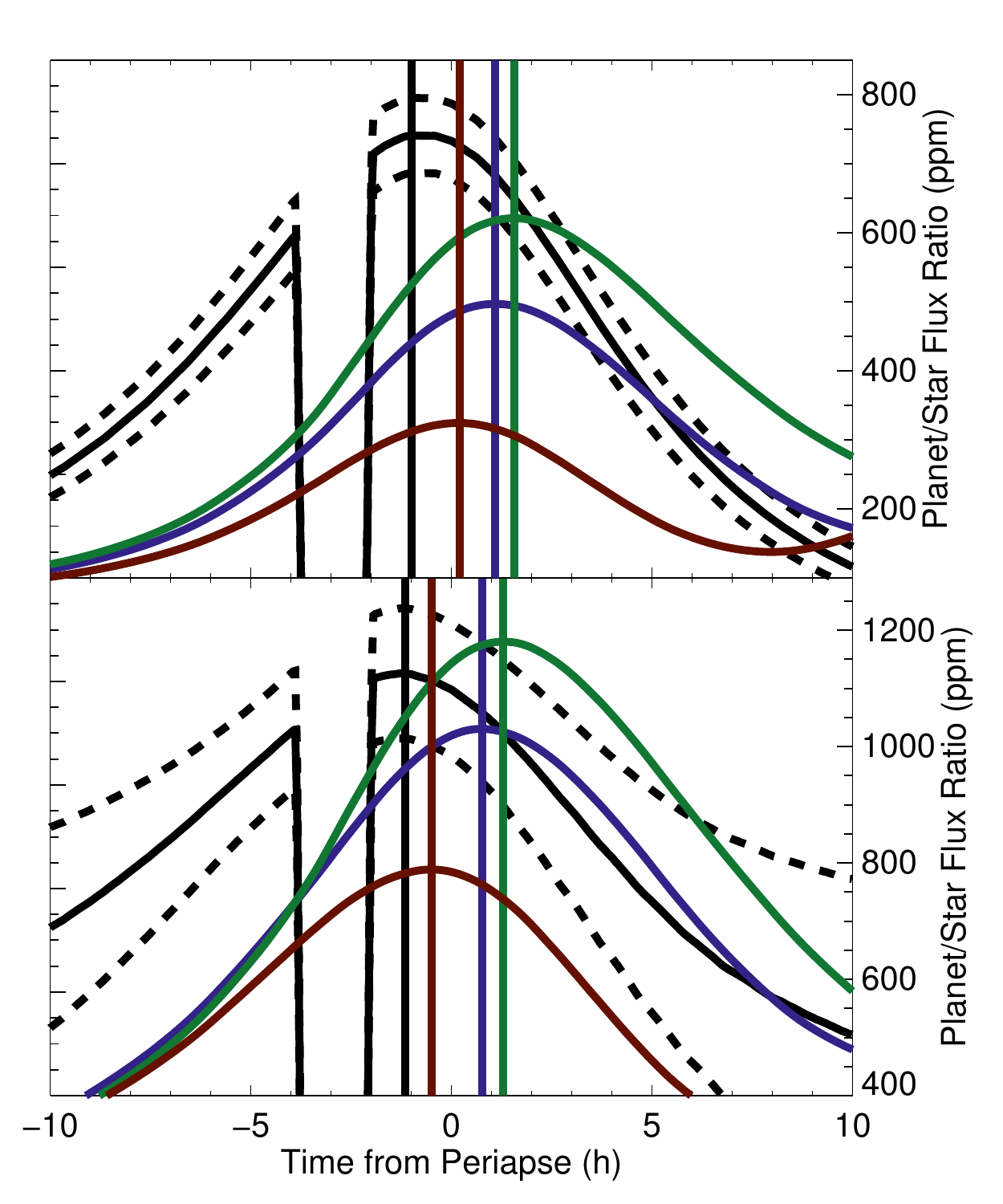}
\caption{Observed Planet/Star flux ratio (in parts per million, ppm) as a function 
of time from periapse passage in the {\it Spitzer} 4.5 and 8.0~$\mu$m channels 
presented in \citet{dewit2016}.  Dashed lines represent the 1$\sigma$ envelopes for 
the observed phase variations.  Theoretical phase curves derived from the 
half-nominal (red), nominal (blue), and twice-nominal (green) rotation period 
cloud-free models of HD~80606b presented in this study are included for comparison.  
Right panel is a zoom-in of the left panel in the region near periapse passage.  
Vertical lines in right panel represent location of the peak of the planetary flux.  
Note that all of our cloud-free models over-predict the time of flux maximum.  The 
twice-nominal (green) model provide the best match of the observed phase amplitude, 
which is consistent with the rotation period derived by \citet{dewit2016}.
         }\label{hd806_lc_ch2_ch4}
\end{figure*}

\subsubsection{Model-Observation Comparisons:  Cloud-Free}

Given the potential for clouds to be sequestered well below the photosphere of 
HD~80606b or rapidly dissipated near periastron passage (see Figures~\ref{806_templon}), 
it is possible that clouds play a negligible role in shaping HD~80606b's time-varying 
flux.  Figure \ref{hd806_lc_ch2_ch4} presents theoretical light curves at {\it Spitzer's}
4.5 and 8~$\mu$m assuming the half-nominal, nominal and twice-nominal rotation period 
for the planet.  The amplitude and shape of the light curves near periapse 
passage is a strong function of the assumed rotation period.  The orbital configuration 
of HD~80606b is such that an earth observer sees the full dayside hemisphere three 
hours before the periapse passage.  As HD~80606b continues to rotate, more and more 
of the nightside hemisphere contributes to the observable flux from the planet.  
We find that peak of the observable planetary flux from our HD~80606b 
models peaks very near periapse as the result of the combination 
of planetary rotation period, radiative timescales, and observing geometry of 
the system.  

The timing of the predicted peak of the planetary flux does vary with the assumed 
planetary rotation period and the observational bandpass being considered.  In 
the 8~$\mu$m bandpass, our models predict the timing of the peak observable 
planetary flux to occur 0.5 hours before, 0.75 hours after, and 1.3 hours after  
periapse passage for the half-nominal, nominal, and twice-nominal rotation periods 
respectively.  In the 4.5~$\mu$m bandpass, our models predict the timing of the peak 
observable planetary flux to occur 0.2, 1.1, and 1.6 hours after periapse passage for the 
half-nominal, nominal, and twice-nominal rotation periods respectively.  In all cases, 
our models over-predict the time of peak observable flux for HD~80606b compared with 
the $\sim$1 hour before periapse passage observed by \citet{dewit2016} in both the 
4.5 and 8.0~$\mu$m channels.

We also find that the magnitude of the peak of the planetary flux is a strong function 
of the assumed rotation period.  This is not surprising since the rotation period of the 
planet determines both the rate at which the substellar point sweeps in longitude, but also 
the fraction of the dayside hemisphere viewed by an earth observer.  We find peak values in 
the planet/star flux ratio at 8~$\mu$m of 790~ppm, 1030~ppm, and 1180~ppm for the 
half-nominal, nominal, and twice-nominal rotation period cases respectively compared with 
the observed peak value of 1130~ppm from \citet{dewit2016}.  We find peak values in 
the planet/star flux ratio at 4.5~$\mu$m of 320~ppm, 500~ppm, and 620~ppm for the 
half-nominal, nominal, and twice-nominal rotation period cases respectively compared with 
the observed peak value of 740~ppm from \citet{dewit2016}.  The twice-nominal rotation 
period model for HD~80606b provides the best match to the amplitude of the observed 
flux variation presented by \citet{dewit2016}. 

In addition to the peak in the planetary flux near periapse, we find that in each 
of our rotation cases secondary peaks in planetary flux occur.  For our half-nominal 
rotation case, the peaks occurs on an interval approximately equal to the assumed 
rotation period of 20.2380~hours.  The second peak in the half-nominal rotation period case, 
occurs slightly before 20.2380~hours from the first peak, which is a result of the 
sub-Earth longitude probing the remnants of a compressional heating event.  The secondary 
peaks in the nominal and twice-nominal rotation period cases occur between one and two days 
after periapse passages are also the result of the sub-Earth longitude crossing the longitude 
region affected by a compressional heating event in each case.  The observed 
phase variations of HD~80606b do no exhibit the predicted `ringing' \citep{cow11, kat13}, 
which was cited by \citet{dewit2016} as further support for the exceptionally 
long planetary rotation period inferred from the observational data ($\sim$93 hours).

\begin{sidewaysfigure}
    \centering
    {\Large Longer Rotation Period $\longrightarrow$}\\
    \includegraphics[width=0.33\textwidth]{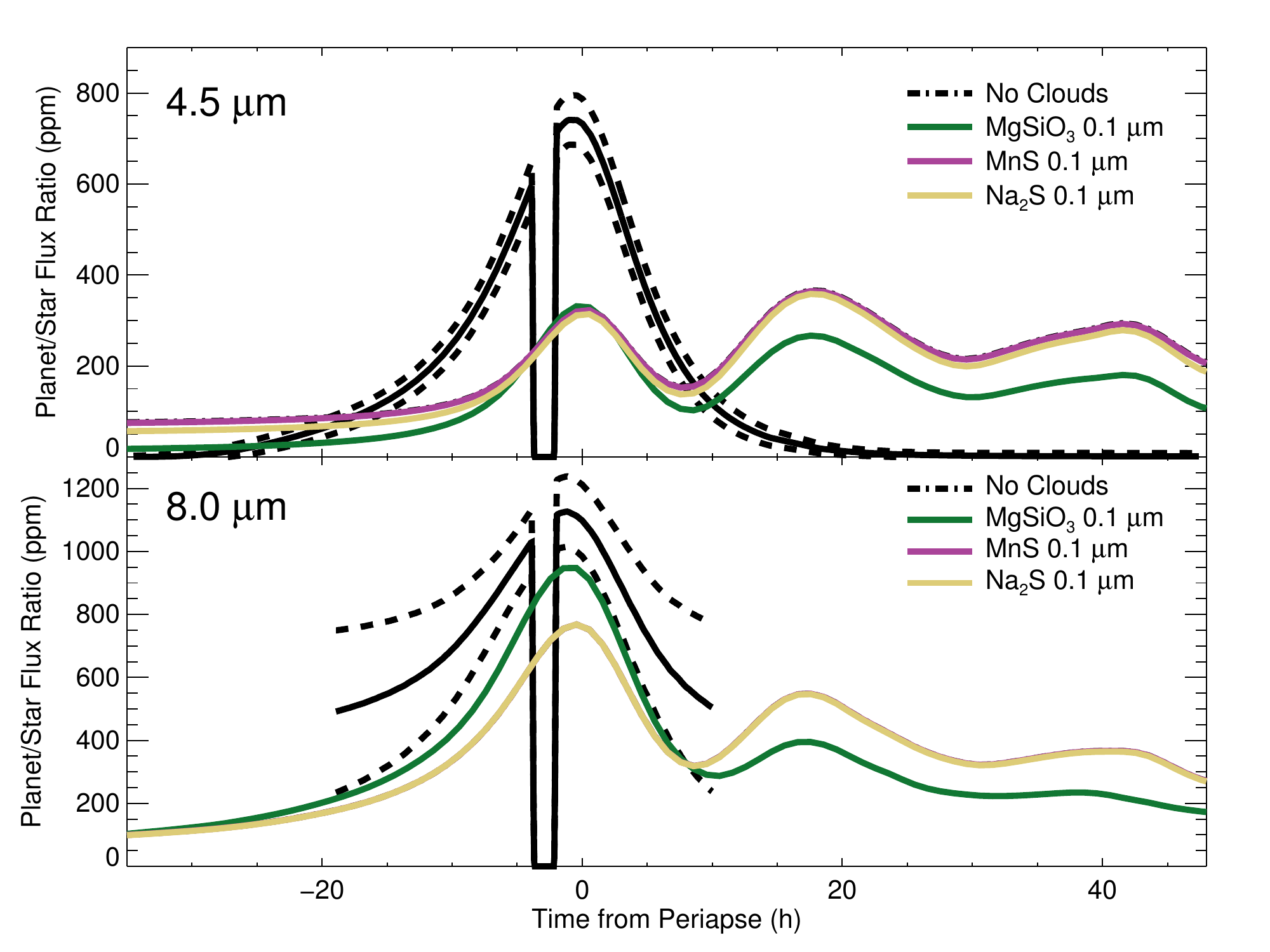}
    \includegraphics[width=0.33\textwidth]{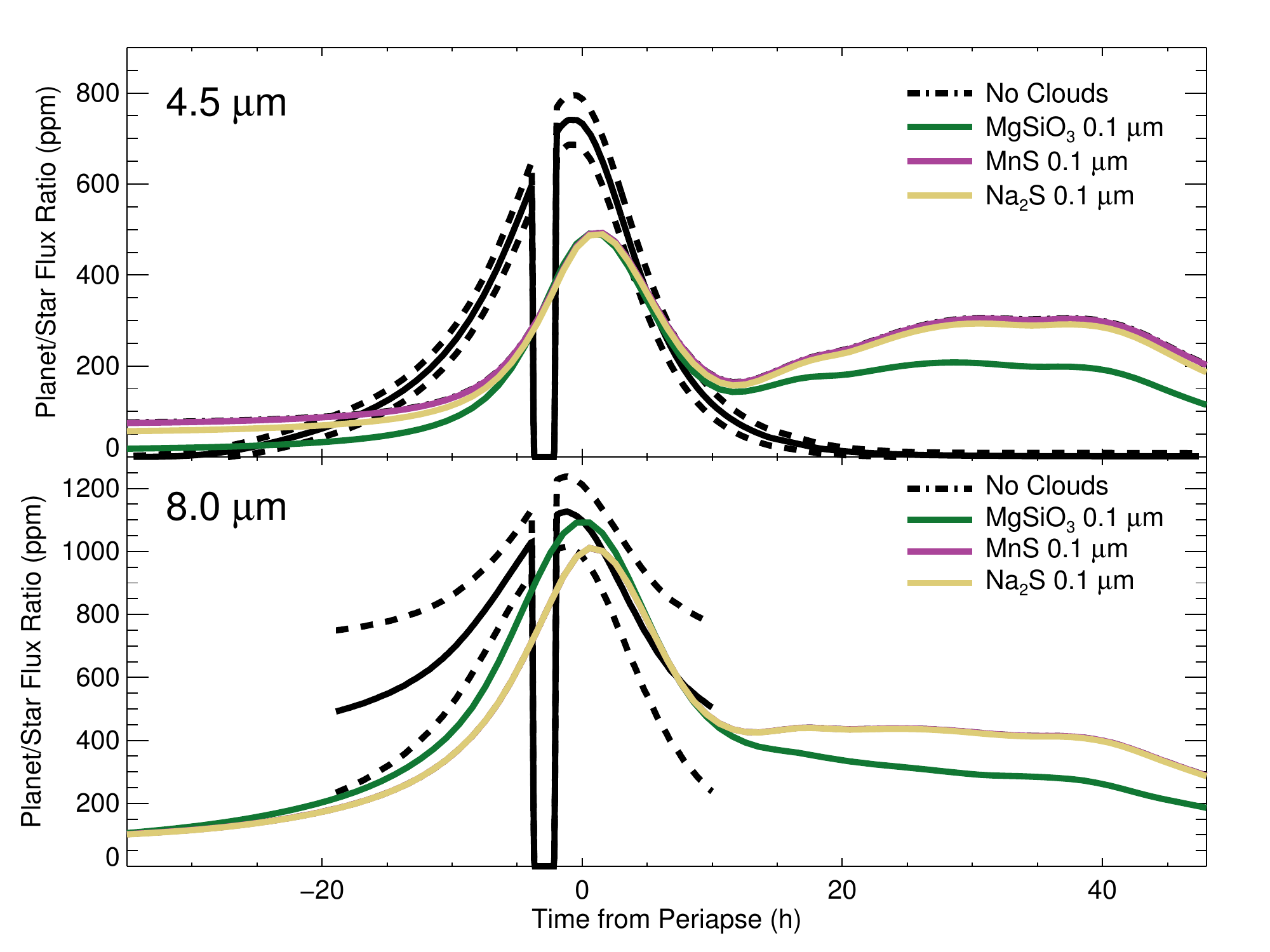}
    \includegraphics[width=0.33\textwidth]{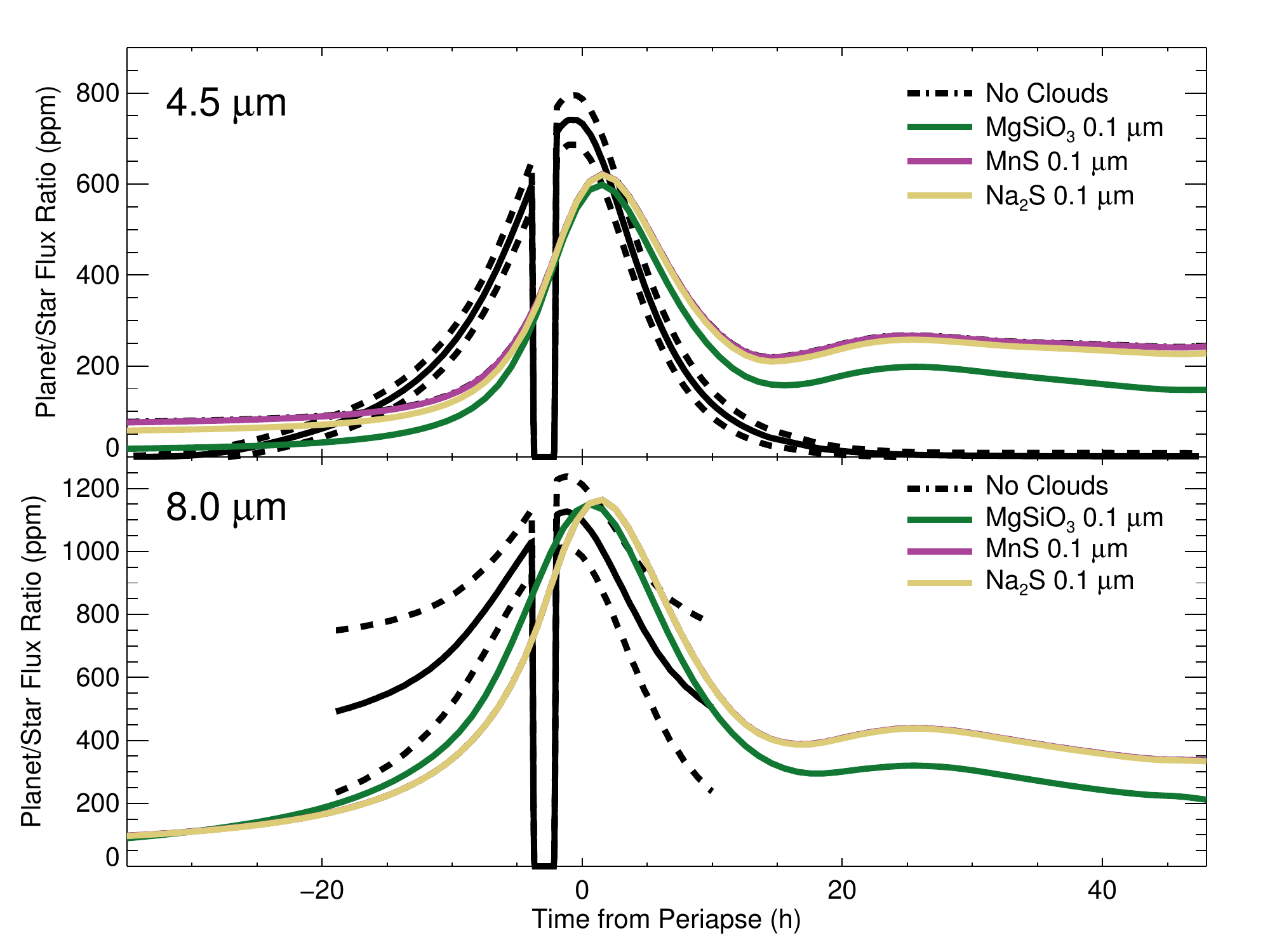}\\
    \includegraphics[width=0.33\textwidth]{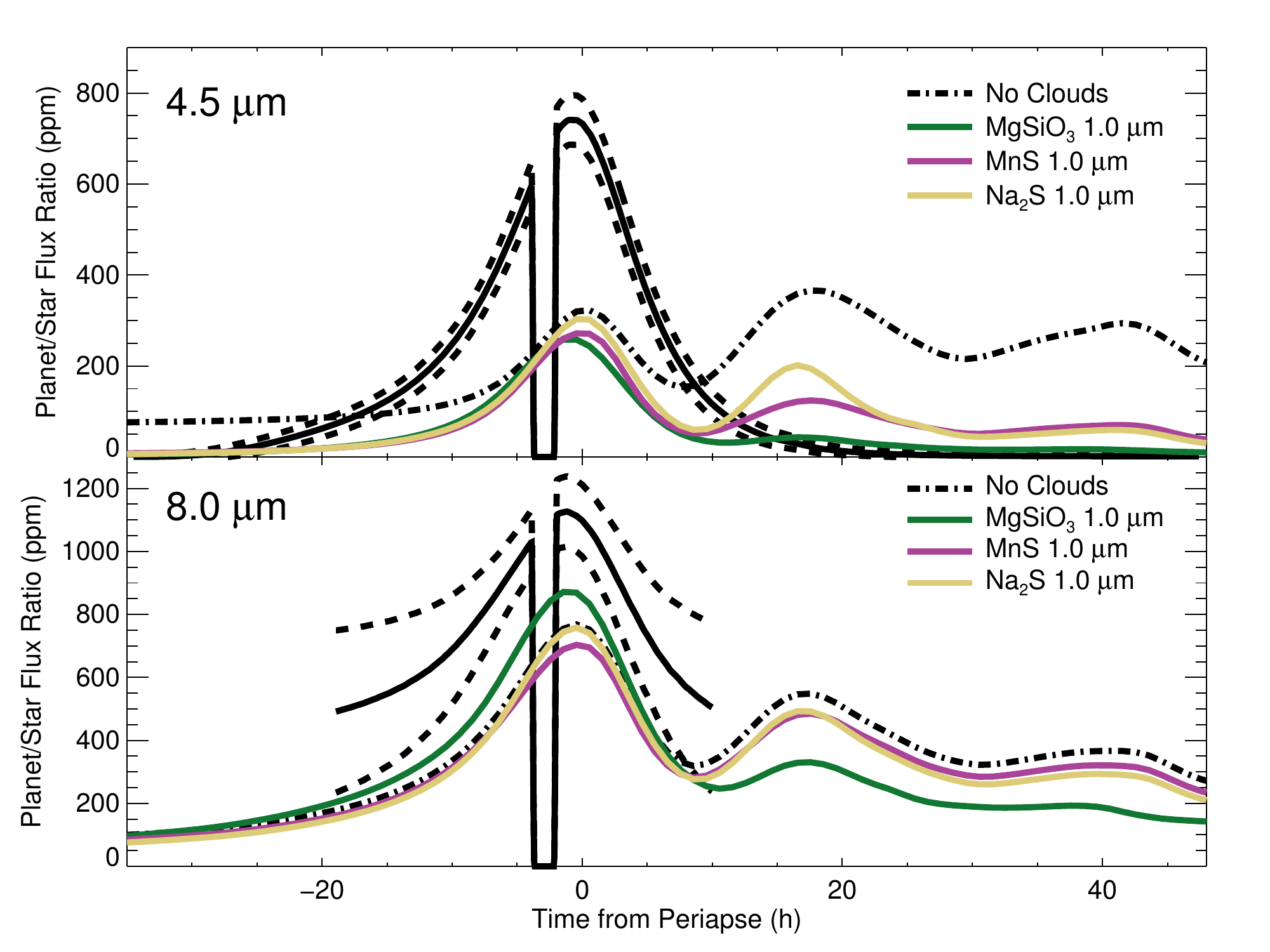}
    \includegraphics[width=0.33\textwidth]{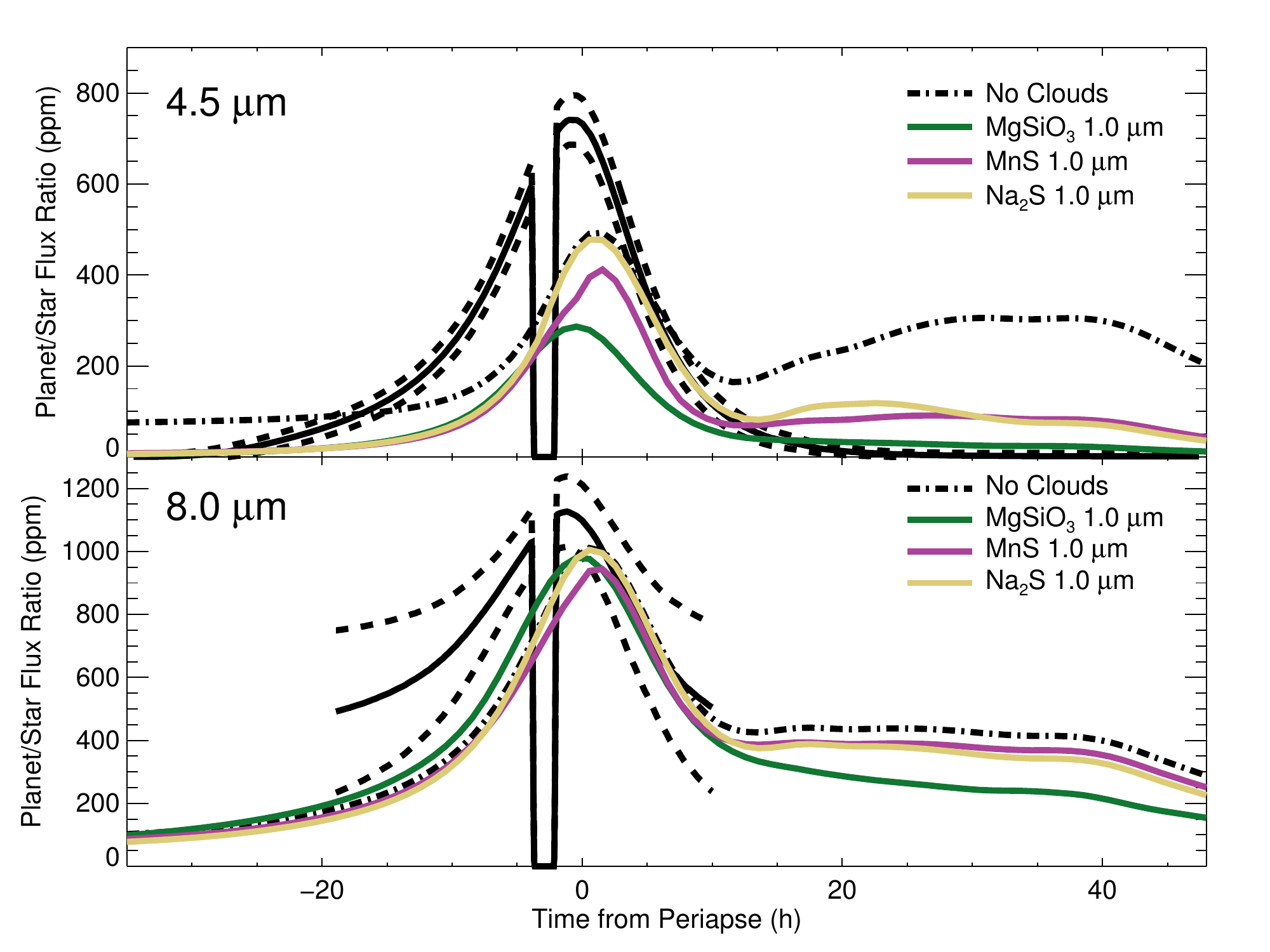}
    \includegraphics[width=0.33\textwidth]{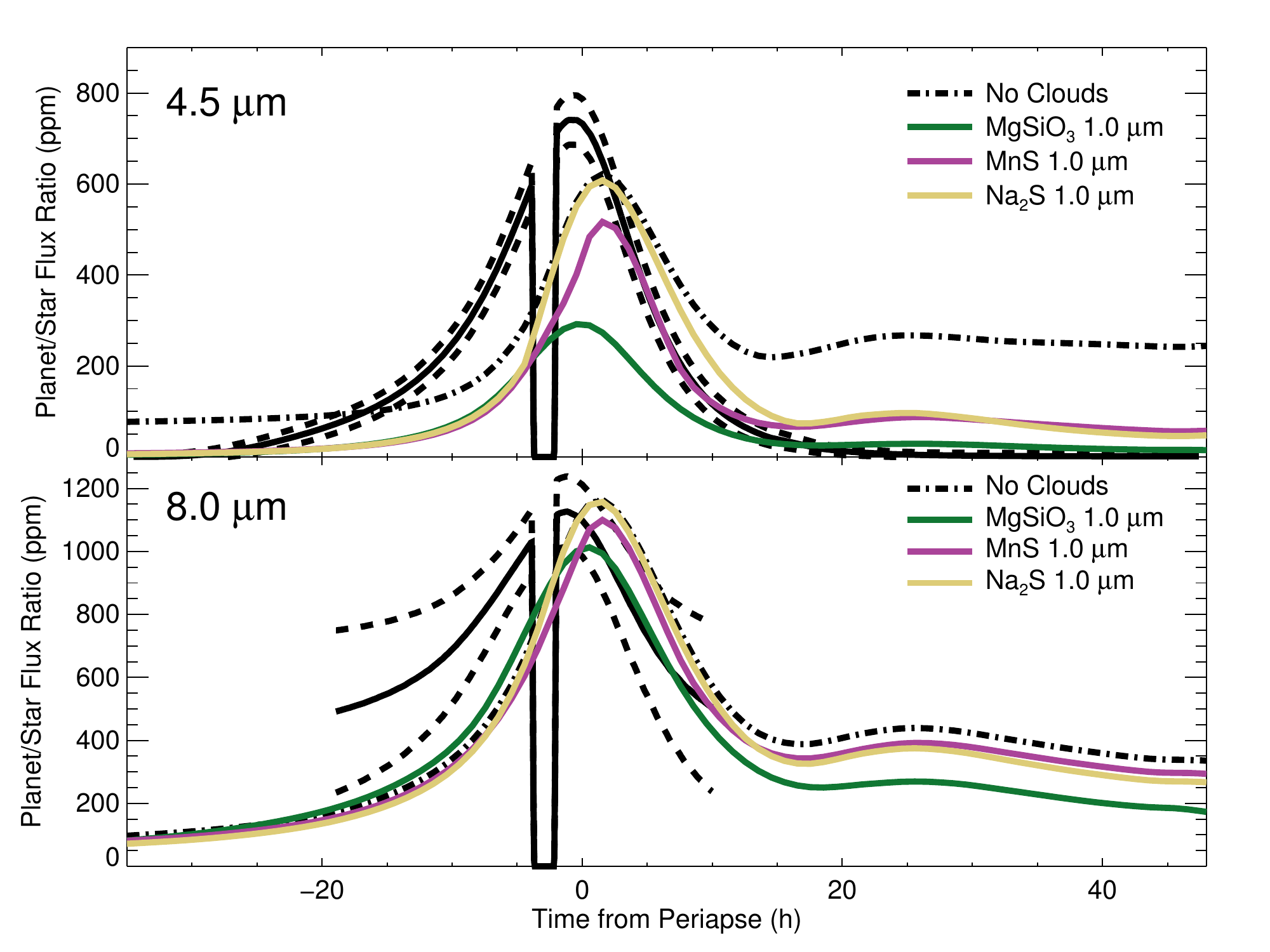}\\
    \includegraphics[width=0.33\textwidth]{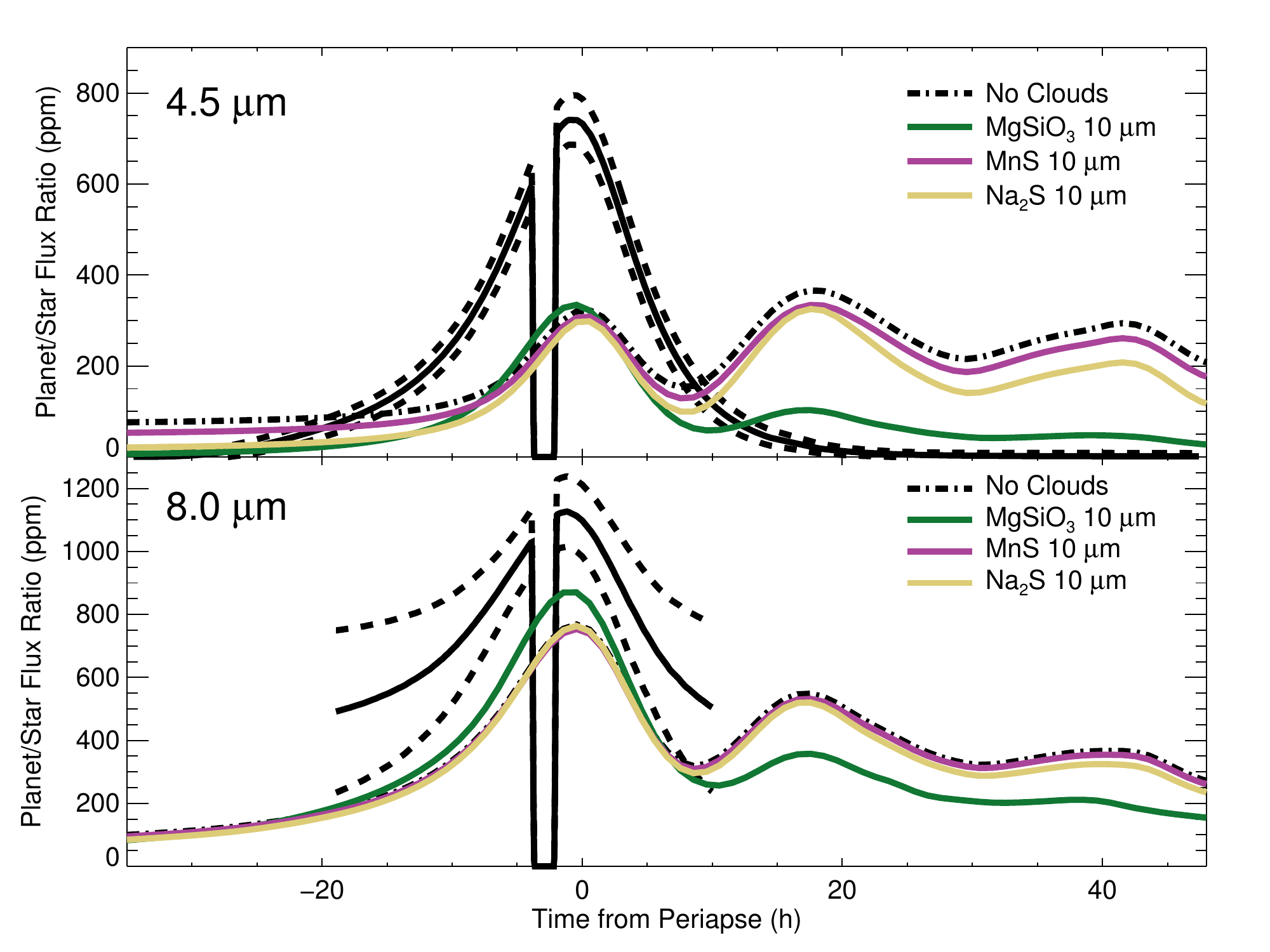}
    \includegraphics[width=0.33\textwidth]{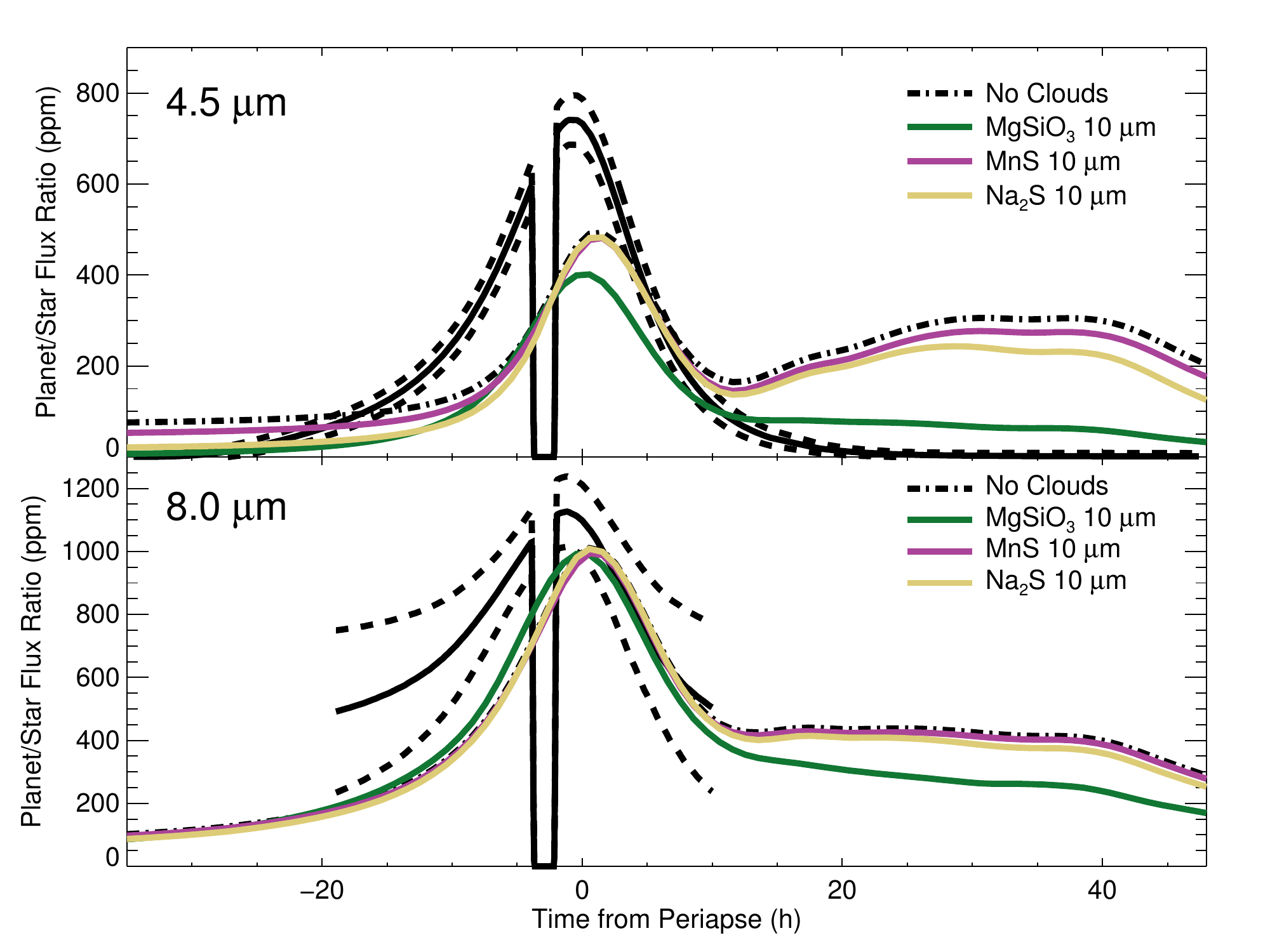}
    \includegraphics[width=0.33\textwidth]{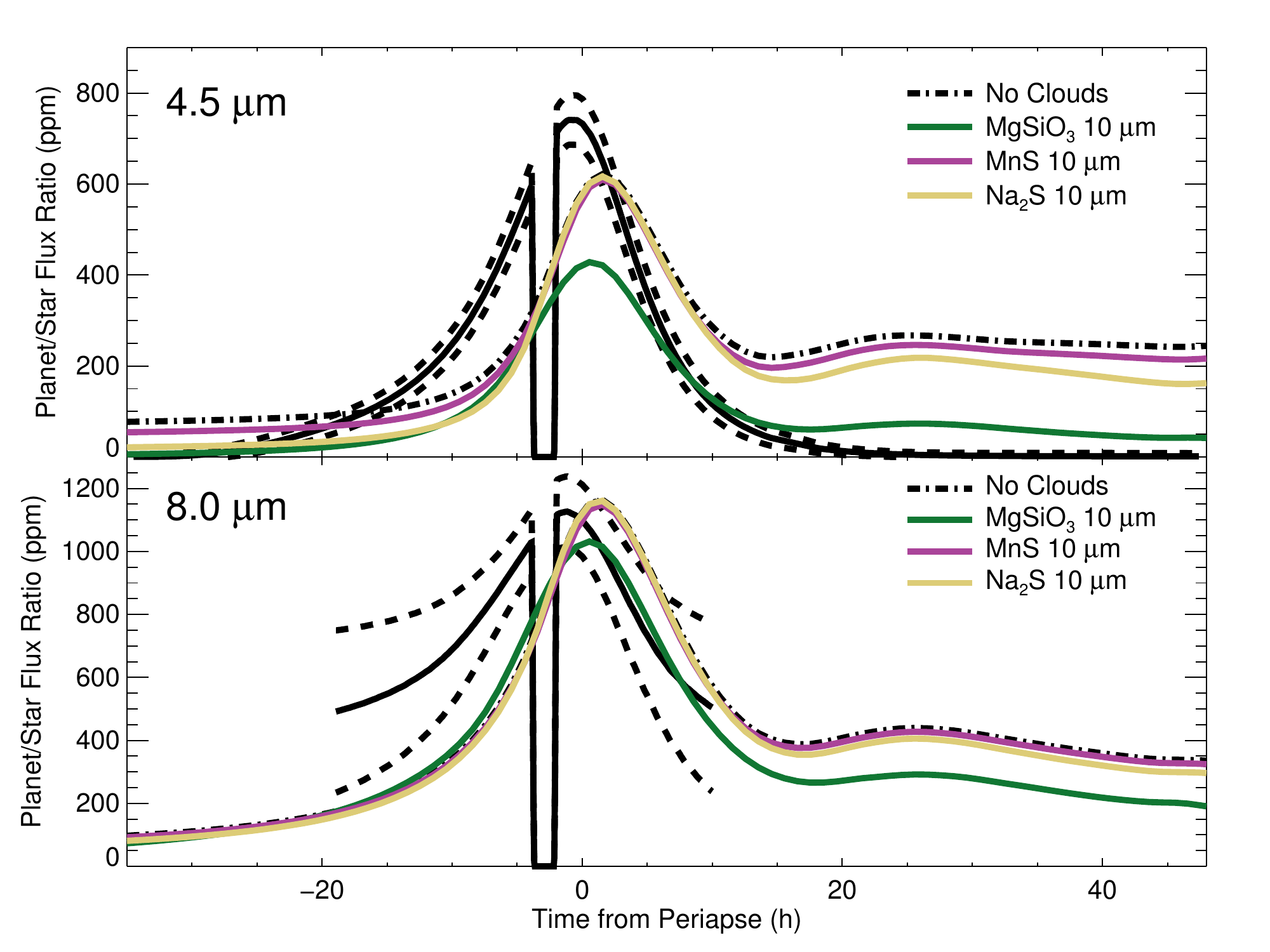}\\
    \caption{Planet/Star flux ratio (in parts per million, ppm) as a function 
    of time from periapse passage as observed with {\it Spitzer} (solid black lines with 
    dashed black line 1$\sigma$ envelopes) and as predicted from our models with 
    half-nominal (left panels), nominal (middle panels), and twice-nominal (right panels)
    with MgSiO$_3$ (green), MnS (magenta), and Na$_2$S (yellow) clouds with particles 
    sizes of 0.1~$\mu$m (top panels), 1.0~$\mu$m (middle panels), 10~$\mu$m (bottom panels).
    Accounting for the presence of clouds does improve the consistency of the model 
    phase curve predictions with the {\it Spitzer} observation in the regions away 
    from periapse passage.}
    \label{hd806_lc_ch2_ch4_cld}
\end{sidewaysfigure}

\subsubsection{Model-Observation Comparisons: Cloudy}

It is possible that clouds will form in or be lofted into the photospheric regions of 
HD~80606b's atmosphere during periapse passage (see Section~\ref{cld_evol}). 
Figure~\ref{hd806_lc_ch2_ch4_cld} 
compares phase variations predicted using our models assuming a range of condensate 
species and particle sizes (see discussion in Section~\ref{cldmod}).  We find that in 
the 8~$\mu$m bandpass assumed particle size (0.1-10~$\mu$m) does not significantly 
change the phase curve predictions for a given cloud species, but in all cases 
the presence of MgSiO$_3$ clouds better suppresses the flux in region outside of 
periapse passage.  A range of of both cloudy and cloud-free models produce 
theoretical phase curves that are consistent with the 8$\mu$m data given the 
uncertainties on the observed flux variations for HD~80606b at that wavelength.  
The 8$\mu$m data are best matched by models that assume either the nominal or 
twice-nominal rotation period for the planet, consistent with the 
rotation period of the planet estimated by \citet{dewit2016} (93$\pm^{85}_{35}$~hours).

The half-nominal rotation period model, while inconsistent with observations, 
does highlight a potentially interesting atmospheric effect that is lacking from our 
other models.  In the half-nominal rotation period model, the presence of 
MnS or Na$_2$S clouds suppresses the flux near periapse passage while 
the presence of MgSiO$_3$ clouds actually enhances the planetary flux near 8~$\mu$m.
This increase in the peak planetary flux when including MgSiO$_3$ clouds 
for the half-nominal rotation period model, as well as the nominal rotation period 
model with 0.1~$\mu$m size MgSiO$_3$ cloud particles, occurs because the significant 
increase in atmospheric opacity causes the 8~$\mu$m photosphere to be 
pushed to shallower pressures that are dominated by the thermal inversion 
seen in the left panels of Figure~\ref{806_templon}.
In general, the presence of MgSiO$_3$ clouds tends to cause the predicted peak in the 
planetary flux to occur earlier than the cloud-free and MnS or Na$_2$S cloud models.  
This behavior in the peak flux timing can also be attributed to enhanced opacity 
causing shallower pressures, with shorter radiative timescales, to be probed.

Much like with the cloud-free cases (Figure~\ref{hd806_lc_ch2_ch4}), the model 
predictions for HD~80606b's phase variations are not able to reproduce the observed 
flux variations at 4.5~$\mu$m.   However, a few key trends can be noted that 
move us toward understanding the atmospheric processes shaping the {\it Spitzer} 
4.5~$\mu$m observation.   First, as postulated in \citet{dewit2016}, clouds 
are necessarily to suppress the flux from the planet in the regions outside of 
periapse passage.  Second, the assumed average particle size 
of a particular cloud species does play a role in the predicted 
phase-curves behaviors at 4.5~$\mu$m, especially for MgSiO$_3$ clouds.
Overall, The twice-nominal rotation model including clouds, in particular 
1~$\mu$m particle size Na$_2$S clouds, best reproduce the observed HD~80606b
phase variations at 4.5~$\mu$m.
The complexities of cloud formation and evolution are not fully captured by 
our current models, but our results indicate that aerosols play a significant 
role in shaping the flux variations observed by {\it Spitzer} for HD~80606b.

\section{Discussion}

Our three-dimensional general circulation models of HD~80606b provide for 
self-consistent treatment of a number of complex atmospheric processes without 
requiring model tuning and/or parametric schemes \citep[e.g.][]{lan08, cow11}.  
However, the complexity of our models does require detailed analysis of the 
circulation patterns, thermo-chemical structure, and cloud coverage that evolves, 
as we've performed in the previous sections.  Here we discuss the key 
atmospheric physical processes that manifest in our models of HD~80606b that 
likely play a significant role in shaping the observed flux variation of 
HD~80606b near the periapse of its orbit.

Two scenarios currently exist to explain the lack of thermal emission from 
HD~80606b outside of periapse passage in the 4.5~$\mu$m {\it Spitzer} observations 
presented in \citet{dewit2016}.
The first scenario is that the visible part of HD~80606b lacks clouds, because they 
could potentially be confined to depths well below the photosphere, and 
also does not have a significant internal luminosity.  The second scenario 
is that HD~80606b has a significant internal luminosity, potentially due to tidal 
heating, and optically thick clouds impeding the internal heat flux to escape at the observed wavelengths. 
Our cloud-free models have a very small internal luminosity 
(T$_{\rm int}=$100~K), yet still predict a significant flux from HD~80606b 
post periapse passage inconsistent with the observations.  Our cloudy models with 
optically thick MgSiO$_3$ clouds provide the best method for suppressing the 
planetary flux outside periapse.  Although none of our models can provide a `perfect' 
fit to current observations, the most plausible scenario for HD~80606b is that an 
optically thick cloud deck is formed via lofting material from depth during periapse 
passage.

In \citet{dewit2016}, the lack of any visible atmospheric `ringing' \citep{cow11, kat13} 
combined with the width of planetary flux `bump' was used to constrain HD~80606b's 
rotation period to be significantly longer than the pseudo-synchronous rotation 
period.  As we've described here, the lack of atmosphere `ringing' could in fact simply 
be due to the development of an optically thick cloud deck post periapse passage.  
However, we find that our models that assume twice the pseudo-synchronous rotation 
period still provide the best explanation of the width and amplitude of the 
flux variations of HD~80606b near periapse.  Our models both with and without 
clouds over-predict the time of peak planetary flux relative to periapse.  We see better 
alignment between the observed and predicted timing of the peak in planetary flux 
when a significant opacity source, such as MgSiO$_3$ clouds, results in 
shallower pressures, with shorter radiative timescales, being probed by a 
given observational wavelength.  In our cloud-free models, the radiative timescale 
we measure near the 4.5 and 8.0~$\mu$m photospheres ($\sim$100-400~mbar) is 
$\sim$8-12~hours, which is longer than the $\sim$4~hour radiative timescale 
measured for HD~80606b by \citet{lau09, dewit2016}. 
The inclusion of MgSiO$_3$ clouds in our models for HD~80606b raise the 4.5 and 8.0~$\mu$m 
photospheres to closer to 10~mbar, where the radiative timescales we measure in our 
models are more consistent with the observed values.
It is likely that the pressures we're probing in generating our synthetic phase 
curves are too deep in the atmosphere, which 
could be compensated for through enhanced opacity from aerosols, a more 
metal-rich atmosphere, and/or a strengthening of the thermal inversion 
that develops in the upper levels of our HD~80606b models (Figure~\ref{806_templon}).

\subsection{Observational Probes Beyond {\it Spitzer}}
Observations at both longer and shorter wavelengths than those probed by {\it Spitzer} 
in \citet{dewit2016} will help to better constrain the physical processes shaping 
HD~80606b atmosphere during its periapase passage.  As seen in Figure~\ref{hd806_lc_ch2_ch4_cld}, 
predicted and observed phase variations for HD~80606b near the periapse of its orbit have 
a strong dependence on both wavelength and assumptions regarding the chemistry and 
bulk rotation period of the atmosphere. In particular, observations at wavelengths 
longward of 8~$\mu$m might better constrain the thermal structure of the planet 
while observations at wavelengths shortward of 4.5~$\mu$m might provide more 
information regarding the formation and evolution of clouds in HD~80606b's atmosphere 
near periastron passage.

\begin{figure*}
\centering
   \includegraphics[width=0.49\textwidth]{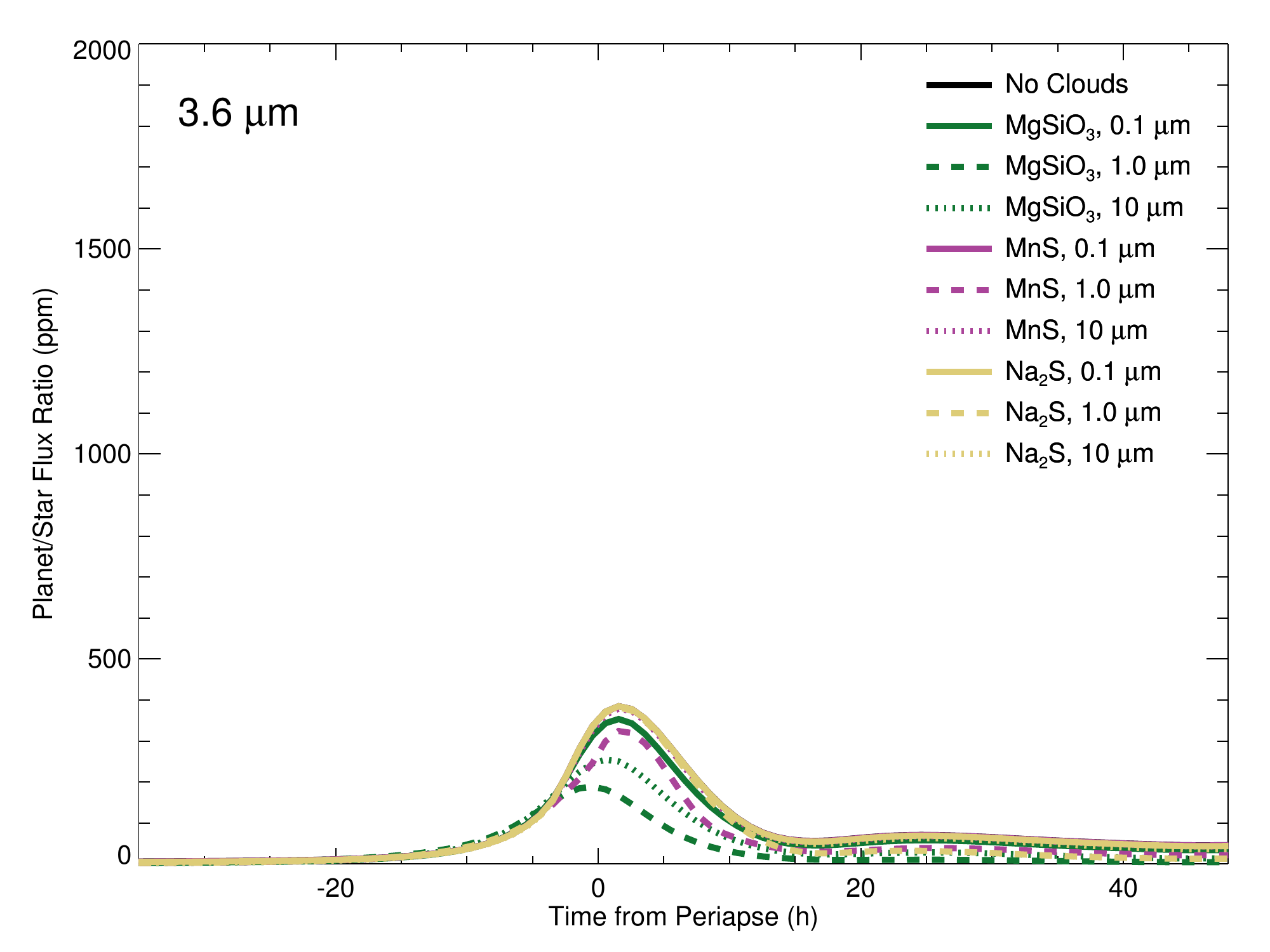}
   \includegraphics[width=0.49\textwidth]{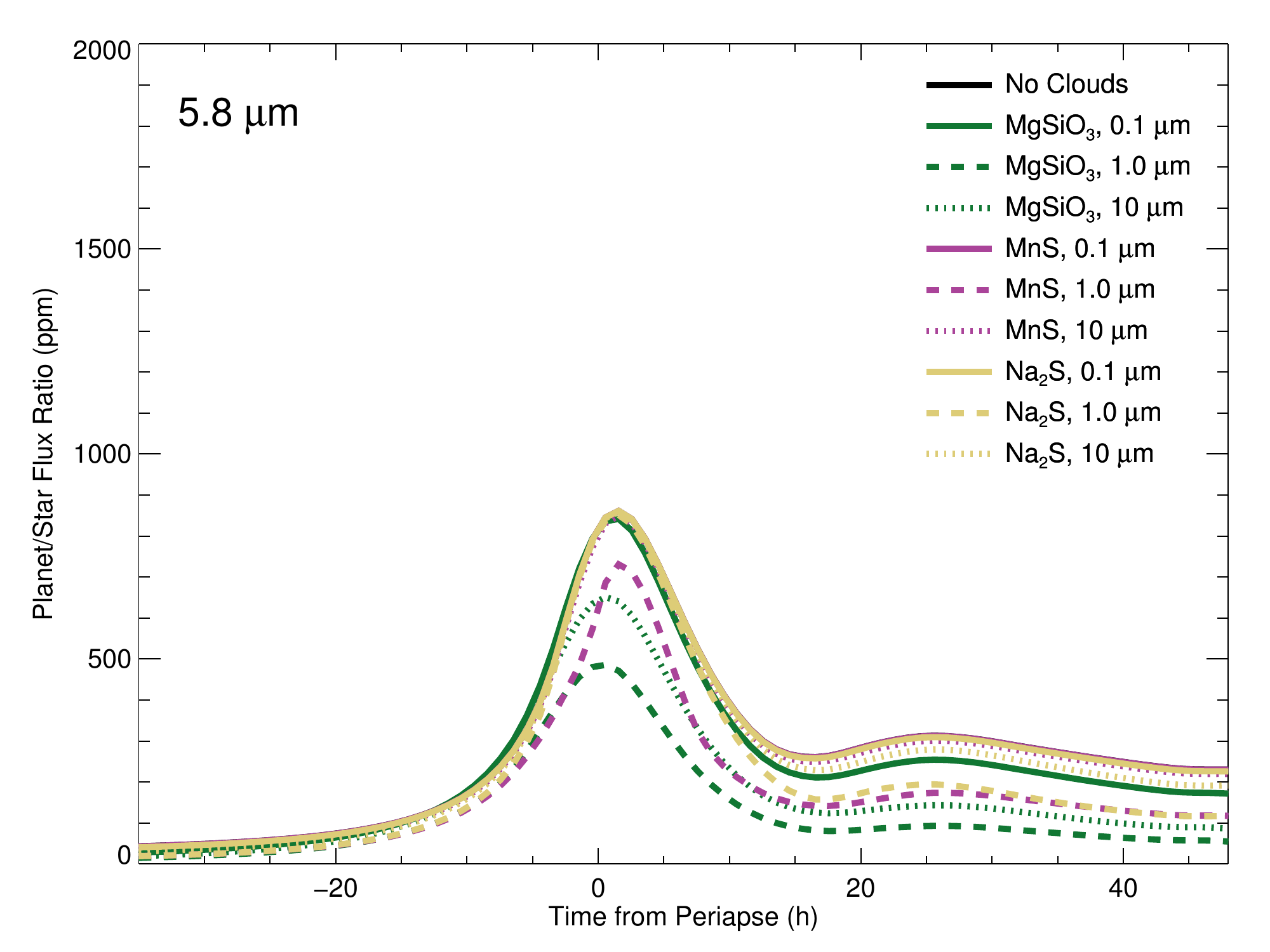}
   \includegraphics[width=0.49\textwidth]{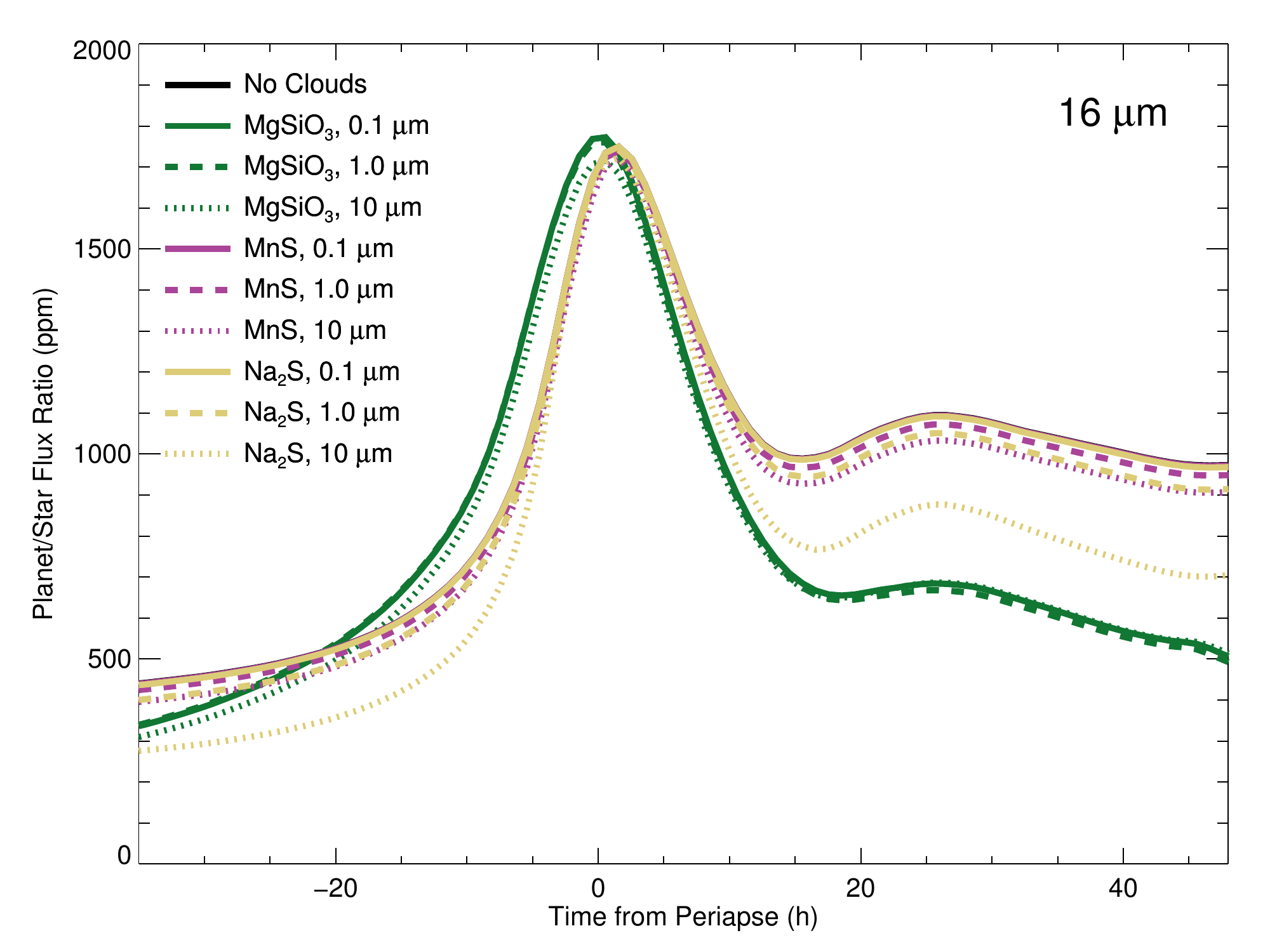}
   \includegraphics[width=0.49\textwidth]{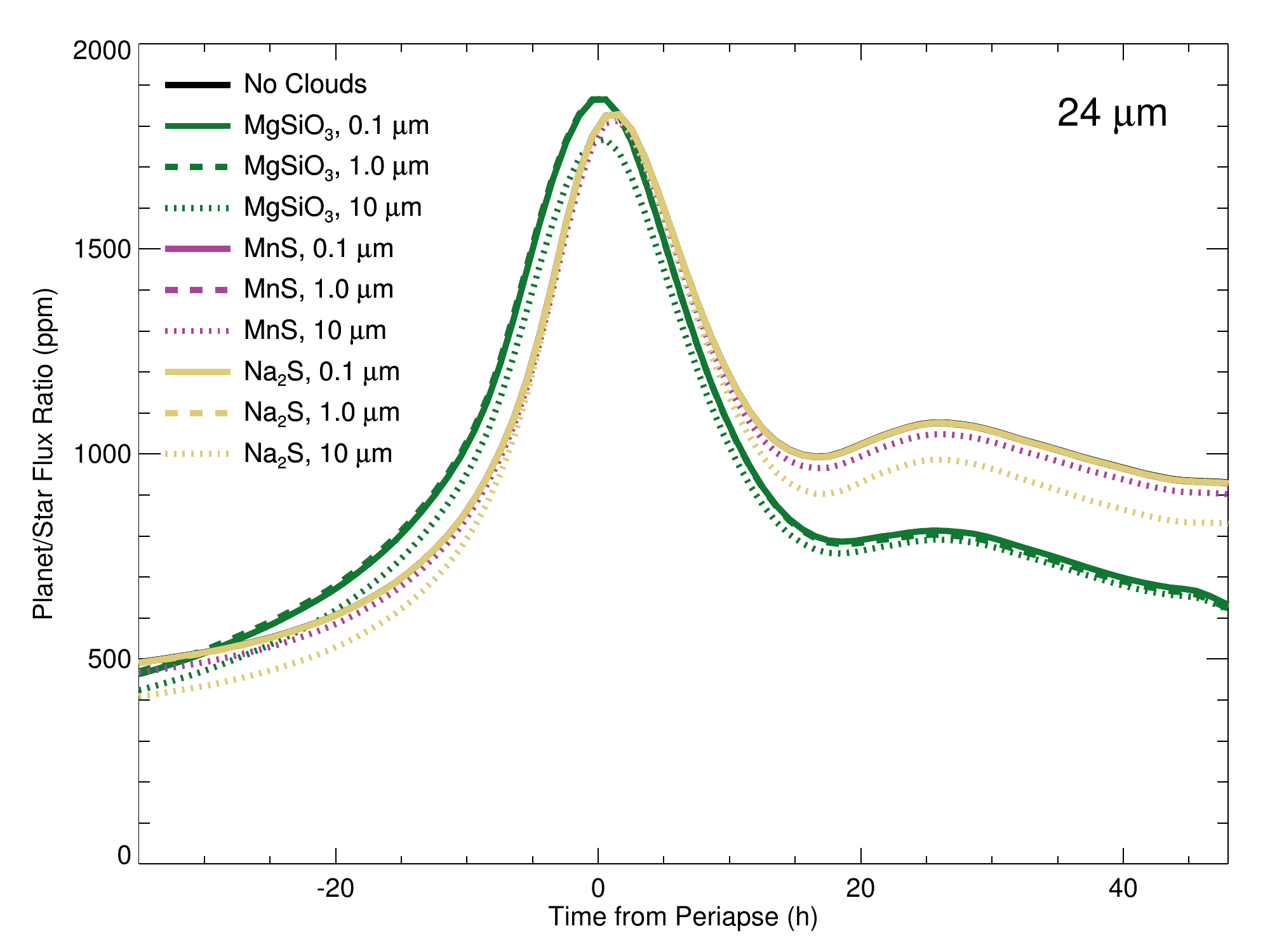}
   \caption{Theoretical Planet/Star flux variations for HD~80606b derived 
   from our twice-nominal rotation period model assuming a range of cloud 
   species and particle sizes for a range of {\it Spitzer} bandpasses.  
   These wavelengths will also be accessible to JWST.}
    \label{fig:jwst}
\end{figure*}

The {\it James Webb Space Telescope} (JWST), 
scheduled for launch in October 2018, will have access to wavelengths longer than 
what is currently available to {\it Spitzer} during the warm phase of its mission.  
Figure~\ref{fig:jwst} shows predicted phase variations using our twice-nominal 
rotation period model for HD~80606b in four {\it Spitzer} bandpasses that will be accessible 
to JWST.  We focus here on the twice-nominal rotation period model because it provides the 
best match to currently available observational data and is in line with the rotation period 
estimate presented in \citet{dewit2016}.  The predicted signal from HD~80606b at 24~$\mu$m 
is a robust 1500 ppm signal with a peak planet/star flux ratio of nearly 2000 ppm.  At 
these long infrared wavelengths, our predicted phase curves do not depend strongly on 
the assumed cloud species or its average particle size, but with 100 ppm precision a 
determination of probable cloud species and rotation period of the planet could be made.  
Similar determinations could be made from 16~$\mu$m observations with stronger variations 
based on composition and average particle size of any clouds present in HD~80606b's atmosphere.
At both 16 and 24~$\mu$m JWST observations will likely reveal this 
secondary `bump', thereby more definitively determining the rotation 
period of HD~80606b, key for understanding tidal dissipation and spin-synchronization for 
exoplanets.

In the warm phase of its mission, {\it Spitzer} can provide exceptionally stable 
long temporal baseline photometery at both 3.6 and 4.5~$\mu$m. Although 3.6~$\mu$m 
observations of HD~80606b were obtained by {\it Spitzer}, the systemmatics present
in the data have prevented a full reduction and analysis.  Figure~\ref{fig:jwst} 
shows that the expected signal from HD~80606b at 3.6~$\mu$m will be on the order of 
a few hundred ppm, which is challenging for {\it Spitzer} but should be achievable 
with JWST.   Both the 3.6~$\mu$m and 5.8~$\mu$m wavelength spectroscopic 
regions could provide valuable insights concerning both the composition and 
average particle size of any clouds present in HD~80606b, although some 
ambiguity may still exist.  JWST will also provide the potential for 
spectroscopic exploration of HD~80606b, which could provide further insights 
into the chemistry, and its relevant timescales, at work in this intriguing atmosphere.

\begin{figure*}
\centering
   \includegraphics[width=0.49\textwidth]{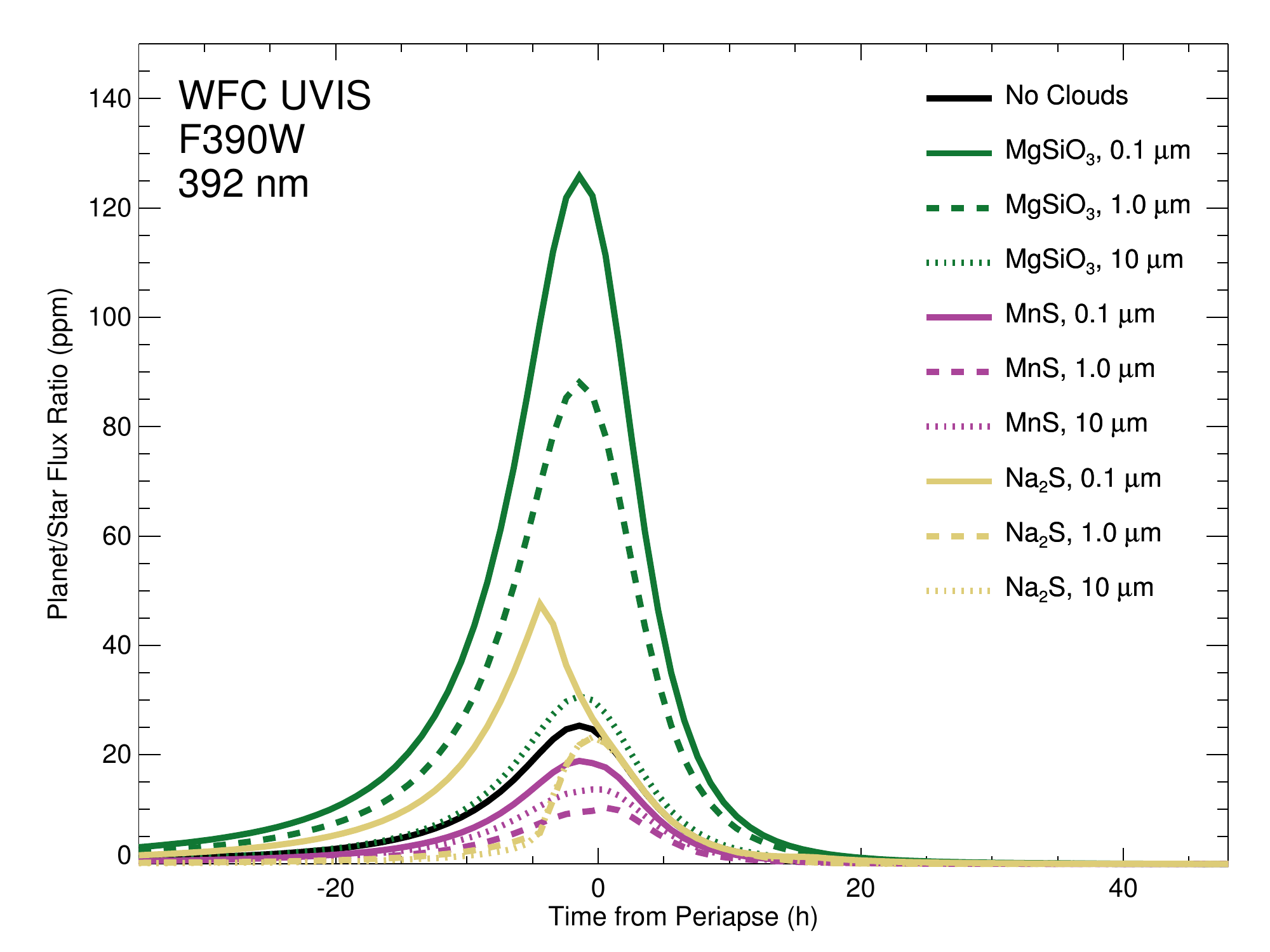}
   \includegraphics[width=0.49\textwidth]{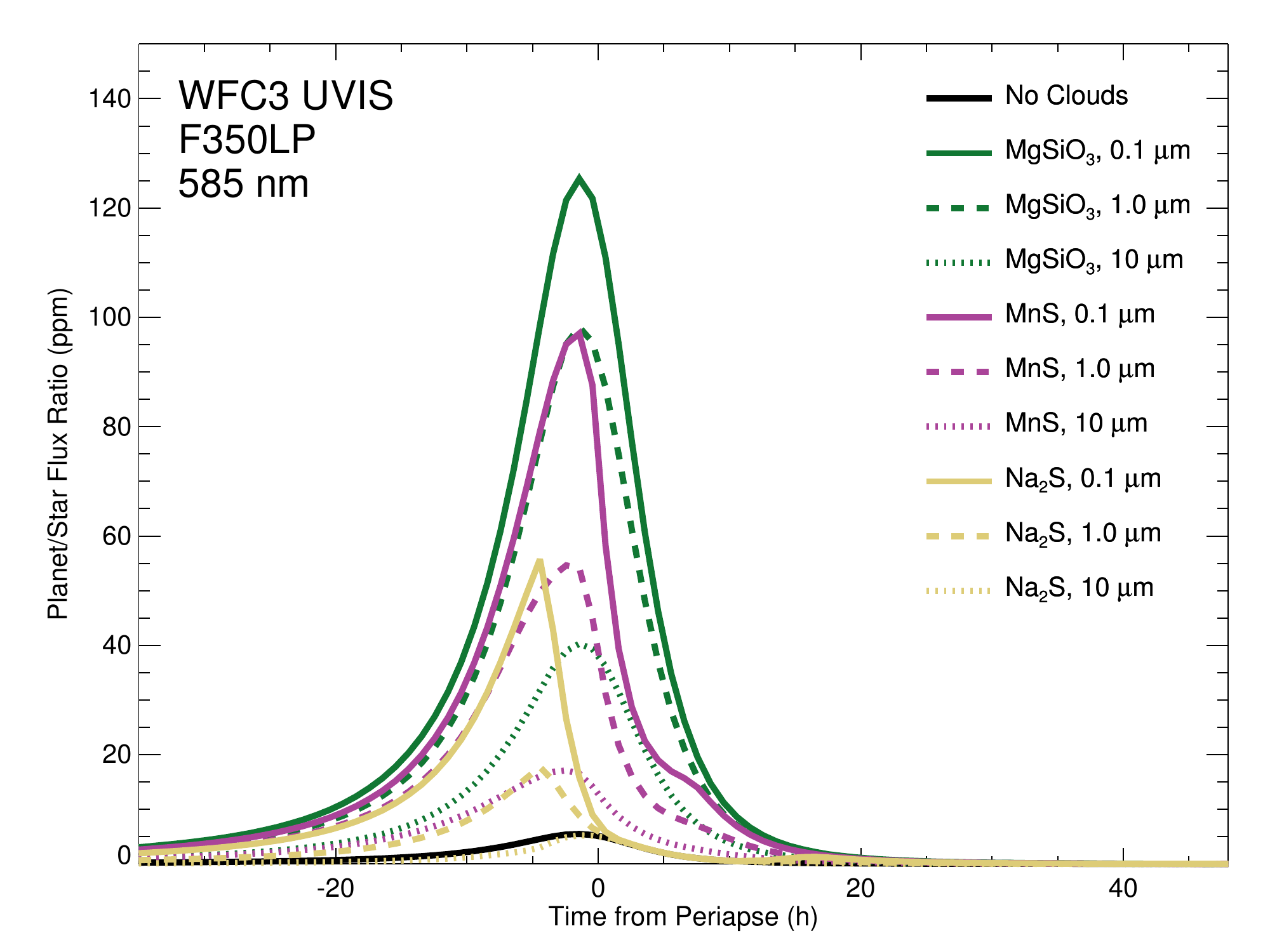}
   \includegraphics[width=0.49\textwidth]{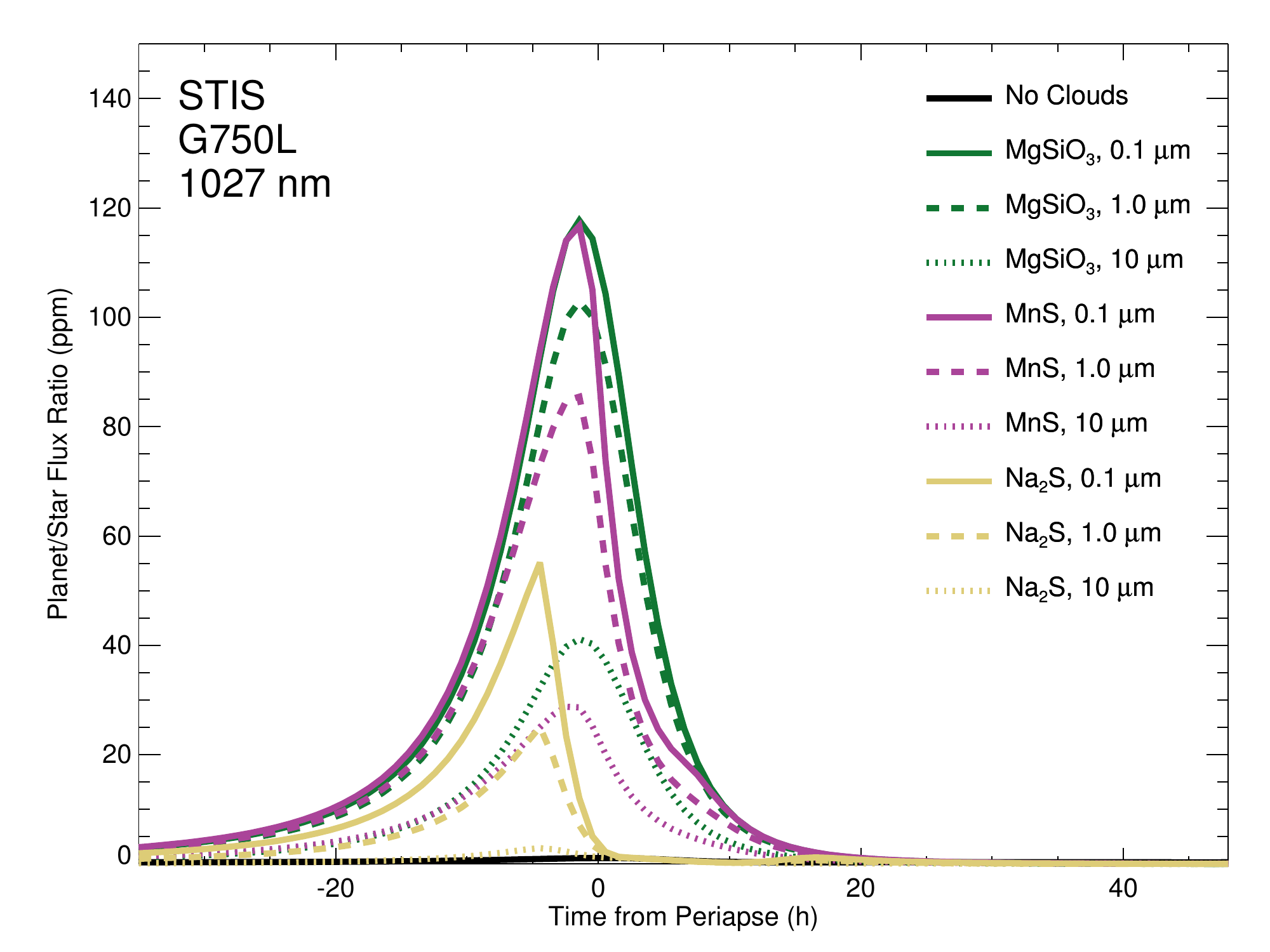}
   \includegraphics[width=0.49\textwidth]{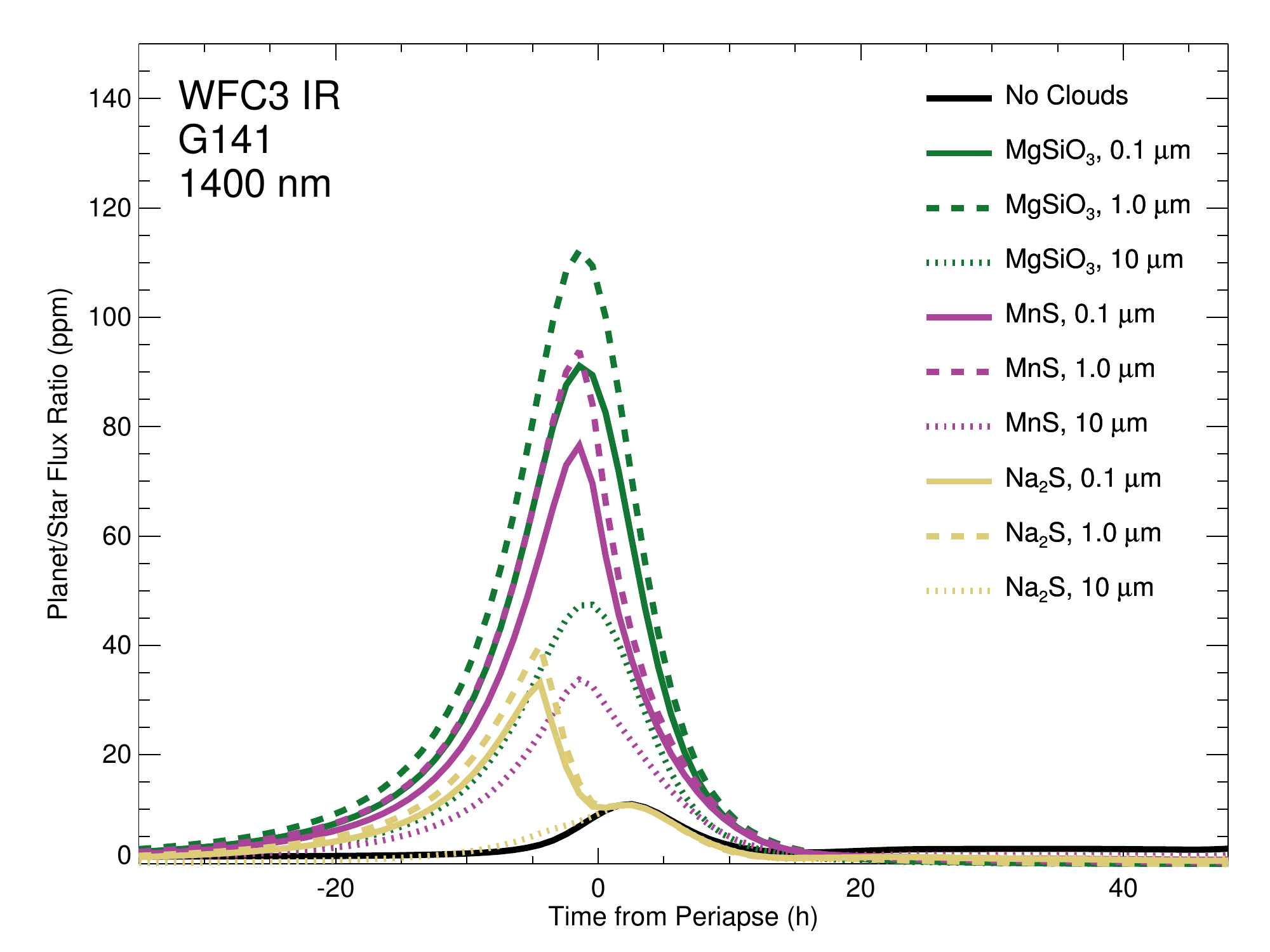}
   \caption{Theoretical Planet/Star flux variations for HD~80606b derived 
   from our twice-nominal rotation period model assuming a range of cloud 
   species and particle sizes for a range of potential HST observations.  
   The no cloud (black lines) theoretical phase curves represent contributions 
   due to thermal emission/Rayleigh scattering.}
    \label{fig:hst}
\end{figure*}

There also exist the potential for observations of HD~80606b near 
its periapse passage with the {\it Hubble Space Telescope} (HST). 
Because HST is in low Earth orbit, it is unable to provide the 
semi-continuous temporal coverage for phase curve observations 
that {\it Spitzer} provides.  Despite this fact, a handful of 
successful exoplanet phase curve studies have been executed with 
HST \citep[e.g.][]{ste14}.  HST provides for both photometric and 
spectroscopic observations spanning the far-ultraviolet to the 
near-infrared.  Figure~\ref{fig:hst} presents a few HST instrument 
and filter/grism combinations that have the potential to 
provide high signal-to-noise observations of HD~80606b near 
secondary eclipse and periastron passage in the optical to near-infrared 
where scattered light from the planet should dominate the signal.  
These predictions for HST were derived using our twice-nominal rotation 
period HD~80606b model for consistency with observations at longer wavelengths, 
but we see only small variations in our predictions based on assumed 
rotation period at HST wavelengths.

The signal amplitudes predicted from our models at optical to near-infrared 
wavelengths are considerably smaller compared with observations at infrared 
wavelengths (Figure~\ref{fig:jwst}), but they have the potential to 
provide a wealth of critical information concerning the formation 
and evolution of clouds in HD~80606b's atmosphere.   The cloud-free 
model predictions presented in Figure~\ref{fig:hst} have almost 
vanishingly small amplitudes, but they represent contributions from 
the planet's thermal emission (near-infrared) and Rayleigh scattering (blue-visible) 
at those relevant wavelengths.  The presence of clouds would likely dramatically 
increase the amplitude of any signal from HD~80606b that could be 
observed with HST.  

In the HST bandpasses presented in Figure~\ref{fig:hst}, differences in 
signals produced by particular cloud species and particle sizes are
readily apparent. 
The amplitude of the predicted HD~80606b phase curve at wavelengths probed by HST
depends on a combination of particle size and the particle composition, 
the case of silicate (MgSiO$_3$) clouds with a particle size of $0.1\mu$m 
presenting the most favorable signal at optical to near-infrared wavelengths. 
The shape of the observed flux variations can give insights into the cloud composition. 
Clouds that would evaporate during periapse passage, such as Na$_2$S clouds, lead to a peculiar triangular-shaped phase curve whereas clouds species that do not evaporate near periapse, 
such as MgSiO$_3$, have a more symmetrical, bell-shaped phase curve 
(see also Figure~\ref{fig::FluxMaps}).  Observations with HST at one or 
more of the bandpasses presented in Figure~\ref{fig:hst} of HD~80606b near 
the periapse of its orbit would provide for the most robust interpretation 
of cloud properties in its atmosphere.

With a V-band magnitude of 9.0, observations of the HD~80606 system with WFC3-UVIS 
should yield a 55~ppm precision per 51~sec frame. Therefore, a program targeting HD~80606b's 
flux modulation around periapse passage at visible wavelengths with HST 
could allow to disentangle between MgSiO$_3$/MnS clouds, Na$_2$S clouds, and a cloud-free 
atmosphere. It is possible that aerosols not considered in this study (e.g. hazes), would 
produce a phase curve signature distinct at visible wavelengths.  An observation window of 
22~hours centered around the periapse passage would yield a sufficient precision on the phase curve shape to constrain the cloud composition. In particular, it should yield a detection of 1$\mu$m size 
MgSiO$_3$/MnS and Na$_2$S clouds at the $\sim$13$\sigma$ and $\sim$7$\sigma$ levels, respectively. 

\section{Conclusions}

Our atmospheric models of HD~80606b present a rich area for investigating 
atmospheric circulation and cloud formation under extreme time-variable 
forcing conditions.  Here we've 
shown that bulk rotation period assumptions can dramatically alter 
the atmospheric circulation of a planet like HD~80606b, especially 
near the periapse of its orbit.  We have shown that a number of 
cloud species could play a significant role in shaping the flux 
variations of HD~80606b near the periastron of its orbit.
In comparing theoretical phase curves 
for HD~80606b derived from our atmospheric models, we find that our 
twice-nominal rotation period model with clouds included as an opacity 
source best match the {\it Spitzer} observations presented in \citet{dewit2016}. 

The \citet{dewit2016} study used a semi-analytic cloud-free model to constrain the 
rotation period, radiative timescale, and baseline brightness temperature
of HD~80606b from {\it Spitzer} 4.5 and 8.0~$\mu$m observations.
\citet{dewit2016} concluded that the lack of significant flux from the 
planet outside of the periapse passage region was an indication that 
HD~80606b's possesses a small internal luminosity.  We find that the lack of 
planetary flux outside of periapse, combined with 
the timing of the peak planetary flux relative to periapse passage, 
is more likely an indication of 
the presence of an optically thick cloud deck composed of material  
lofted from deep within HD~80606b's atmosphere during periapse passage.  
This cloud formation and lofting mechanism would work most efficiently 
in scenarios where HD~80606b has significant internal flux, which 
provides an important constraint on the tidal and internal evolution 
of the planet.

In future work, we will include more self-consistent treatment of cloud 
formation and evolution in our models and an exploration of a larger 
range of values for the internal temperature and rotation period 
of HD~80606b.  Observations at both shorter and longer wavelengths 
than those currently available for HD~80606b with JWST and HST would 
provide more robust constraints on the physical processes at work in 
HD~80606b's atmosphere.  Such atmospheric constraints can then be 
extended to our global understanding of exoplanet atmospheres and 
allow us to further refine our models as we begin to explore cooler 
and smaller worlds beyond our solar system.

\acknowledgments

NKL thanks H.~Wakeford, K.~Stevenson, J.~Fraine, J.~Valenti, S. H\"orst, and M.~Lewis for their support during the 
writing of this manuscript.  This work was performed in part under contract with the California 
Institute of Technology (Caltech) funded by NASA through the Sagan Fellowship Program executed by 
the NASA Exoplanet Science Institute.

\bibliography{hd806_circulation}

\end{document}